\documentclass{article}
\usepackage{ijcai26}

\usepackage{times}
\usepackage{soul}
\usepackage{url}
\usepackage[hidelinks]{hyperref}
\usepackage[utf8]{inputenc}
\usepackage[small]{caption}
\usepackage{graphicx}
\usepackage{amsmath}
\usepackage{amsthm}
\usepackage{booktabs}
\usepackage{algorithm}
\usepackage{algorithmic}
\usepackage[switch]{lineno}

\usepackage{wrapfig}
\usepackage{comment}
\usepackage{subcaption}
\usepackage{booktabs}
\usepackage{multirow}
\usepackage{newfloat}
\usepackage{listings}

\usepackage{adjustbox} 
\usepackage{booktabs}
\usepackage{multirow}
\usepackage{xcolor}

\usepackage{mathtools}
\usepackage{amsthm}
\usepackage{amssymb}
\usepackage{xspace}
\newcommand{\noise}{\text{noise}}
\newcommand{\PRF}{\text{PRF}}
\newcommand{\softplus}{{\sf softplus}}
\usepackage{packages/macros-common}
\usepackage{packages/macros-project-specific}
\usepackage{packages/color-edits}
\addauthor{ss}{magenta} 
\addauthor{kr}{red} 
\addauthor{eb}{purple} 
\DeclareCaptionStyle{ruled}{labelfont=normalfont,labelsep=colon,strut=off} 
\floatstyle{ruled}
\newfloat{listing}{tb}{lst}{}
\floatname{listing}{Listing}

\usepackage{tikz}
\usetikzlibrary{positioning, arrows.meta, shapes.misc}

\usepackage{caption}

\usepackage{soul}

\newtheorem{theorem}{Theorem}

\title{Can Watermarking Techniques Help Prevent LLM Model Stealing?}

\author{
Elette Boyle$^1$
\and
MohammadTaghi Hajiaghayi$^2$\and
Keivan Rezaei$^{2}$\and
Suho Shin$^2$\And
Amos Stern$^1$
\\
\affiliations
$^1$Reichman University\\
$^2$University of Maryland\\
}

\begin{document}

\maketitle

\begin{abstract}
    Model stealing attacks have recently been introduced, enabling the extraction of precise information from black-box commercial language models. 
In this work, we propose defense methods against a recent attack of \cite{carlini2024stealing} and extensions for extracting the hidden layer dimension of production language models.
Our methods are inspired by watermarking techniques that perturb the logits layer of these models to prevent such attacks.
We provide empirical experiments demonstrating the effectiveness of the proposed defense versus model quality degradation across various configurations, and propose an effective defense against such attacks while preserving model utility.
\end{abstract}

\section{Introduction}
Commercial language models (LMs) such as Gemini, GPT-4, and Claude~\cite{achiam2023gpt,team2023gemini,anthropic}, which are publicly accessible, require substantial time and resources for training. 
Consequently, details about these models, including training data, methodologies, architectures, and parameters, are typically not disclosed. However, recent reports highlight the risks of {\em model stealing}, where unauthorized actors attempt to extract proprietary knowledge.  For example, recent article~\cite{reuters2025deepseek} indicate the possibility of knowledge being extracted from OpenAI's API output to produce a competing product.


These models operate through API access, allowing users to input prompts and receive predictions.
\cite{carlini2024stealing} proposed techniques to extract precise information from various black-box LMs, such as their hidden state dimensions and final projection layers.
This raises serious concerns that extension of these attacks could potentially reveal even more sensitive information.

Specifically, \cite{carlini2024stealing} present a scalable approach targeting production LMs.
Unlike earlier model stealing methods~\cite{carlini2020cryptanalytic,carlini2024polynomial,rolnick2020reverse}, it does not depend on specific activation functions in the model or network structures, making it both versatile and a significant concern. This highlights the need for robust defenses against such attacks.
The method described in~\cite{carlini2024stealing} leverages a model’s API to extract logit vectors of next-token-prediction on a set of prompts.
These logits are then used to construct a matrix, whose properties are used for extracting hidden dimension, and later the last projection layer in a black-box LMs. While the specific API queries that enable logit vector reconstruction in the~\cite{carlini2024stealing} attack have since been removed from major models,\footnote{Still, one may recover the entire logits using a binary search-like approach~\cite{morris2023language}.} their attack indicates a clear danger and potential vulnerability.
In particular, the hidden latent dimension of the target model is a critical component for \cite{carlini2024stealing},
as it serves as the foundation for stealing the model’s last layer.




\begin{figure}
  \centering
  \includegraphics[width=0.45\textwidth]{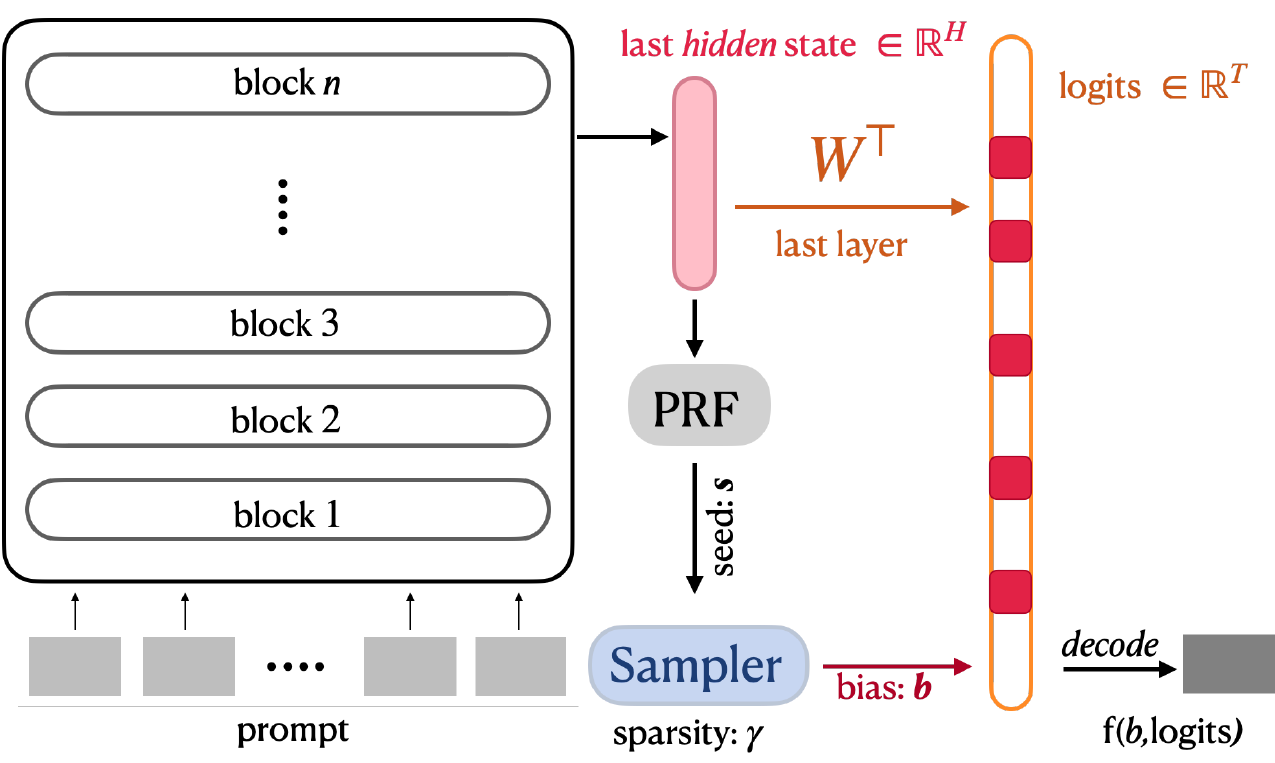}
  \caption{Our framework of perturbing logits to prevent dimension-extraction model stealing attacks.
   \vspace{-.2in} }\label{fig:tease}
\end{figure}

In this work, we propose a structured perturbation of logits to counter rank-estimation attacks, inspired by watermarking techniques~\cite{kirchenbauer2023watermark}. Unlike prior methods that sample and add independent noise, we seed targeted perturbations using a private hash of the model’s hidden latents. Our approach significantly increases the complexity of attacks attempting to break the defense, while preserving the model's quality.
Our overall framework is given in Figure 1.

We remark that watermarking has been used for post-hoc detection of behavioral cloning~\cite{zhao2023protecting}. Our method takes a very different approach, proactively disrupting rank-based extraction attacks of architectural details such as hidden dimension. (See Related Work in Appendix~\ref{sec:rel}.)

\subsection{Summary of results}

We study rank-estimation attacks based on principal component analysis (PCA) and its variants, and develop defense mechanisms that are robust to \emph{any such attack}.
We first summarize the main attack methods considered, then describe our defenses.
Further details appear in Sections~\ref{sec:rank} and~\ref{sec:defense}.

\subsubsection{Attack methods}
We briefly describe the main classes of attacks, noting that we additionally consider stylized variants tailored to specific defenses.

\paragraph{PCA}
Following \cite{carlini2024stealing}, the attacker constructs a matrix whose columns are logit vectors obtained from controllable prompts, and applies PCA to estimate the numerical rank, corresponding to the hidden dimension of the LLM.
PCA is robust to small additive perturbations: if the observed matrix is $A+E$ with $\|E\|_{\mathrm{op}}$ small, Weyl’s inequality guarantees recoverability of the true rank.

\paragraph{PCA with averaging}
In practice, PCA can be sensitive to noise with large spectral norm, such as i.i.d.\ Gaussian noise.
In the additive-noise setting, an attacker can mitigate this by querying the same prompt multiple times and averaging the resulting matrices, thereby reducing the variance and spectral norm of the noise and enabling rank recovery.

\paragraph{Robust PCA}
If the defender injects sparse but large-magnitude noise, classical PCA may fail.
However, robust PCA \cite{candes2011robust} shows that exact rank recovery remains possible under reasonable conditions via convex optimization, even with gross but sparse corruptions.

\subsubsection{Defense mechanisms}
A naive defense is to add independent noise to each logit coordinate.
This approach is undesirable due to non-reproducibility, and is fundamentally vulnerable to averaging attacks, which can effectively remove independent noise.

To counter this, the injected noise should be \emph{prompt-dependent} yet \emph{deterministic}: identical prompts should always induce the same perturbation, preventing averaging, while different prompts induce different noise.

\paragraph{Prompt-dependent noise via pseudorandom functions}
We generate noise deterministically using a cryptographic pseudorandom function (PRF) applied to the model’s last hidden states, with a secret key held by the model owner.
This ensures reproducibility while appearing random across prompts.
This idea is inspired by LLM watermarking \cite{kirchenbauer2023watermark}, but unlike token-based seeding, our construction avoids trivial attacks where an adversary forces identical seeds across prompts.

\paragraph{Additive versus nonadditive noise}
Given prompt-dependent noise, a natural approach is additive perturbation of logits.
We empirically study additive noise across distributions, finding it often effective but potentially vulnerable to sophisticated averaging-style attacks.
We therefore also consider nonadditive alternatives.
Simple multiplicative noise significantly degrades model quality and still fails to prevent rank recovery (Appendix~\ref{app:multiplicative-noise}).

Preserving the relative ordering of logits is critical for maintaining LLM quality.
This motivates an order-preserving additive variant that reorders sampled noise to match the logit ranking before addition.
Despite its appeal, this method is ineffective: the reordered noise matrix itself lies in a low-dimensional subspace, enabling rank recovery (Appendix~\ref{app:sorted-gaussian-noise}).

\paragraph{Softplus-then-perturb mechanism}
We therefore propose the \emph{softplus-then-perturb} (STP) mechanism.
It first applies an order-preserving softplus transformation to skew the logits, then adds prompt-dependent perturbations.
The initial skewing helps preserve token ordering under perturbation, while the subsequent noise prevents rank-estimation attacks.

We evaluate multiple perturbation schemes and characterize their trade-offs between quality and robustness.
Gaussian perturbations combined with STP are robust to all considered attacks while maintaining high output quality.

\section{Problem Setup}\label{sec:model}
We investigate model stealing attacks \cite{carlini2024stealing} that exploit the logits generated by models. 

We assume the adversary is able to make adaptive black-box prompt queries to the model and receives the corresponding logit vectors.

By collecting logit vectors corresponding to the model’s predictions for next-token generation across multiple prompts, the attackers construct a \textit{logits matrix} and estimate its rank to infer the hidden dimension.
This serves as a stepping stone toward further stealing of the model's last layer’s parameters.
Our defense, inspired by watermarking techniques of \cite{kirchenbauer2023watermark}, aims to perturb the logits in a way that prevents accurate rank estimation, thereby mitigating rank-based model stealing attacks.
Importantly, these perturbations must not degrade the language model’s performance or render it unusable.
Thus, evaluating the language model’s utility is essential to validate the practicality of the defense.




\subsection{Preliminaries}
Let $[n] = \{1,2,\ldots, n\}$.
Let $\cT$ be the set of tokens in a LLM and $T$ be its cardinality.
For a set $X$, we write $X^*$ to denote the family of every finite sequence over $X$.
To describe the overall behavior of the LLM token generation process,
once a user writes a prompt $p \in \cT^*$ into the LLM,
the LLM generates the corresponding hidden state $\h \in \R^H$.
Then it maps each hidden state to vocabulary logits $\g \in \R^V$ by matrix multiplication $\g = W^\top \h$ where $W \in \R^{H \times T}$. 
Finally, it (possibly randomly) samples a vocabulary based on the logit vector $\g$
according to the LLM's decoding algorithm, then the user sees the next token written by LLM.

\cite{carlini2024stealing} propose a model stealing attack where an adversary tries to infer the hidden state dimensionality $H$ from $N$ controllable prompts.
Let such prompts be $\{p_i\}_{i\in [N]}$, and corresponding hidden states be $\{h_i\}_{i\in [N]} $ with $h_i \in \mathbb{R}^H$.
For each $y_i$, the LLM generates the logit vector $\g_i = W^\top \h_i$,
where $W$ refers to the projection matrix that maps the latent space to the vocabulary space.
The adversary can access the logit vector $\{\g_i\}_{i \in [N]}$, and use this information to infer $H$.
\cite{carlini2024stealing} suggest that $H$ can simply be obtained by computing the rank of the matrix constructed by stacking the \( \g_i \) vectors and compute its rank, \ie
$
    H = \rank 
    \left( \left[
        \g_1, \g_2, \dots, \g_N
    \right]
    \right) .
$
Indeed, $H$ intrinsically serves as an upper bound on this rank; in practice,
the hidden latent generates a full-rank matrix (and $W$ itself is full rank), resulting in equality.
We aim to design a perturbation mechanism that perturbs each logit vector $\g_i$ to $\g'_i$ to prevent such attacks, while preserving the  model's quality.

\section{Rank Estimation of Noisy Matrices}\label{sec:rank}
A crucial step in the attack of~\cite{carlini2024stealing} is computing the {\em rank} of the generated logits matrix $(\ell_1,\dots,\ell_N)$.
We next provide a deeper treatment of the rank estimation procedure used in~\cite{carlini2024stealing} and extensions.


\paragraph{Baseline: Numerical rank from~\cite{carlini2024stealing}}
Practically, since one can only access every numeric values over floating-point numbers that prevents us from computing the exact number of linearly independent rows, one can instead compute the \emph{numerical rank} by principal component analysis (PCA) to obtain a sorted list of singular values $\sigma_1 \ge \sigma_2 \ge \ldots \ge 0$, and identify the index $i$ with the largest multiplicative gap $\sigma_i/\sigma_{i+1}$ (alternatively, logarithmic difference).
This index $i \in [N]$ constitutes the output of their attack: namely their prediction for the hidden dimension.

\paragraph{Beyond~\cite{carlini2024stealing}}Consider instead a perturbed logit matrix, e.g.\ where each logit vector is offset by additive noise. Intuitively, since the magnitude of the introduced noise should be sufficiently small to guarantee that the perturbed model does not deviate too much from the original, one may ask whether the adversary can still recover the rank of the matrix by conducting PCA and investigating the singular values.
Whether this approach could successfully recover the rank, however, largely depends on the density and magnitude of the noise.

In what follows, we introduce some seminal results on matrix perturbation theory to overview possible approaches to recover the rank of a noised matrix beyond the model (rank) stealing attack by~\cite{carlini2024stealing}.

\subsection{PCA: small but possibly dense noise}

%

Given a matrix $M$, let $\norm{M}_2 = \inf\{c \ge 0: \norm{Av}_2 \le c\norm{v}_2, \forall v \in \R^{n}\}$ be operator norm, $\norm{M}_* = \sum_{i}\sigma_i(M)$ be nuclear norm, and $\norm{M}_1 = \sum_{i,j} |M_{i,j}|$ be $\ell_1$-norm of $M$,
where $\sigma_i(M)$ denotes $i$-th largest singular value of $M$.

Suppose we have a matrix \(A \in \mathbb{R}^{m \times n}\) of rank \(r\). Then it possesses \(r\) nonzero singular values. We consider a noise matrix \(E\) and the perturbed matrix \(A + E\). We want to understand how large \(E\) (in operator norm) must be in order to make \(A + E\) change rank.

\paragraph{Weyl's inequality}
A core fact from matrix perturbation theory is that if $\|E\|_2 < \sigma_{\min}(A)$, then \(A + E\) must still have rank \(r\).
Intuitively, the operator norm \(\|E\|_2\) cannot push a nonzero singular value of \(A\) to zero unless it is at least as large as that singular value.
Equivalently, to modify a low-rank matrix into a higher-rank one, the noise must have operator norm exceeding the relevant singular values.

Formally, a classical result is the following version of \emph{Weyl’s inequality}:
\begin{theorem}[\cite{weyl1912asymptotische}]\label{thm:weyl}
    Let \(A, E \in \R^{n \times m}\), and let $\sigma_i(\cdot)$ denote the $i$-th largest singular value.
    Then, for each \(i \ge 1\),
    \[
    	\bigl|\sigma_i(A + E) - \sigma_i(A)\bigr|
	    \;\le\;
	    \|E\|_2 .
    \]
\end{theorem}
In particular, no singular value of \(A\) can increase by more than \(\|E\|_2\), implying that substantial noise is required to alter the numerical rank.

Given these stability guarantees, PCA (or SVD) is a natural tool for estimating the rank of a matrix contaminated by moderate noise.
In practice, one computes the SVD of $A+E$ and identifies a spectral gap: a large drop between \(\sigma_r\) and \(\sigma_{r+1}\) typically signals rank \(r\).

\subsection{PCA \& averaging: reducing spectral norms}
From Theorem~\ref{thm:weyl}, analyzing the operator norm of the noise matrix is a key step in understanding when PCA-based rank recovery succeeds.

For sub-Gaussian noise matrices (including Bernoulli and Gaussian), the operator norm scales with dimension:
\begin{theorem}[Chapter 4, \cite{vershynin2018high}]
    Let $E$ be an $m\times n$ random matrix with independent sub-Gaussian entries.
    Then $\norm{E}_2 = \Omega(\sqrt{m} + \sqrt{n})$.
\end{theorem}
This follows since the operator norm is bounded below by the norm of any row or column.
As a result, Weyl’s inequality alone provides only weak guarantees for rank recovery in high dimensions.

However, an adversary can significantly reduce the effective noise magnitude if multiple copies of the \emph{same} logit matrix are available with independent noise.
The key idea is averaging: querying each prompt multiple times and averaging the resulting matrices reduces noise variance.

Formally, let $A+E$ denote the $N\times T$ logit matrix for prompts $\p_1,\ldots,\p_N$, and suppose independent noise is added across $k$ repeated queries.
This yields matrices $A'_j = A + E_j$ for $j\in[k]$, where the $E_j$ are independent.
Averaging gives
\[
\tilde A' = \frac{1}{k}\sum_{j=1}^k A'_j
= A + \frac{1}{k}\sum_{j=1}^k E_j,
\]
so the operator norm of the noise decreases as $k$ grows.
This argument relies critically on independence of the $E_j$.\footnote{In contrast, our defenses use deterministic pseudorandom noise, so repeated queries yield identical perturbations.}

\paragraph{Discussion}
Due to the structured nature of logits, averaging can be even more effective than this analysis suggests.
Each noise-free logit vector lies in the low-dimensional column space induced by the projection matrix $W$, so averaged matrices remain close to a low-rank subspace.
If the noise $E_j$ is weakly correlated with $A_j$, the averaged matrix can again be expressed as a low-rank signal plus reduced-variance noise.
Thus, the effectiveness of averaging crucially depends on the dependence structure between $A_j$ and $E_j$.
Empirically, we observe that independent noise degrades under averaging around $k=40$, while dependent (embedding-based) noise remains robust beyond $k=50$ (Appendix~\ref{app:pca-with-averaging}).

\subsection{Robust-PCA: sparse but possibly gross noise}\label{sec:rpca}
What if the noise is large enough and the adversary cannot rely on the averaging argument due to potential dependency between the noise matrices?
In this case, one might employ robust PCA~\cite{candes2011robust} (RPCA) instead of the standard PCA to estimate the numerical rank of matrices.
RPCA seeks to robustly estimate the original matrix given a perturbed matrix, when the noise matrix is sparse yet could have large values in element-wise.
Surprisingly, such estimation can be done by solving the following convex optimization problem, termed \emph{principal component pursuit} (PCP), given an observed matrix $A'$: minimize $\norm{A}_{*} + \lambda \norm{E}_1$ subject to $A + E = A'$.
\cite{candes2011robust} show that this exactly recovers the low-rank matrix $A$ and sparse noise $E$ even if the rank of $A$ grows near-linearly in the dimension of the matrix and the noise in each element of $E$ is up to a constant factor of the original matrix $A$.
See Appendix~\ref{app:rpca} for more details.



\newcommand{\bias}{\text{bias}}

\section{Defense against dimension extraction}\label{sec:defense}

In this section, we propose defense strategies against rank estimation attacks.
Let $p$ represent a prompt input to the LLM,
where the model’s standard forward pass generates the logit vector $\g$,
used for sampling the next token.
Our methodology involves perturbing the logits layer,
transforming $\g$ into a perturbed vector $\g’$.
These perturbations alter the properties of the logits matrix, that the adversary constructs by stacking logits corresponding to different prompts.

Formally, inspired by \cite{kirchenbauer2023watermark},
given an input prompt $\p \in \cT^*$ and corresponding logit vector $\g \in \R^V$,
our method finds a bias vector $\bias(\p) \in \R^T$ as a function of the prompt $\p$ and computes $\g' = f(\g,\bias(\p))$ for perturbation function $f$.
Then it decodes the next token based on the perturbed logit vector $\g'$.

\subsection{Various noises and their vulnerability}
We first discuss why simple noise structures are vulnerable to attack, providing insights behind our proposed mechanism.

\paragraph{Warm-up: fixed noise}
We first explain why the noises should be varying with respect to the prompt.
Suppose we run a perturbation scheme that deterministically uses a fixed bias function $\bias(\p) = \vec{\text{b}} \in \R^{T}$.
Then, the adversary can input prompts $\{\p_1, \p_2, ..., \p_N\}$, stack perturbed logits,
and recover the rank of the hidden layer as the rank remains the same.
Namely, $\rank([\ell_1, \ldots, \ell_N]) = \rank([\ell_1 + \vec{b}, \ldots, \ell_N + \vec{b}])$.
Thus, a reasonable defense method should vary the noise added to the logit vector $\g$ to avoid such scenario.

\begin{figure}[h!] %
    \centering    \includegraphics[width=0.36\textwidth]{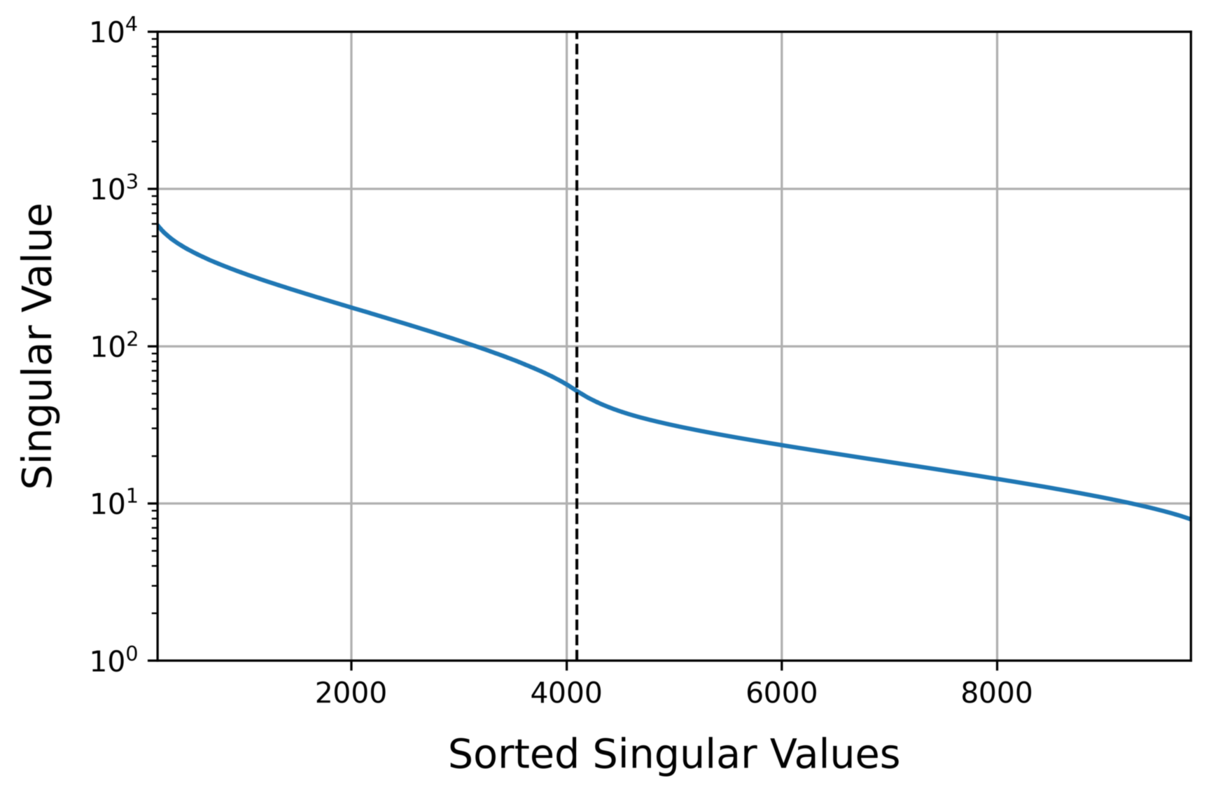}
    \caption{Dimension-extraction attack on Mistral-7B. Each row of the matrix was constructed by sampling Gaussian noise $\sim \mathcal{N}(0, 1)$, then reordering the noise values to match the rank order of the original logits. Singular Value Decomposition (SVD) was performed on this sorted noise matrix—without using the original logits—to estimate the hidden dimension.}
    \label{fig:sorted-gaussian-noise}
\end{figure}

\paragraph{Sorted Gaussian noise}\label{app:sorted-gaussian-noise}
We constructed a matrix $E$ by sampling Gaussian noise as follows: for each row, we sampled $T$ independent values from $\mathcal{N}(0, 1)$ and then sorted these values according to the ordering of the original logits. Remarkably, we discovered that the matrix $E'$ of reordered Gaussian samples alone exhibits similar low-dimensional structure as the LLM's hidden representations. Singular value decomposition of $E'$ reveals a signal corresponding to the actual hidden dimension, as demonstrated in Figure~\ref{fig:sorted-gaussian-noise}.

\begin{figure}[h!] %
    \centering    \includegraphics[width=0.36\textwidth]{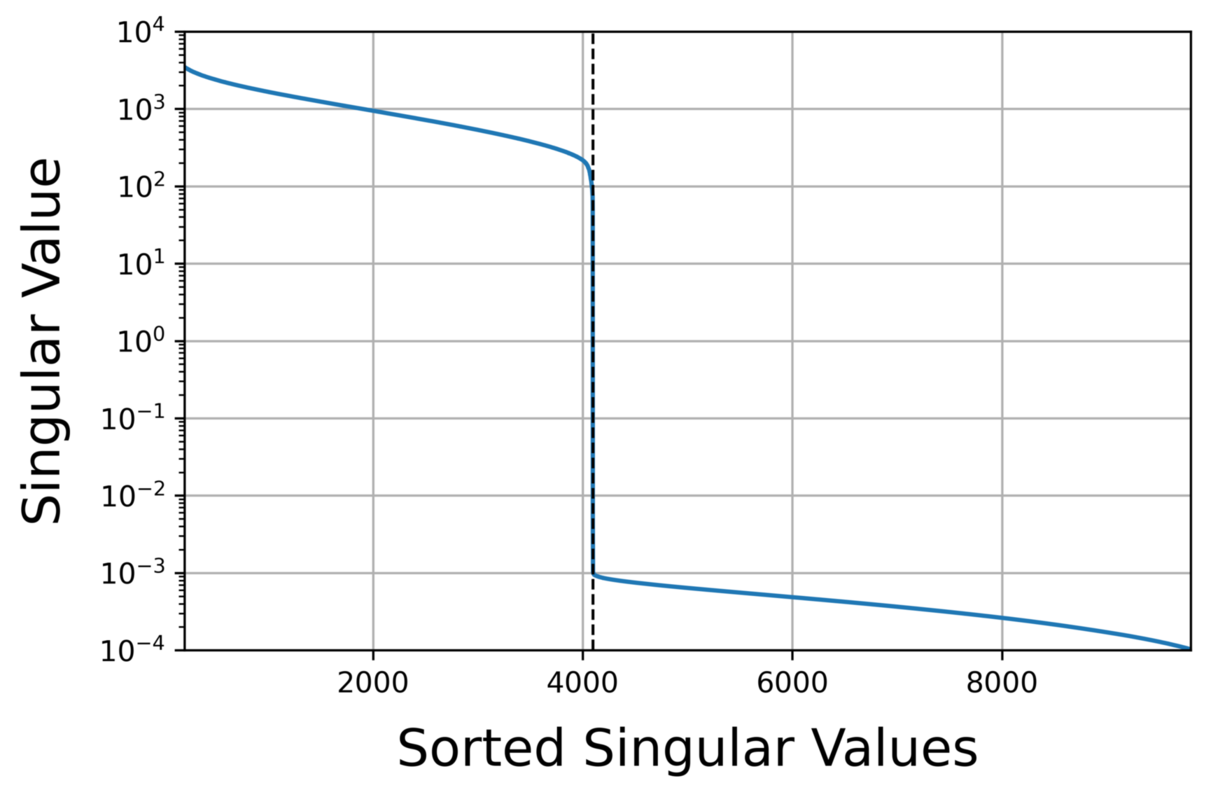}
    \caption{Dimension-extraction attack on Mistral-7B. Each row of the matrix was generated by sampling $\epsilon \sim \mathcal{U}(0, 10)$ and multiplying the logits by $1+\epsilon$. While this row-wise multiplicative noise significantly degrades the model’s output quality, it does not impede the attack. Thus, this type of noise is ineffective as a defense.}
    \label{fig:multiplicative-noise}
\end{figure}

\paragraph{Multiplicative noise}\label{app:multiplicative-noise}
Given that additive noise is vulnerable to PCA with averaging attack, a promising approach to make it robust is to impose nonadditive noise.
We consider a multiplicative noise such that we sample noises $\eps$ i.i.d. from uniform distribution over $[0,10]$,\footnote{This is an extremely strong version of the defense mechanism. Any weaker perturbation would still be vulnerable to the dimension extraction attack.} and then multiply the logits by $1+\eps$.
Interestingly, however, we observe that this scheme is still significantly vulnerable to simply performing PCA over the final matrix, as shown in Figure~\ref{fig:multiplicative-noise}. 
This suggests that more complicated noise structure is necessary to make it robust against PCA.

\subsection{Proposed Method}

We consider a defense mechanism that introduces a structured pseudorandom noise to the logit vector for each query prompt, 
generalizing the additive noise in the watermarking method of~\cite{kirchenbauer2023watermark}.
\begin{figure} 
  \begin{minipage}{0.5\textwidth}
\begin{algorithm}[H]
\textbf{Input:} LLM forward function $F$, prompt $p$, noise distribution $\mathcal D$, $\text{PRF}$, and perturbation function $f$.\\
\textbf{Output:} Biased logits
\caption{Embedding-Dependent Noise Addition}
\begin{algorithmic}[1]
\STATE $\g, \h \gets F(p)$ \hfill \textit{// logits $\g$ and last hidden state $\h$}
\STATE $s \gets \text{PRF}(\h)$ \hfill \textit{// seed $s$ as a function of $\h$}
\STATE $\noise(p) \gets {\mathcal D}(\text{seed} = s)$ \hfill \textit{// sample noise vector}
\STATE \textbf{return:} $f(\g,\noise(p))$
\end{algorithmic}\label{alg:nonlsh}
\end{algorithm}
  \end{minipage}
\end{figure}

More concretely, the noise values are derived from a random sample from a given noise distribution $\mathcal D$, with the sampling process determined by a seed.
Importantly, this seed is computed as a function of the input prompt, ensuring that the perturbation process is deterministic. 
We design the seed as a cryptographic \emph{pseudorandom function} (PRF) applied to the model’s last hidden states.  A PRF is an efficiently computable keyed function family such that for a random and secret key $k$, the function $\PRF_k(k, \cdot)$ is computationally indistinguishable from a truly random function. 
In particular, any set of evaluations $\PRF(k,x_i)$ on different inputs $x_i$ appear as i.i.d.\ sampled outputs; however, on the same input $x$, $\PRF(k,x)$ will always return the same output.
This design ensures that the added noise depends on the prompt-specific hidden states,
which encode the features the language model extracts based on the input tokens to predict the next token.
Finally, the sampled noise and the logit vector outputs a final perturbed logit vector with respect to a perturbation function $f$.
The pseudo-code for generic defense mechanism is provided in Algorithm~\ref{alg:nonlsh}.  In the system setup, we assume the model owner samples and maintains a secret PRF key $k$ once for all time, and omit $k$ from PRF notation.

\paragraph{Additive noise}
We first consider additive noise, analogous to~\cite{kirchenbauer2023watermark}, \ie $f(\g, \noise(p)) = \g + \noise(p)$.
We investigate two types of noise distribution families.  {\em Discrete noise} of magnitude $\delta$ and sparsity $\gamma$ corresponds to a random vector in which $\gamma$ fraction of the coordinates are set to $\delta$, and the remaining are 0.
{\em Gaussian noise} with standard deviation (std) $\sigma$ and sparsity $\gamma$ will correspond to a random vector in which $\gamma$ fraction of coordinates are sampled as i.i.d.\ Gaussian with mean 0 and std $\sigma$, and the remaining are 0.
We note that for constant choices of magnitude and sparsity, the $N \times T$ matrix formed by $N$ samples from either distribution will have high row norm (approximately $\gamma \delta T$ in the case of discrete noise, and $\sqrt {\gamma T}$ for Gaussian noise), and in turn high spectral norm. This means they stand as valid candidates for increasing the rank of the LLM logits matrix.

\begin{figure*}[h!] 
    \centering
    \begin{subfigure}{0.4\textwidth} 
        \centering
        \includegraphics[width=\textwidth]{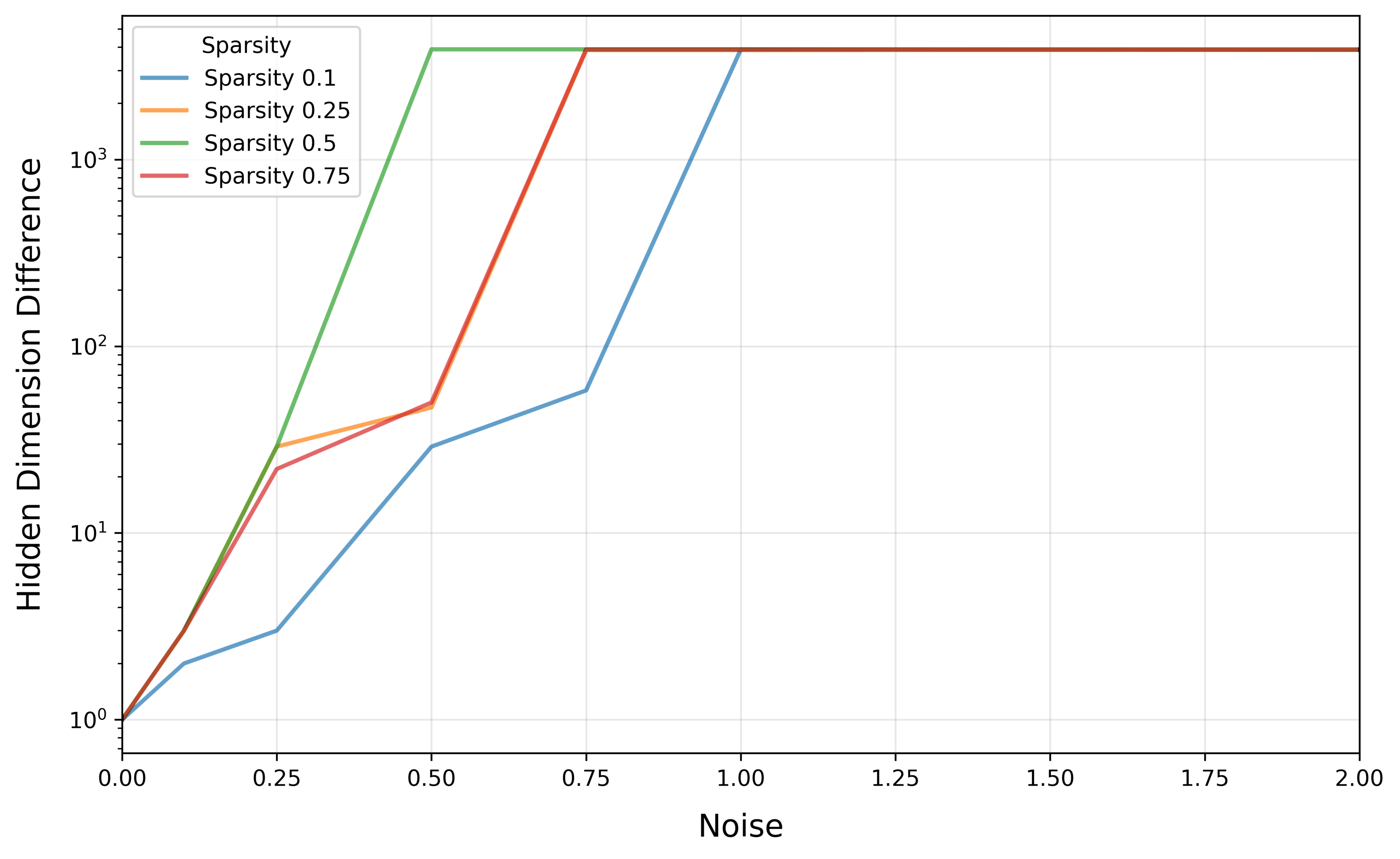}
        \caption{Discrete noise}
    \end{subfigure}
    \begin{subfigure}{0.4\textwidth}
        \centering
        \includegraphics[width=\textwidth]{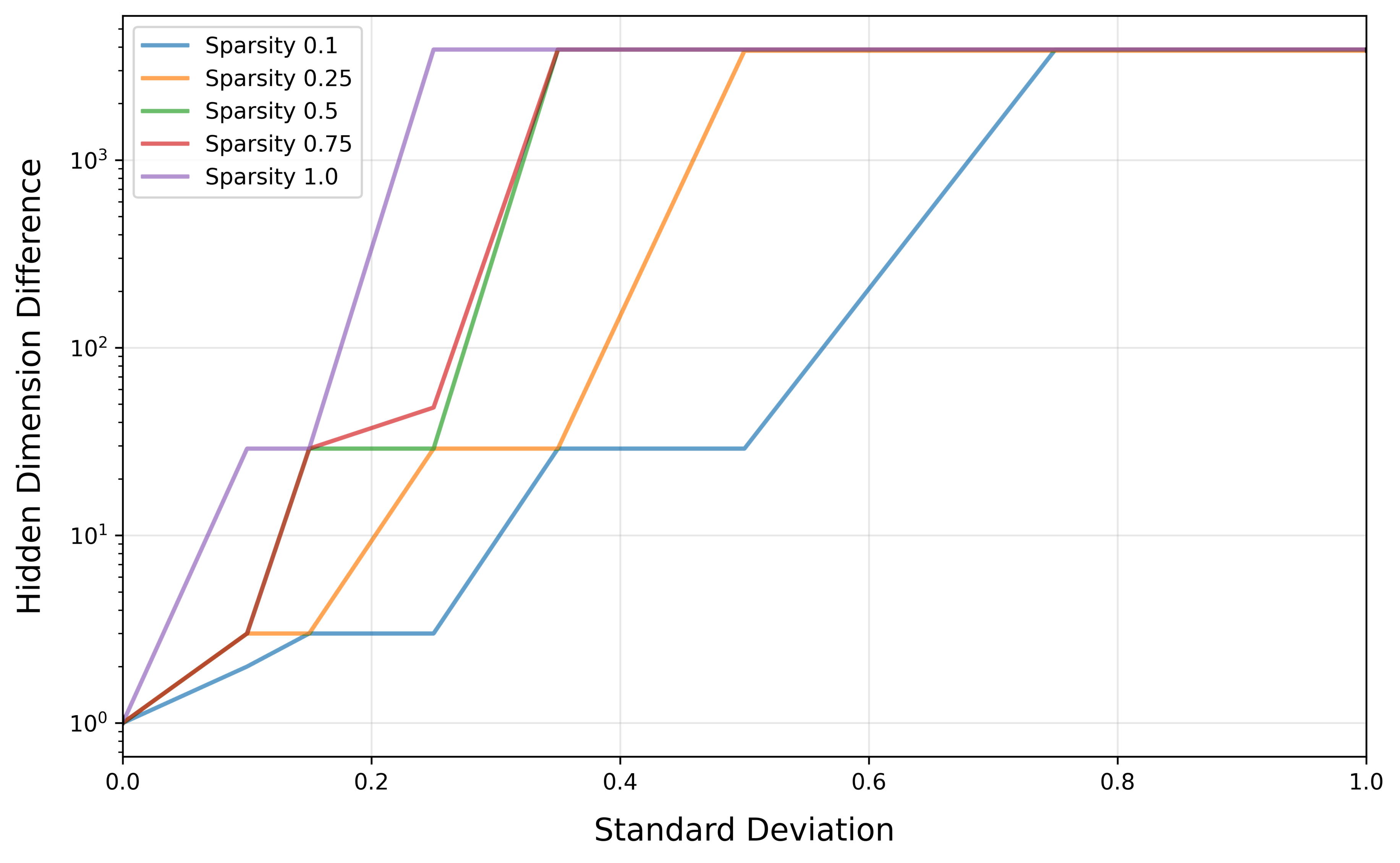}
        \caption{Gaussian noise}
    \end{subfigure}
    
    \caption{Summary of attack success on Mistral-7B given additive discrete or Gaussian noise as a function of noise magnitude ($x$ axis) and sparsity (curves).  Hidden dimension difference ($y$ axis) indicates distance between the attack's predicted hidden dimension and ground truth, 4,096. Reaching the top value $y=4,096$ indicates complete failure of the attack.}
    \label{fig:attack-summary}
\end{figure*}
\paragraph{Softplus noise}
We next consider further approaches toward noise perturbation, with the aim of improving the effect on the model output.  We observe that maintaining the {\em ordering} of each token within the logit vector plays a significant role in model quality, which is likely to be reordered given additive noise to individual logit vector components.

We thus propose a generic mechanism, dubbed \emph{softplus-then-perturb} (STP) mechanism, that first applies the softplus transformation to the logit vector to make it more skewed which would help preserve the original order after further modification, then apply random perturbation to prevent attacks. More concretely, in the above notation, $f(\ell,\noise(p)) = \softplus(\alpha \ell) + \noise(p)$, where $\alpha$ is either a fixed or per-query randomized scale factor.

We remark that while $\softplus(\alpha \ell)$ itself already yields a matrix with high rank, this alone is insufficient for defense, as an attacker can simply invert the softplus computation to retrieve a matrix whose rows lie within the original low-rank subspace.
In turn, we investigate variations of the discrete and Gaussian noise families above for $\noise(p)$ distribution.

\paragraph{Other options for random seed}
For the embedding-dependent noise, we select the seed for noise to be a function of the model’s hidden state when prompted with an input $p$. 
We emphasize that this design choice crucially differs from that of \cite{kirchenbauer2023watermark}, who added noise seeded by a sliding window of the last few tokens.



The reason for this modification is that it is critical in our setting for an adversary to be unable to generate several different prompts yielding the same seed. Otherwise, this enables
producing identical noise that fails to alter the rank of the logits matrix.
For example, if the noise is seeded by the last $m$ tokens, an adversary could simply append $m$ fixed tokens to a collection of prompts,
resulting in identical noise for each and effectively compromising the defense.

To remove this vulnerability, we design our noise addition to be dependent on the model’s inner states. Relying on hidden states (together with the security of the PRF) ensures that the noise generation becomes significantly harder to replicate, thereby making the defense more robust.  
If one wishes to remove dependence on the model, the seed can instead rely on a hash of the complete input prompt.




\section{Experiments}\label{sec:exp}

In this section, we provide empirical evidence of the effect of various distributions of additive noise in preventing the hidden dimension extraction attack of~\cite{carlini2024stealing}, in the case of Mistral-7B. Complimentarily, we demonstrate that our noise modifications do not significantly degrade the quality of the underlying models by providing evaluations of resulting model perplexity. In Appendix~\ref{appendix:gpt2}, we provide analogous experiments for GPT-2.

\paragraph{Implementation}
All experiments were performed in the Google Colab environment, using the A100 GPU runtime type.
We instantiate the PRF output for sampling noise vectors via BLAKE3~\cite{Blake3} in keyed-hash mode. In some cases, the resulting pseudorandom output is extended further using the numpy python package RNG, seeded by the BLAKE3 output.


\subsection{Model Stealing Prevention}
\label{sec:exp-stealing-prevention}

We consider the effectiveness of the~\cite{carlini2024stealing} attack on extracting hidden model dimension for an assortment of noise parameters, including discrete noise and i.i.d.\ Gaussian  with varying magnitude and sparsity.  

In each experiment with Mistral-7B (token vocabulary size $T=32,000$, hidden dimension 4,096), we query $N=10,000$ prompts, consisting of randomly sampled token pairs from the model vocabulary (following~\cite{carlini2024stealing}).  Each resulting output logit vector is additively perturbed by a noise vector sampled using randomness that is computed by applying a PRF to the last hidden state (see Algorithm~\ref{alg:nonlsh}).  The resulting $N$ perturbed logit vectors are appended into an $N \times T$ matrix $A = (\tilde \ell_1,\dots,\tilde\ell_N)$.  
We perform SVD decomposition of the matrix $A$; sort the resulting $N$ singular values in decreasing order; compute the difference between each pair of adjacent singular values; and output index $i \in [N]$ for which this distance is maximal as the predicted dimension.




\begin{figure*}[h!] 
    \centering
    \begin{subfigure}{0.36\textwidth} 
        \centering
        \includegraphics[width=\textwidth]{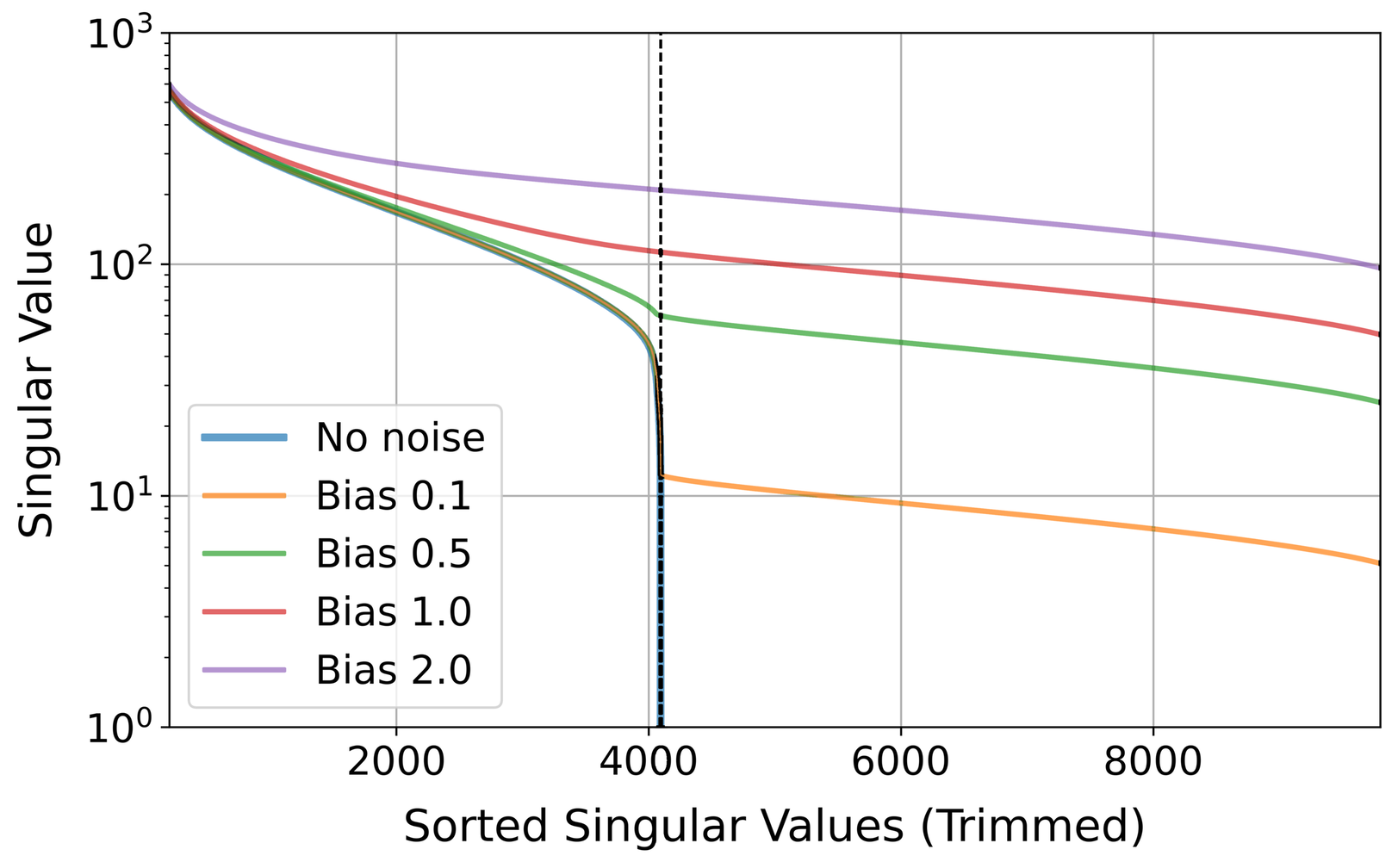}
        \caption{Constant bias comparison}
    \end{subfigure}
    \hspace{0.02\textwidth}
    \begin{subfigure}{0.36\textwidth}
        \centering
        \includegraphics[width=\textwidth]{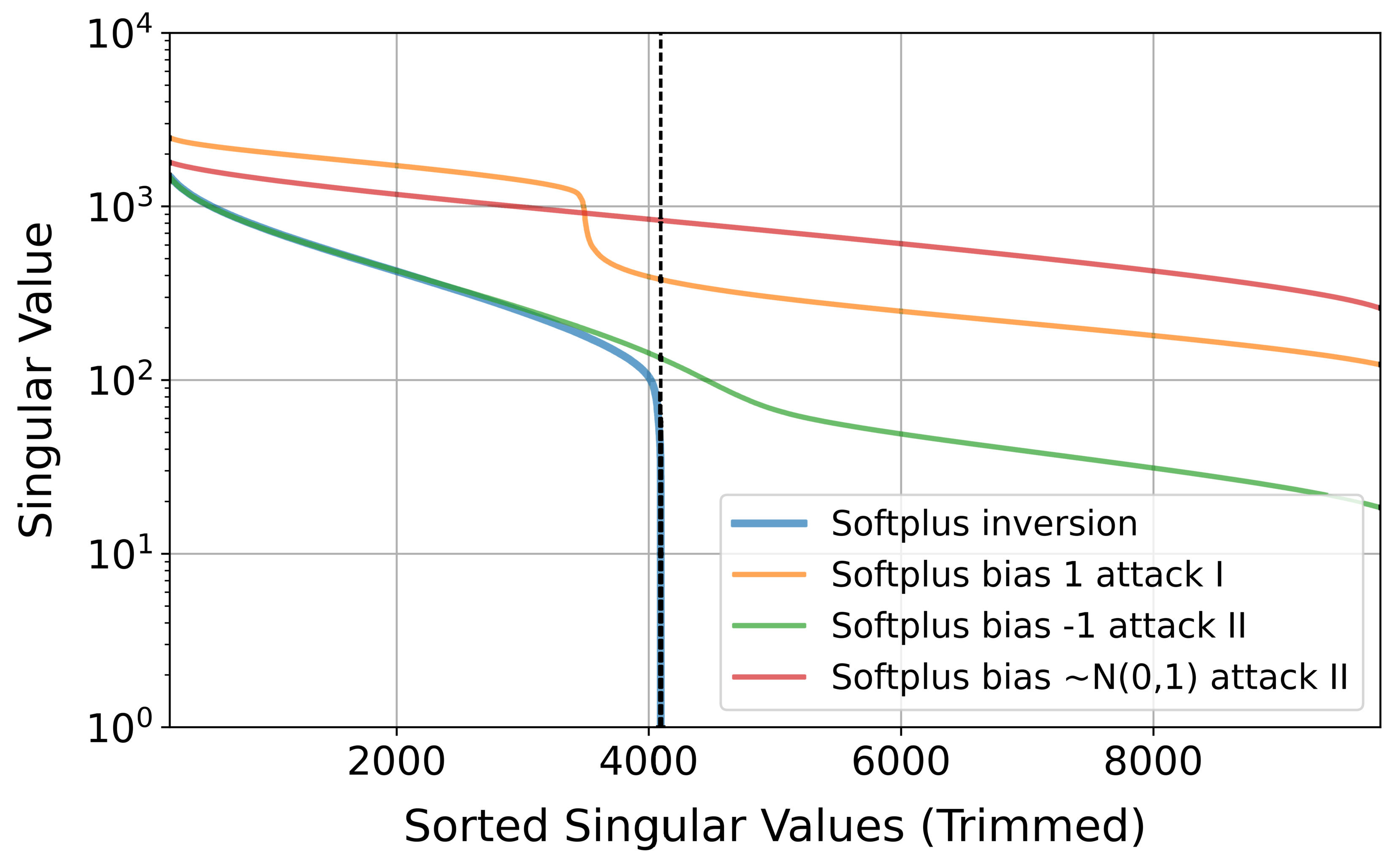}
        \caption{STP comparison}
    \end{subfigure}
    
    \caption{Dimension-extraction attack on Mistral-7B with (a) Discrete additive noise for a progression of noise magnitudes (0, 0.1, 0.5, 1.0, and 2.0) and sparsity 0.5. Each plot depicts the sorted singular values of the corresponding noisy logits matrix, demonstrating a significant jump at correct position 4,096 for no noise  and low noise 0.1; an inflection point at 4,096 for mid-level noise 0.5; and no identifiable anomaly for the case of larger noise 1.0 and 2.0. (b) STP comparison showing different noise perturbations and attacks. The pure softplus inversion (blue) shows a sharp drop at position 4,096, clearly revealing the model dimension. The STP with bias 1 attack I (orange)  shows a notable drop at around position 3,600, while the STP with bias -1 attack II (green) shows more gradual degradation. When using the attacks on STP with bias $\sim N(0,1)$, there is no identifiable anomaly. We provide analogous plots for further settings in Appendix~\ref{appendix:more-noise}. \vspace{-.1in}}
    \label{fig:main-plots}
    \end{figure*}

\paragraph{RPCA attack}
We implement an extension of the SVD-based attack, using the Robust PCA's PCP algorithmic procedure from~\cite{candes2011robust}.  
Interestingly, the RPCA process did not appear to significantly enhance the attack. In each of our tested experimental noise parameter settings, the standard PCA attack and the RPCA algorithm for rank estimation either both succeeded or both failed to provide a realistic estimate of the matrix rank. See Appendix~\ref{app:rpca_expirements} for additional information.

\paragraph{Softplus inversion attacks}
We explore tailored extensions of the~\cite{carlini2024stealing} attack against variations of the softplus-then-perturb mechanism (Figure~\ref{fig:main-plots}(b)).  As mentioned, applying $\softplus(\alpha \ell)$ alone to (even randomly $\alpha$-scaled) logit vectors does not suffice as a defense mechanism, as this operation can be inverted by the adversary to yield $\alpha$-scaled copies of the original logit vectors, (yielding the same original matrix rank).  
When further perturbation is added, however, inversion does not work directly.  

We identified new attacks for softplus with discrete noise. Under positive bias $+b$, we recovered rank approximations (e.g., 3,600 for $+b=1.0$, sparsity $0.5$) by applying PCA to the $N \times T$ matrix constructed via attack I: (1) shifting each perturbed logit vector $\tilde\ell_i$ (for each row $i \in [N]$) component-wise by the difference between its mean and the global mean; (2) replacing non-positive elements with $\epsilon = 10^{-10}$; and (3) softplus inversion.
For negative bias $-b$, many elements became negative, invalidating softplus inversion. Instead, attack II exploits the constant bias: the most negative post-perturbation value estimates $-b$, which we subtract from all non-positive elements, effectively removing the bias.

Softplus plus {\em Gaussian} noise, on the other hand, withstood all attack attempts. The varying Gaussian offsets disrupted the softplus-inversion without evident means of removal. Further, the nonlinear skew via softplus disrupted attack methods on pure additive noise such as PCA with averaging.
We leave further investigation of the attack space on the softplus-then-perturb class (including Gaussian and beyond) as an important direction of study.


    

\subsection{Quality Evaluation}
Overall, we validate various quality metrics for LLMs including perplexity (PPL), massive multitask language understanding (MMLU) as well as several distributional shift metrics between the original logit vector and the final perturbed logit vector that capture how much our defense mechanism changes the logit vector.
These distances include total variation distance (TVD), Kullback–Leibler divergence (KL), Jensen-Shannon divergence (JS), and top $50$ token's agreement percentage (Top-50) between two logit vectors.

\paragraph{Perplexity evaluation}
By design, our discrete noise perturbation (specifically, our relatively conservative magnitude 2.0 and sparsity 0.5) corresponds identically to the perturbation induced by the soft watermarking procedure in~\cite{kirchenbauer2023watermark}.  In turn, we directly inherit the corresponding conclusions regarding benign impact on quality of generated text.
We use Meta-Llama-3-8B~\cite{Llama3} as \emph{larger oracle model} to evaluate perplexity.

To generate samples, we took a random selection of 500 texts each of length 500 tokens from the news-like subset of the C4 dataset \cite{c4dataset}, and removed the final 100 tokens.
Each remaining 400 tokens served as a prompt to the model being tested, where completions of length 100 were generated using greedy sampling with a temperature of 0.1.  These prompts were additionally given to the oracle Meta-Llama-3-8B model, and perplexity was computed as the exponential of the average token-wise loss with respect to the predicted next-token distribution of the oracle model for each token output.
The resulting calculated perplexity measurements are given in Table 1.

We observe that while some noise-perturbed models exhibit only modest increases in perplexity compared to the original model, others show more significant degradation, especially with stronger perturbation settings. These results indicate that the perturbations do not uniformly affect text quality, with some configurations maintaining near-original performance. More sophisticated sampling procedures, such as beam search~\cite{kirchenbauer2023watermark}, can result in further improvements in perplexity.

\begin{table}[h!]\label{tab:mistral-ppl}
    \centering
    \begin{adjustbox}{width=\columnwidth} 
    
    \setlength{\tabcolsep}{6pt}
    \renewcommand{\arraystretch}{1.2}
    \small{
    \begin{tabular}{c|cc|cccccc||c}
        \toprule
        {Defense Mechanism} & $\gamma$ & {Bias} & {MMLU} & {PPL} & {TVD} & {KL} & {JS} & {Top-50} & {Breaks attacks} \\ 
        \midrule
        \multirow{6}{*}{Additive} & 0.25 & 1 & 54.61 & 3.73 & 0.18 & \textbf{0.09} & \textbf{0.02} & 0.81 & Yes \\
        & 0.1 & 2 & 49.52 & 4.47 & 0.27 & 0.23 & 0.06 & 0.71 & Yes \\
        & 0.5 & 2 & 49.28 & 4.49 & 0.34 & 0.38 & 0.08 & 0.66 & Yes \\
        & 0.1 & 3 & 44.75 & 6.43 & 0.47 & 0.59 & 0.15 & 0.51 & Yes \\
        \cmidrule{2-10}
        & 0.25 & $\sim N(0,1)$ & 55.28 & 3.58 & \textbf{0.14} & 0.12 & 0.03 & 0.85 & Yes \\
        & 0.5 & $\sim N(0,1)$ & 53.65 & 3.74 & 0.23 & 0.22 & 0.05 & 0.76 & Yes \\
        \midrule
        \multirow{4}{*}{\shortstack[c]{Softplus \\ $\alpha \sim \text{Uni}[1,4]$}} & \multicolumn{2}{c|}{No Perturbation} & 57.78 & 3.26 & 0.48 & 3.22 & 0.20 & 1.00 & \textcolor{red}{No: Inv+SVD} \\
        & 0.25 & 1 & 56.67 & 3.35 & 0.50 & 3.25 & 0.21 & 0.92 & \textcolor{red}{No: attack I} \\
        & 0.25 & -1 & 56.87 & 3.39 & 0.49 & 3.26 & 0.21 & 0.93 & \textcolor{red}{No: attack II} \\
        & 0.25 & $\sim N(0,1)$ & \textbf{56.55} & \textbf{3.37} & 0.50 & 3.27 & 0.21 & \textbf{0.94} & Yes \\
        \midrule
        \multicolumn{3}{c|}{Original} & 57.78 & 3.40 & 0.0 & 0.0 & 0.0 & 1.0 & \textcolor{red}{No: SVD} \\
        \bottomrule
    \end{tabular}
    }
    \end{adjustbox}
    \caption{
    Evaluation scores for Mistral-7B. Original refers to the original Mistral-7B model.
    $\gamma$ denotes the sparsity parameter of the noise, and bias denotes the magnitude of the bias (\eg $1$ denotes constant bias).
    The last four rows denote the softplus-then-perturb mechanisms with corresponding perturbations, where $\alpha$ is sampled randomly for each prompt.
    Distributional shift metrics are calculated as the average from 10,000 distinct prompts.
    }
\end{table}

\paragraph{MMLU Evaluation}
We evaluate the performance of both the unmodified and perturbed Mistral-7B models on the Massive Multitask Language Understanding (MMLU) benchmark~\cite{hendrycks2021measuringmassivemultitasklanguage}. MMLU is a widely adopted benchmark comprising 57 diverse tasks that test language model's reasoning ability across subjects such as mathematics, science, and the humanities. Under the zero-shot setting, we assess the extent to which noise perturbations degrade performance in knowledge-intensive tasks.

The original model achieves an MMLU accuracy of 57.78\%. The additive noise shows significant degradation (44.75-55.28\%).

Softplus-based noise functions, while inducing high distributional divergence, preserve the top-k logit ordering and thus maintain strong MMLU performance. The pure Softplus variant matches the original model’s accuracy (57.78\%), and even with additional constant or Gaussian noise, the performance drop remains modest (around 56-57\%). This suggests that preserving the relative ranking of high-confidence logits can mitigate the impact of perturbations on task accuracy, even with significant shifts in the probability distribution.

These results highlight that the impact of noise perturbations on model performance is not solely determined by the magnitude of the perturbation. Instead, the functional characteristics of the noise play a significant role. Specifically, Softplus-based perturbations demonstrate that it is possible to introduce substantial distributional shifts while still preserving strong performance on benchmarks like MMLU, due to their ability to maintain the relative ranking of logits.

\paragraph{Distributional Shift Metrics}
As expected, stronger constant bias noise creates larger divergence, with the most aggressive configuration ($\gamma=0.1$, bias=3) exhibiting highest KL (0.59) and JS divergence (0.15), and a significant drop in top-50 agreement (0.51). Gaussian perturbations result in moderate divergence and more stable output agreement.

Softplus-based perturbations behave differently. Despite inducing comparatively high divergence values (\eg $\text{KL} > 3.2, \text{JS} > 0.21$), they maintain high top-$50$ accuracy agreement ($\ge 0.92$) and negligible degradation in perplexity or MMLU accuracy. This behavior arises because the Softplus function primarily affects lower-magnitude logits while preserving the relative ordering of top-scoring logits. As a result, the model’s most confident predictions remain largely unchanged, which explains the stability in accuracy despite measurable distributional shifts.

\section{Conclusion}
Our work proposes watermarking-inspired defenses against model stealing attacks targeting hidden dimension extraction via rank estimation ~\cite{carlini2024stealing}, achieved by perturbing the logits layer of large language models. Guided by theoretical analysis and validated through extensive experiments, our technique proves robust against adversaries seeking to extract hidden latent dimensions. The results show that the proposed approach remains effective across different configurations, highlighting its potential to bolster the security and integrity of commercial language models.



\bibliographystyle{named} 
\bibliography{ref}

\appendix
\setcounter{secnumdepth}{2} 
\onecolumn
\section*{Appendix}

\section{Related Work}\label{sec:rel}
\paragraph{Model Stealing}
We consider the setting of {\em model stealing \cite{TramerZJRR16}} (or model extraction) attacks, where an adversary who has black-box access to a model aims to recover information about the model's internal architecture or parameters, or tries to recover the functionality of that model. 
Prior work has primarily studied in ReLU-based networks, \cite{milli2019model} showed that gradient access enables near-exact model recovery, while \cite{jagielski2020high} used finite differences for gradient approximation.
Later efforts extended these attacks to deeper models \cite{carlini2020cryptanalytic,rolnick2020reverse,carlini2024polynomial}.
However, these methods do not scale to production language models due to token-based inputs, non-ReLU activations, complex architectures.
\cite{carlini2024stealing} proposes a method to extract the hidden latent dimensions and last-layer projection of commercial language models using API access, enabling model stealing even for large and expensive models.

\paragraph{Watermarking}
Broadly, watermarking is a form of embedding signature in a hidden way.
It is categorized into format-based \cite{Rizzo2016ContentpreservingTW,UniSpaCh},
lexical-based \cite{Topkara2006TheHV,yang2023watermarking,munyer2023deeptextmark},
syntatic-based \cite{wordTopkara,Meral2009NaturalLW,Atallah2001NaturalLW},
and generation-based \cite{lau2024waterfall,zhang2024remark,abdelnabi2021adversarial} watermarking. 
Recently, \cite{kirchenbauer2023watermark} propose a watermarking scheme where watermark is efficiently injected into the logits layer of a language model.
\cite{panaitescu2025can} find that watermarking can reduce the likelihood of generating copyrighted content.
The most relevant to ours is~\cite{zhao2023protecting}, which perturbs logits to detect model distillation. While similarly based on logit perturbation, our method targets a different threat model. Rather than detecting behavioral cloning after the fact, we proactively disrupt rank-based attacks to prevent extraction of architectural details, such as the hidden dimension.

\paragraph{Rank estimation}
The theory of matrix perturbation~\cite{stewart1998perturbation} (see also~\cite{konstantinides1988statistical}) indicate that the smallest noise needed to reduce a full-column-rank matrix’s rank is precisely its smallest nonzero eigenvalue. 
In its simplest form, Weyl's inequality ensures that the singular value decomposition (SVD) can reliably recover the numerical rank\footnote{See Section~\ref{sec:rank} for detailed definitions.} of a matrix under small perturbations. 
In particular, one may truncate at the $r$-th largest singular value when it exhibits a significant drop, ensuring accurate rank detection. 
On the other hand, if the noise's magnitude becomes larger, one cannot rely on PCA to recover the exact rank.
\cite{candes2011robust} suggests the robust PCA (RPCA) that exactly recovers the numerical rank if the noise is sparse but possibly gross by constructing a convex programming, dubbed principal component pursuit (PCP).
\cite{zhou2010stable} present a variant of PCP which is simultaneously stable against small entry-wise noise and gross sparse error.


\paragraph{Distillation}
While not directly related to model stealing, knowledge distillation techniques have been widely studied as a method for transferring knowledge from large models to smaller ones. Distillation is typically used to improve efficiency, enabling smaller models to retain much of the performance of their larger counterparts while reducing computational costs. However, these techniques also pose risks to model security and intellectual property, as they can be used to replicate proprietary models without direct access to their architectures or training data. Several surveys, e.g.~\cite{gou2021knowledge,xu2024survey} provide an overview of distillation methods, covering strategies, evaluation metrics, and ethical considerations, along with their benefits and potential risks in AI development.

\section{Omitted Details in Section~\ref{sec:rpca} for RPCA}\label{app:rpca}
We here provide the technical result for RPCA proven by~\cite{candes2011robust}.
Let $A = U\Sigma V^\top = \sum_{i=1}^r \sigma_i u_i v_i^\top$ be the SVD of $m \times n$ dimensional matrix $A$ with rank $r$.
Then define the following two conditions parameterized by $\mu$:
\begin{align}
    \max_i \norm{U^\top e_i}^2 \le \frac{\mu r}{m}, \quad \max_i \norm{V^\top e_i} \le \frac{\mu r}{n}\label{eq:rpca-cond1}
    \\
    \norm{UV^\top}_{\infty} \le \sqrt{\frac{\mu r}{mn}},\label{eq:rpca-cond2}
\end{align}
where $e_i$ has one in $i$-th coordinate and otherwise zero.
Formally, \cite{candes2011robust} prove the following theorem:
\begin{theorem}[\cite{candes2011robust}]\label{thm:rpca}
    Let $A$ be $m \times n$ matrix obeying~\eqref{eq:rpca-cond1} and~\eqref{eq:rpca-cond2}.
    Let the support set, \eg the nonzero elements, of the noise matrix $E$ is uniformly distributed among all sets of cardinality $S$.
    Then, there is a constant $c$ such that with probability at least $1-c\max(m,n)^{-10}$, solving (PCP) with $\lambda = 1/\sqrt{m}$ exactly returns the original matrix $A$ and the noise $E$, provided that
    \begin{align*}
        \rank(A) \le \rho_r \min(m,n) \mu^{-1}(\log \max(m,n))^{-2}
    \end{align*}
    and $S \le \rho_s mn$, where $\rho_r$ and $\rho_s$ are absolute constants.
\end{theorem}
They also conduct numerical experiments to validate that RPCA succeeds in recovering the exact rank up to the sparsity of $10\%$ with gross errors.

\section{Experimental results for RPCA}\label{app:rpca_expirements}
We further tested our perturbed Mistral-7B instances against an extension of the SVD-based dimension-extraction attack, using the Robust PCA Principal Component Pursuit (PCP) procedure from~\cite{candes2011robust}.  

Overall, our observation was that the RPCA process did not appear to significantly enhance the attack. In each of our tested experimental noise parameter settings, the standard PCA attack and the RPCA algorithm for rank estimate either both succeeded or both failed to extract the matrix rank. 

\begin{figure*}[h!] 
    \centering
    \begin{subfigure}{0.33\textwidth} 
        \centering
        \includegraphics[width=\textwidth]{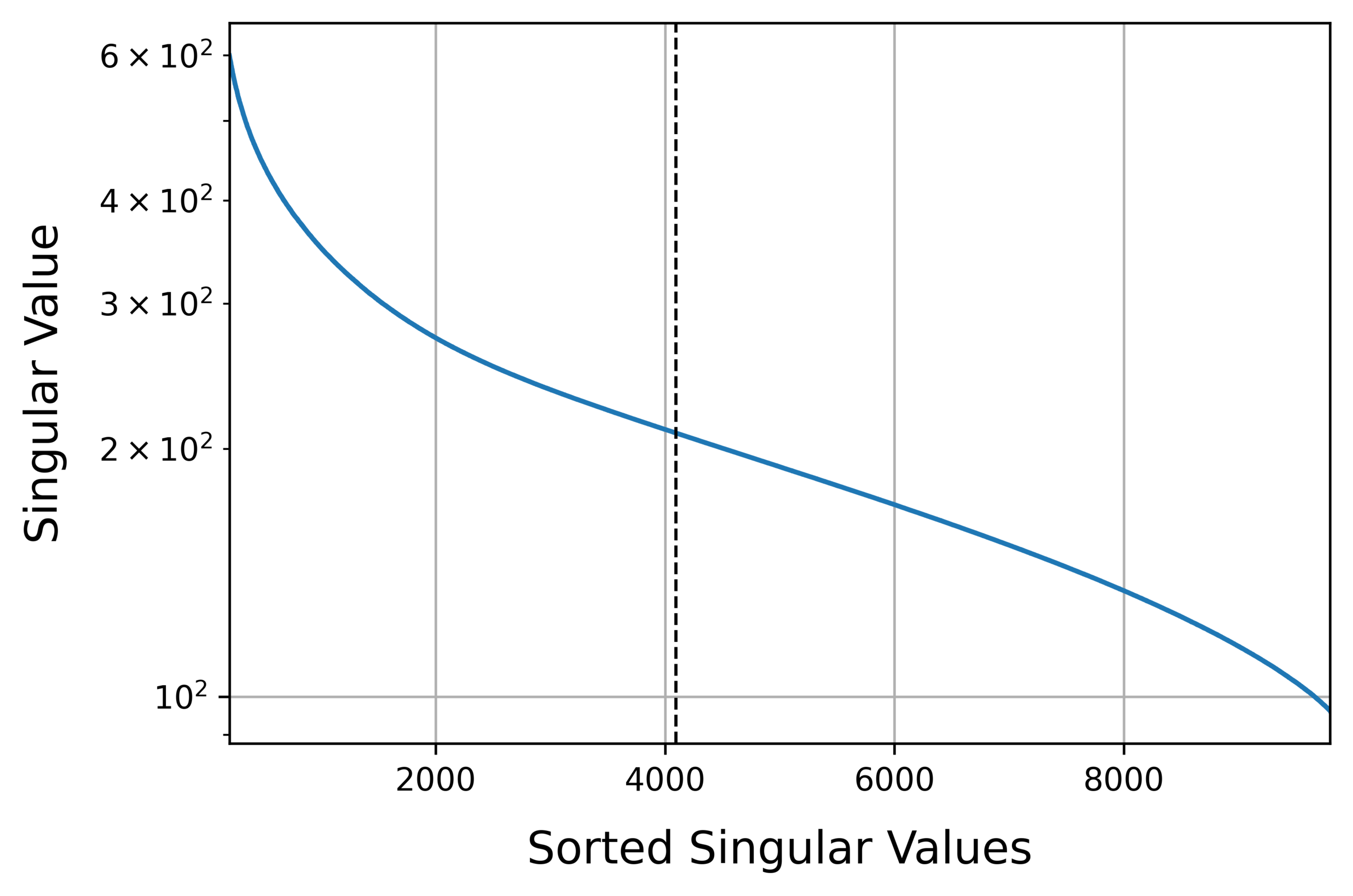}
    \end{subfigure}
    \begin{subfigure}{0.33\textwidth}
        \centering
        \includegraphics[width=\textwidth]{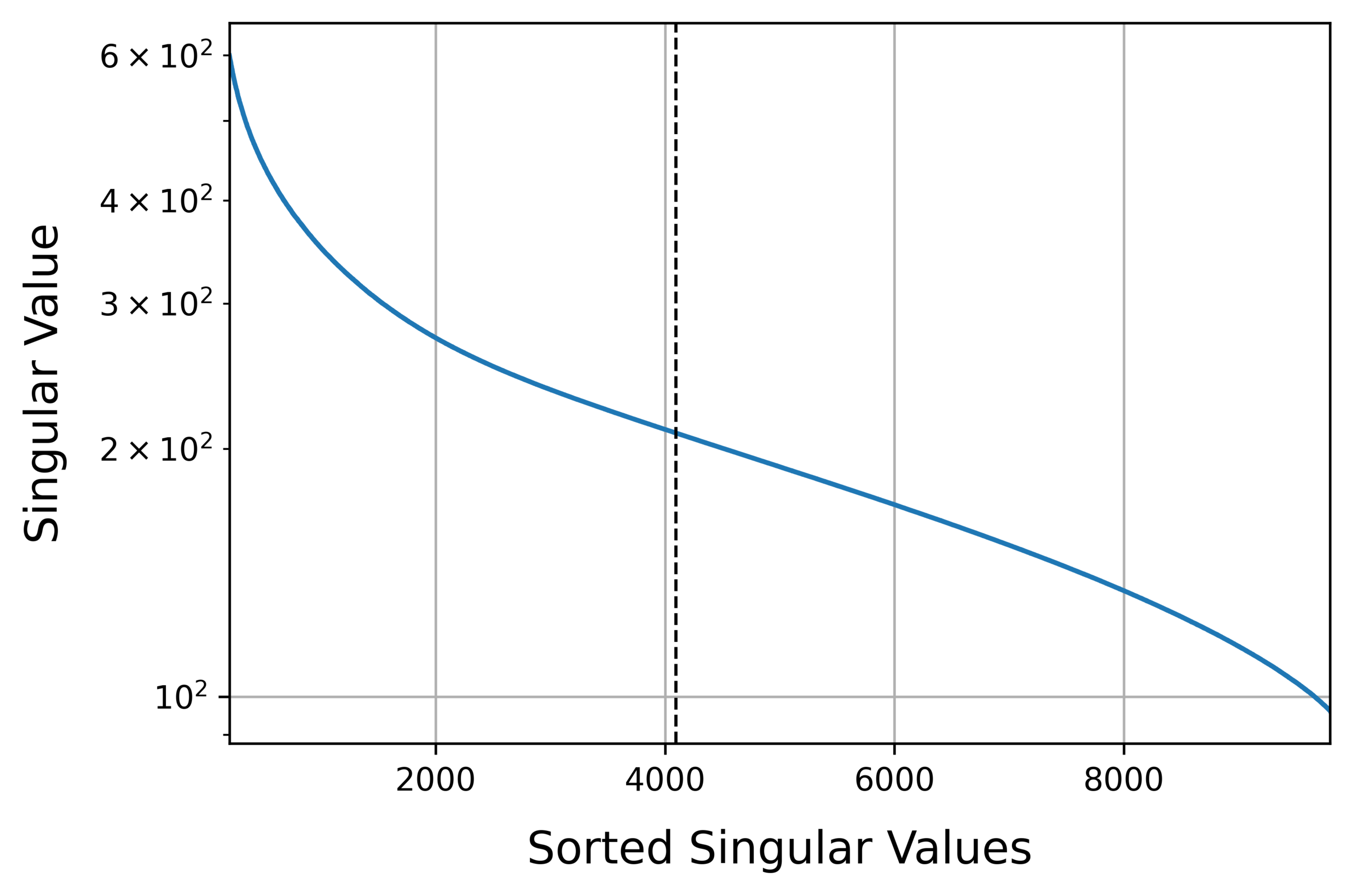}
    \end{subfigure}
    
    \vspace{0.1in}
    
    \begin{subfigure}{0.33\textwidth}
        \centering
        \includegraphics[width=\textwidth]{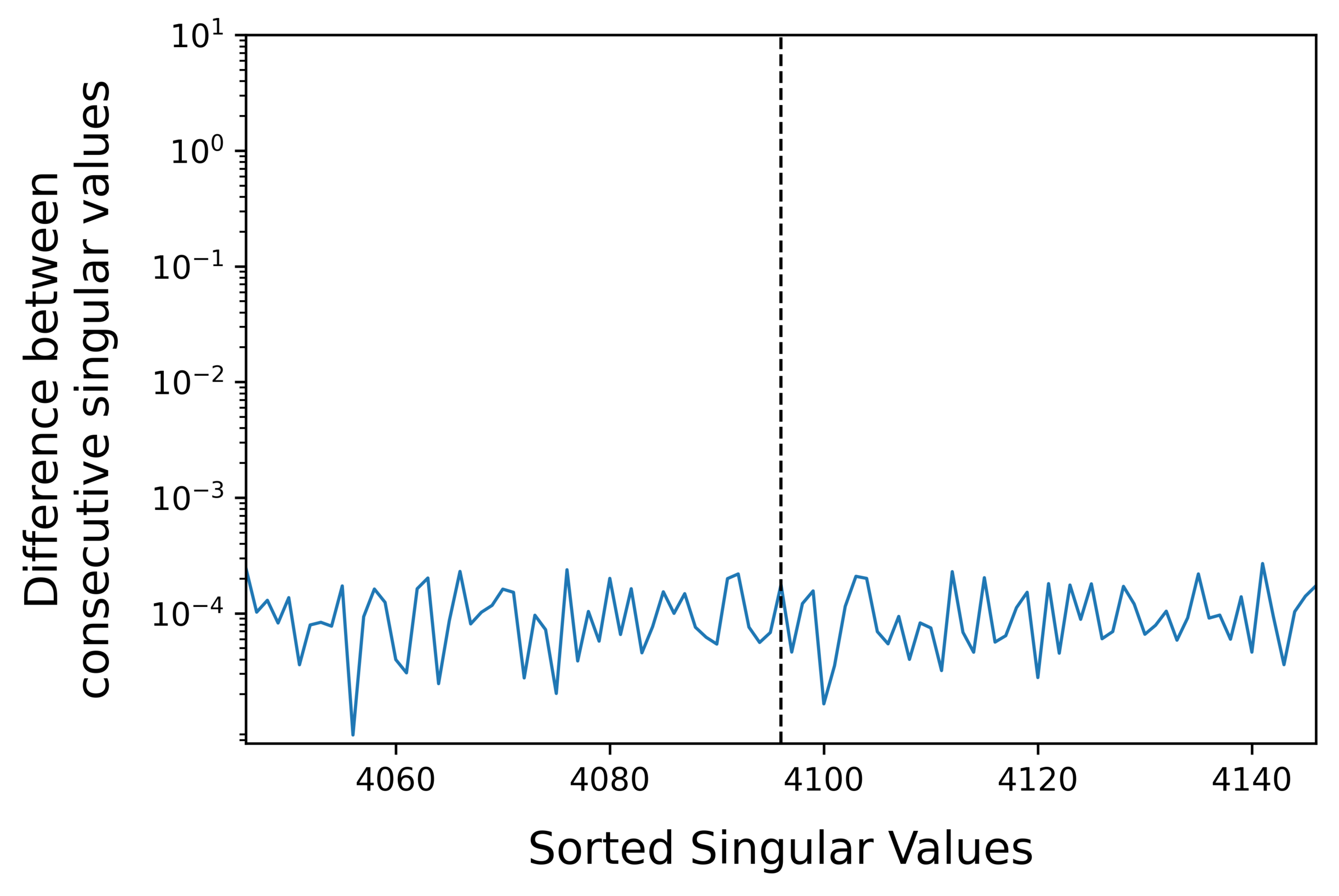}
        \caption{SVD-based attack}
    \end{subfigure}
    \begin{subfigure}{0.33\textwidth}
        \centering
        \includegraphics[width=\textwidth]{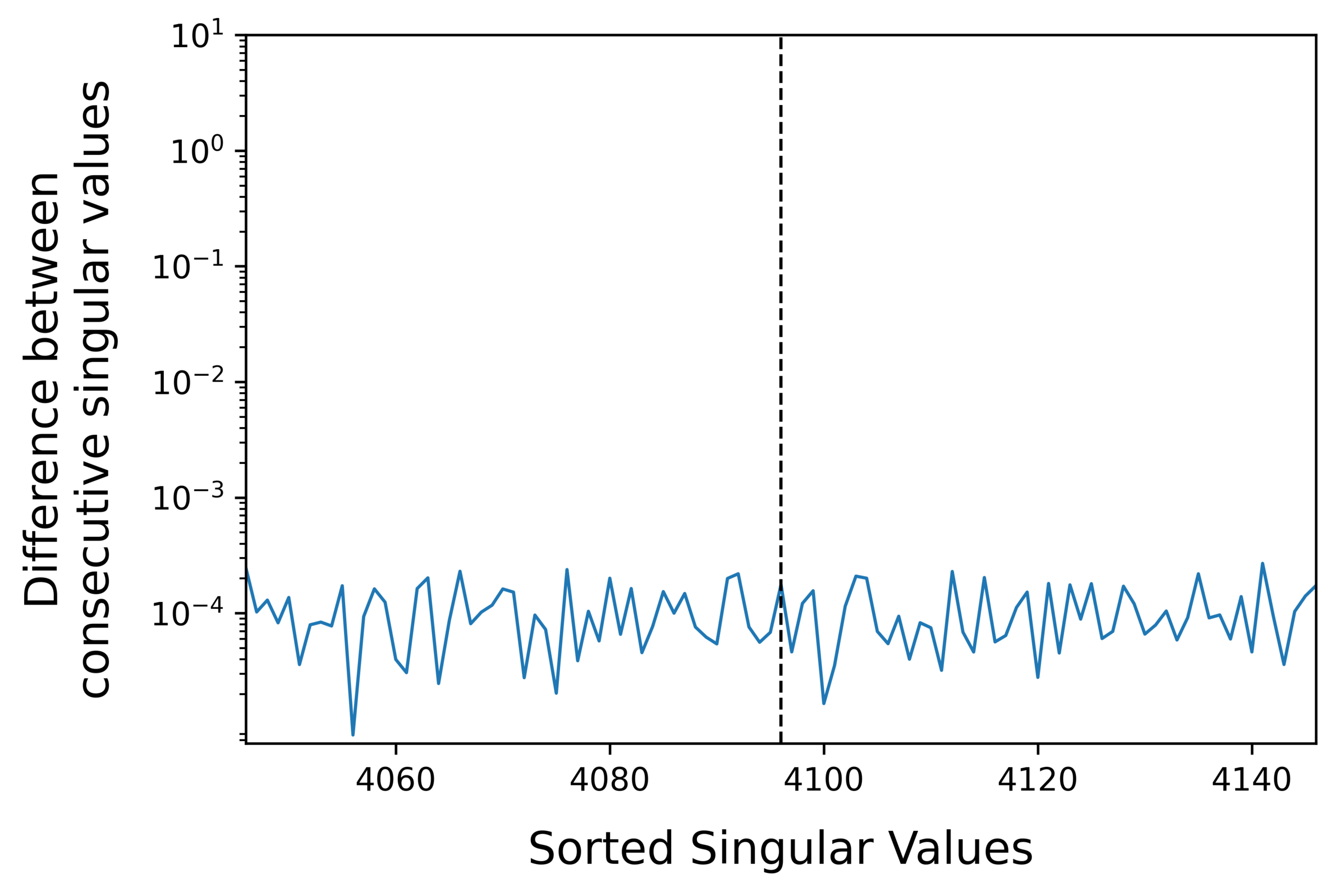}
        \caption{Robust PCA attack}
    \end{subfigure}
    \caption{(Mistral-7B RPCA). Comparison of SVD-based attack versus extended Robust PCA attack on Mistral-7B with discrete additive noise, for the case of magnitude 2.0 and sparsity 0.5. In particular, there is no significant distinction.
    }
    \label{fig:RPCA}
    \end{figure*}

We provide an example comparison between SVD and RPCA resulting outputs in Figure~\ref{fig:RPCA} for the case of discrete noise of sparsity 0.5 and magnitude 2.0.

\section{Further Experimental Results}
\subsection{PCA with averaging}\label{app:pca-with-averaging}
Recall the standard SVD-based attack forms a single matrix $A$ of size $N \times T$ from $N=10,000$ prompts.  In the averaging attack, $k$ such matrices $A_,\dots,A_k$ are independently generated, via $k \cdot N$ prompt queries, and then rank estimation is performed on the {\em average} of the matrices, $\frac{1}{k} \sum_{i=1}^k A_i$.

We tested the effectiveness of the averaging attack up to $k=50$ instances, focusing on the case of discrete noise with sparsity 0.5 and magnitude 2.0. Each matrix was generated in the same manner as the standard attack, using an independent random token pair for $k \cdot N$ total prompts.

We also simulated the same experiment in a modified setting where every prompt is given {\em independently} sampled noise, as opposed to deterministically determined (via PRF) by the hidden layer as in our construction. This was recreated by first generating $N$ {\em noise-free} logit vectors via random token pair prompts as above. Then, $A_i$ was formed by adding independently sampled noise to each row.

\begin{figure}[h!] 
    \centering
    \includegraphics[width=0.48\textwidth]{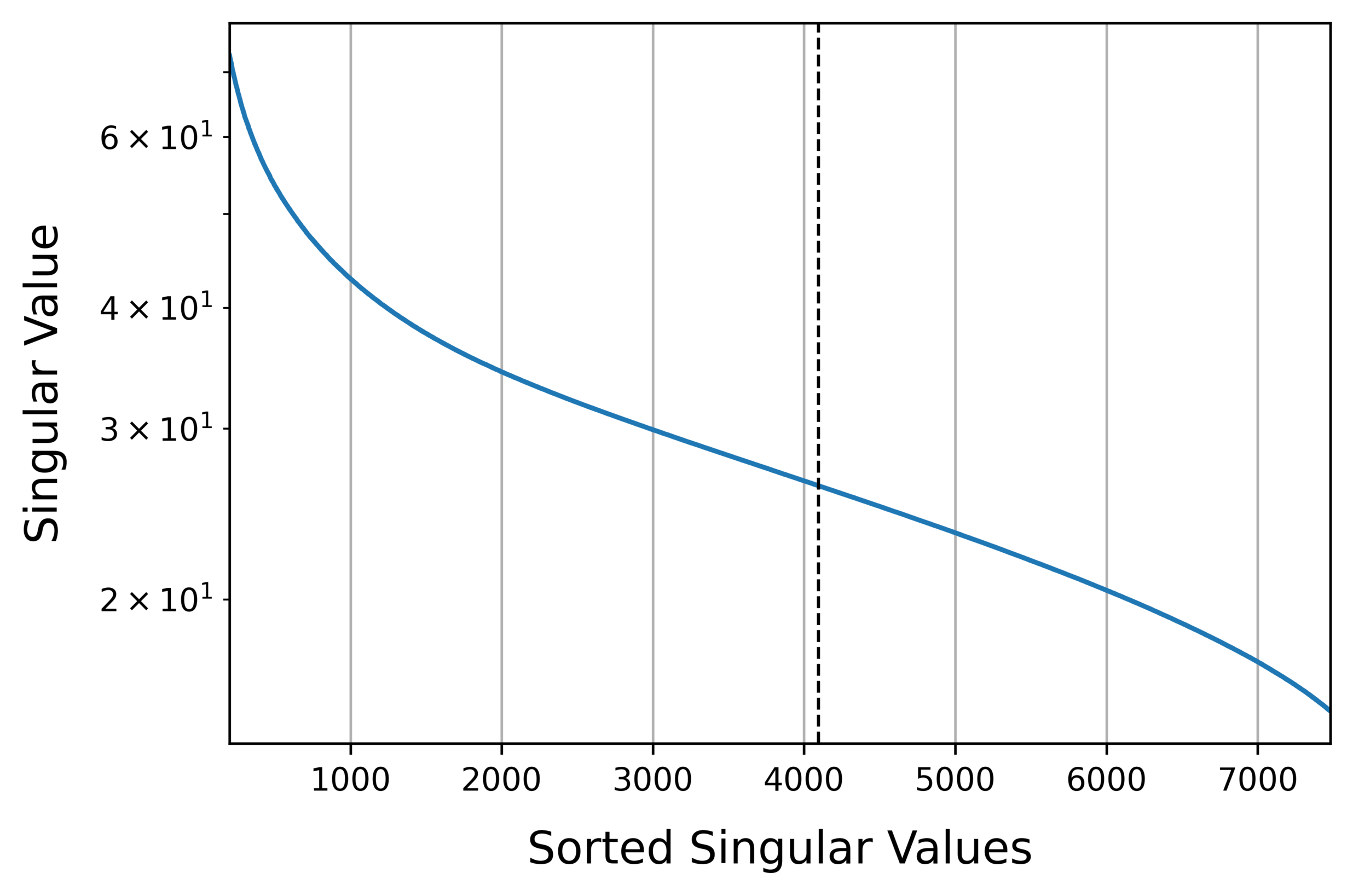}
    \caption{Sorted singular values as a result of {\em averaging} attack for $k=50$ instances on Mistral-7B with discrete noise magnitude $b=2.0$, $\gamma = 0.5$. Smoothness of the curve indicates failure of the attack to extract hidden dimension.}
    \label{fig:averaging}
\end{figure}

Our experiments indicate our (embedding-based) dependent noise withstood the averaging attack up through $k=50$; see Figure~\ref{fig:averaging}. In contrast, the independent noise version appeared to break down around $k=40$. We conjecture that this distinction comes from loss in signal precision within the dependent noise setting given $k$ different matrices as opposed to $k$ identical matrices as the starting point. 
We leave further investigation of this subject, including identifying if or when there is a breakdown point for the dependent noise setting, to future work.

\begin{table}[h!]\label{tab:mistral_perplexity}
    \centering
    \begin{minipage}{0.8\linewidth}
      \centering
      \resizebox{\textwidth}{!}{
        \begin{tabular}{cccccccc}
        \toprule
               & Original & Constant & Constant & Constant & Constant & Gaussian & Gaussian\\ \midrule
            Bias/Std dev & -  & $1$ & $2$ &$2$ & $3$ & $1$ & $1$\\ 
            Sparsity    & -  & $0.25$ & $0.1$ & $0.5$& $0.1$ & $0.25$ & $0.5$\\
            Perplexity    & $19.34$  & $20.59$ & $40.16$ & $38.8$& $66.72$ & $31.63$ & $39.85$\\
            \bottomrule
        \end{tabular}
        }
    \end{minipage}%
    \caption{Perplexity results for GPT2. Original refers to the original GPT2 model, Constant denotes constant noise, and Gaussian refers to Gaussian noise added to the original model. Bias/Std dev denotes the magnitude of the noise bias for discrete noise and standard deviation for Gaussian noise.
    Sparsity denotes the sparsity parameter $\gamma$ of the noise, and Perplexity is evaluated against Meta-Llama-3-8B as the larger oracle model.}
    \label{tab:gpt2_perplexity_tab}
\end{table}

\subsection{Mistral-7B with additional noise parameters} \label{appendix:more-noise}

\begin{figure*}[h!] 
    \centering
    

    \begin{subfigure}{0.24\textwidth} 
        \centering
        \includegraphics[width=\textwidth]{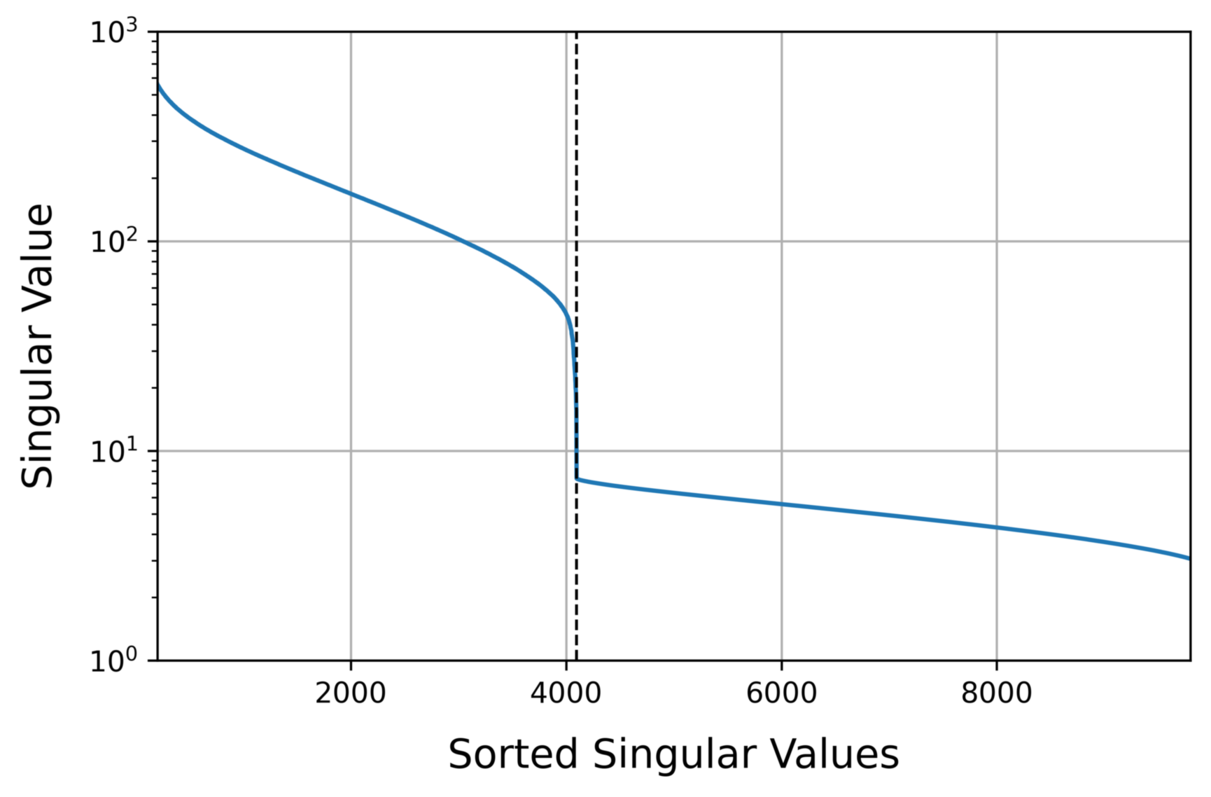}
    \end{subfigure}
    \begin{subfigure}{0.24\textwidth}
        \centering
        \includegraphics[width=\textwidth]{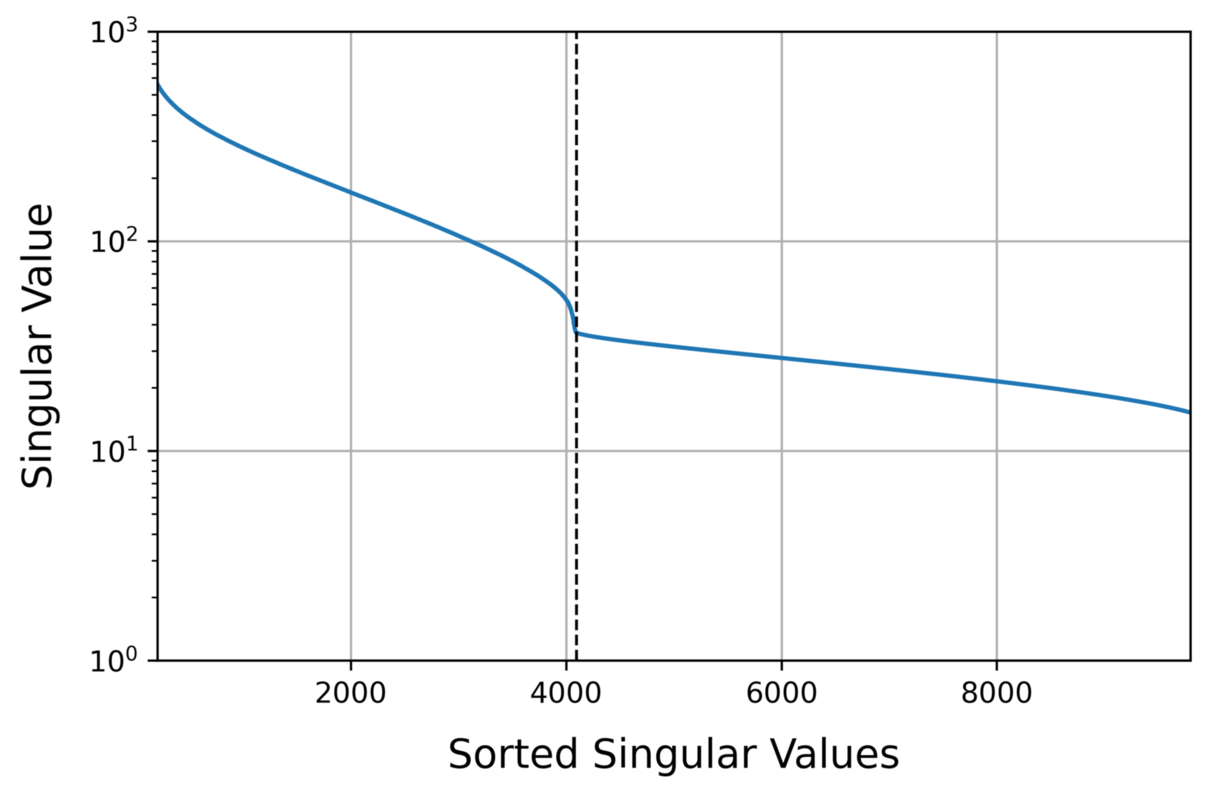}
    \end{subfigure}
    \begin{subfigure}{0.24\textwidth}
        \centering
        \includegraphics[width=\textwidth]{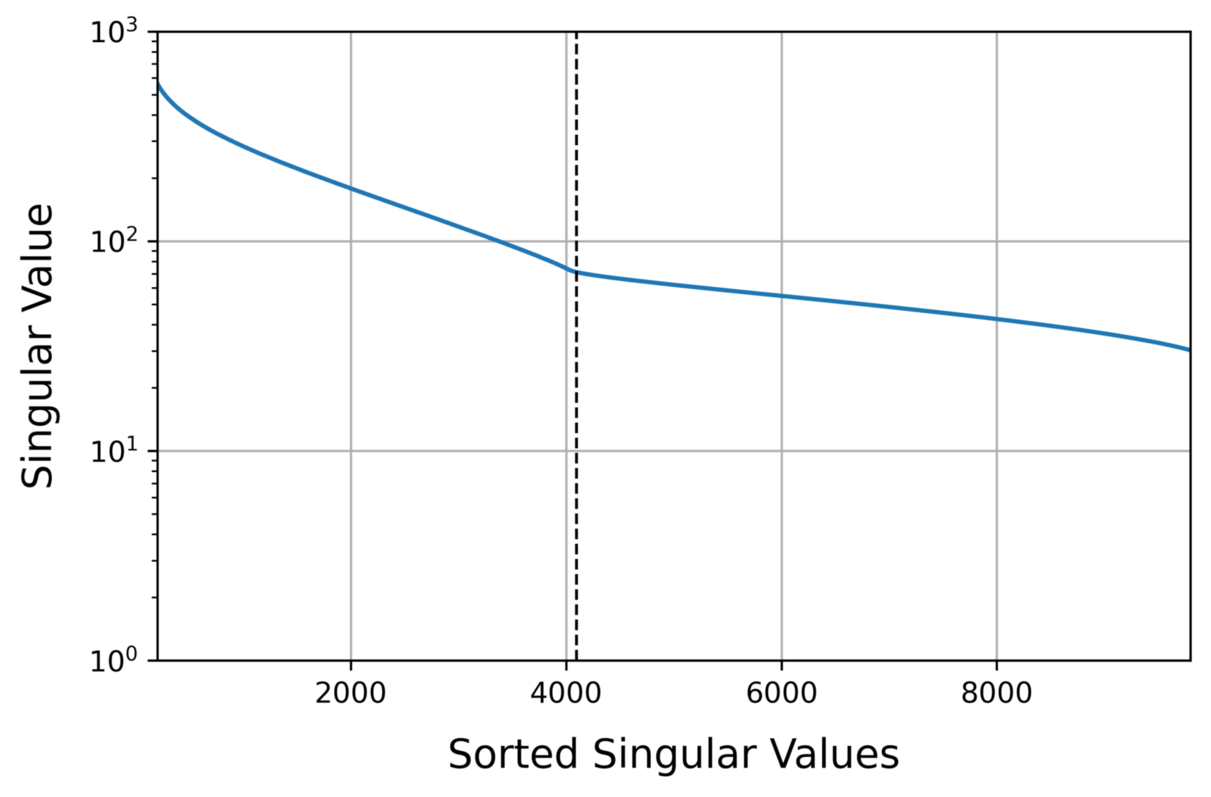}
    \end{subfigure}
    \begin{subfigure}{0.24\textwidth}
        \centering
        \includegraphics[width=\textwidth]{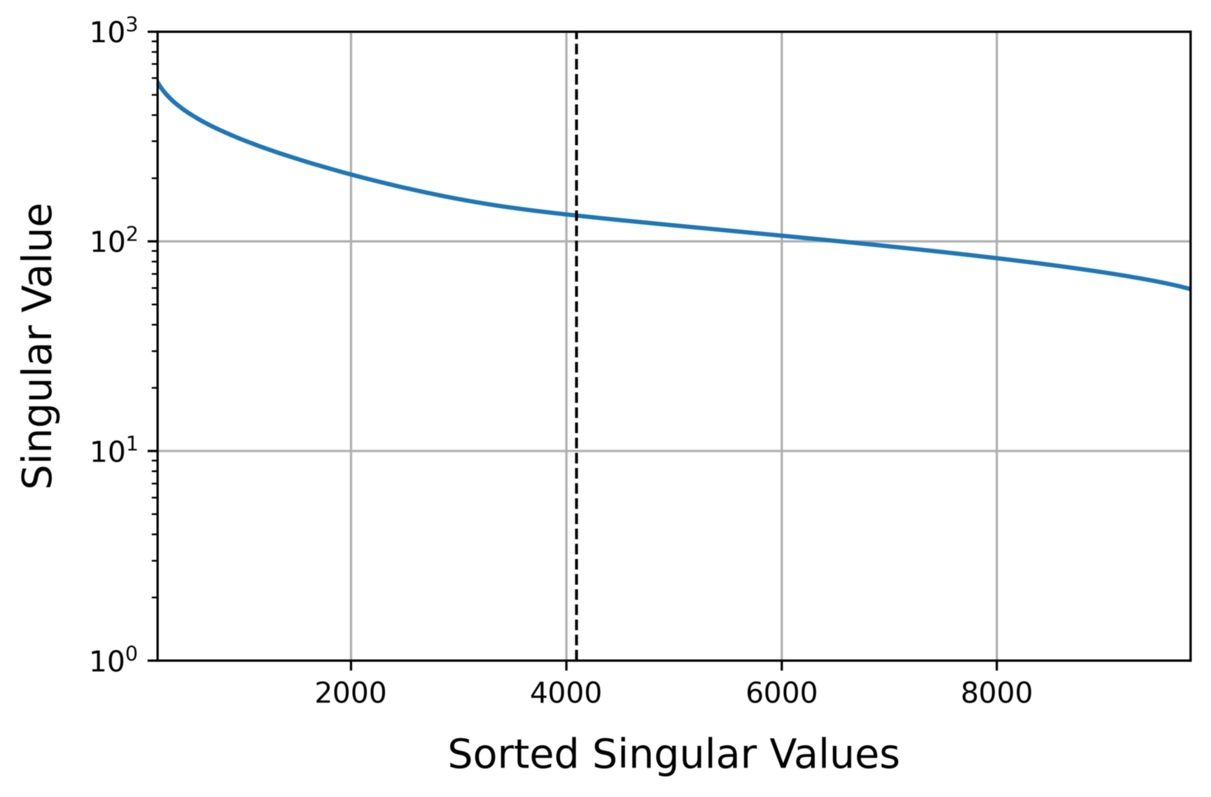}
    \end{subfigure}


    \begin{subfigure}{0.24\textwidth} 
        \centering
        \includegraphics[width=\textwidth]{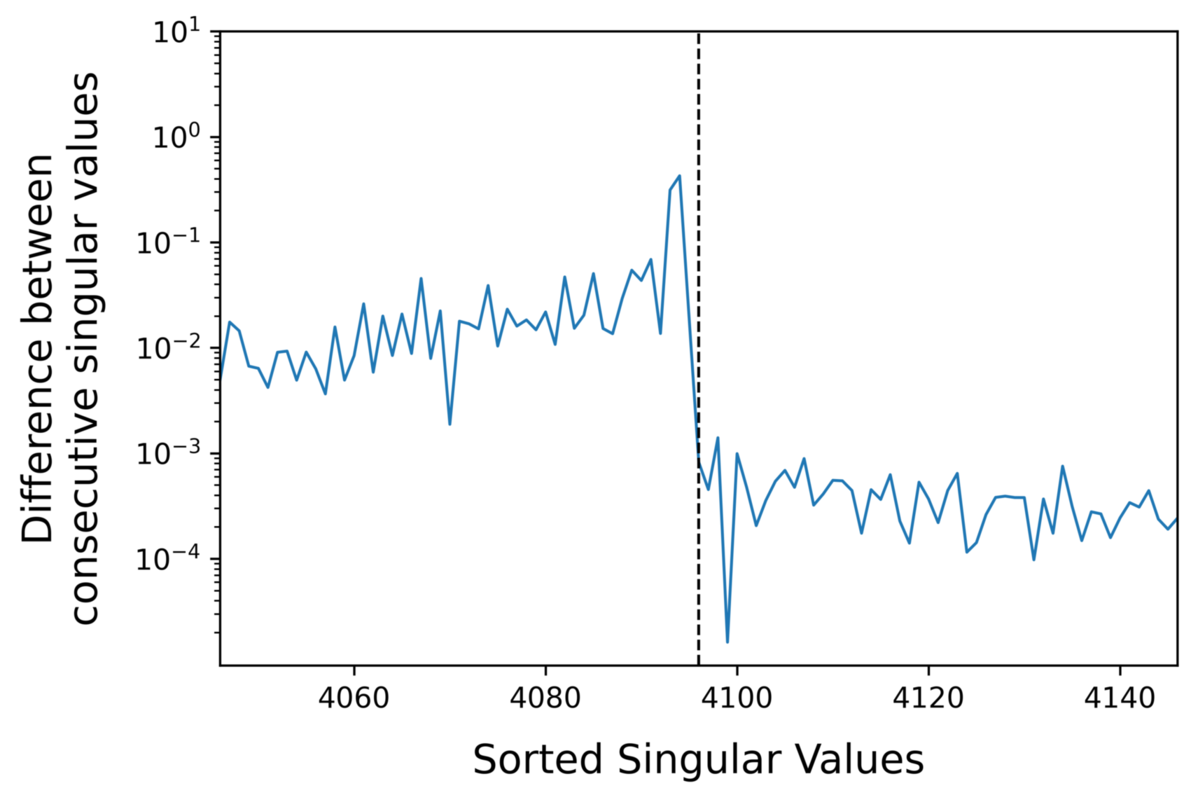}
        \caption{Noise 0.1, Sparsity 0.1}
    \end{subfigure}
    \begin{subfigure}{0.24\textwidth}
        \centering
        \includegraphics[width=\textwidth]{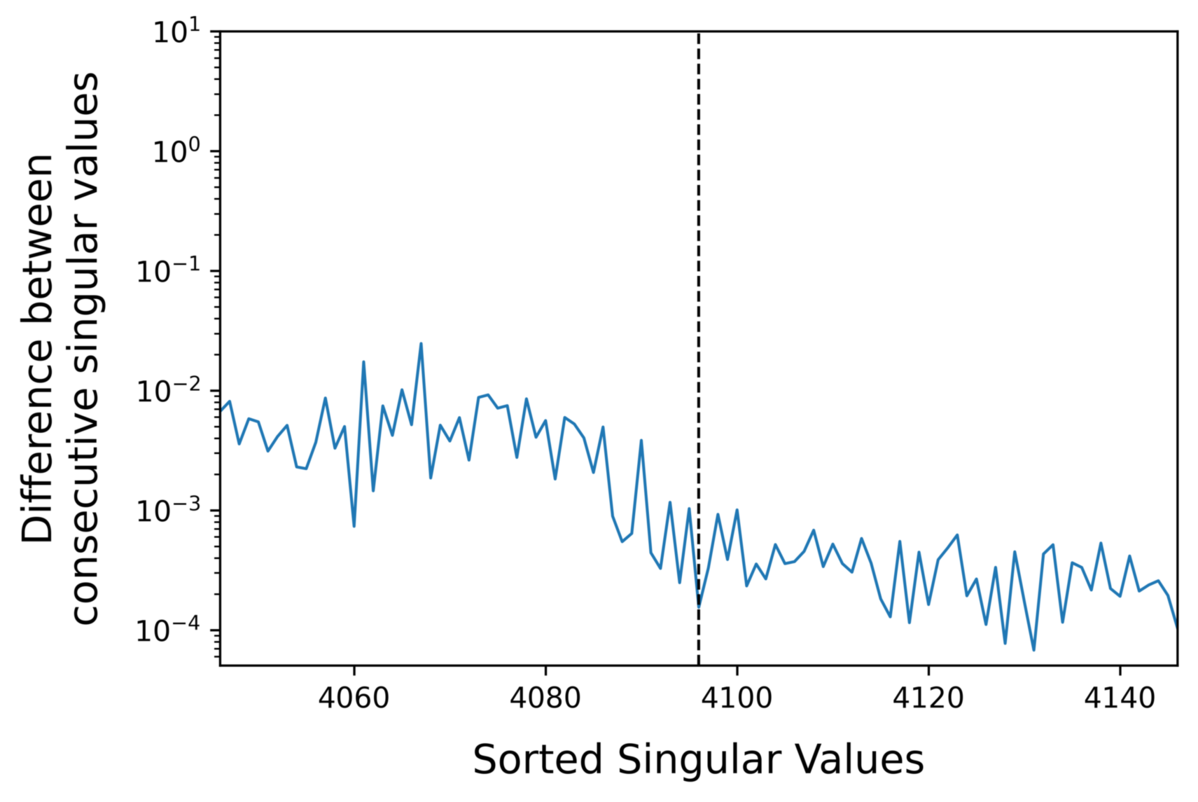}
        \caption{Noise 0.5, Sparsity 0.1}
    \end{subfigure}
    \begin{subfigure}{0.24\textwidth}
        \centering
        \includegraphics[width=\textwidth]{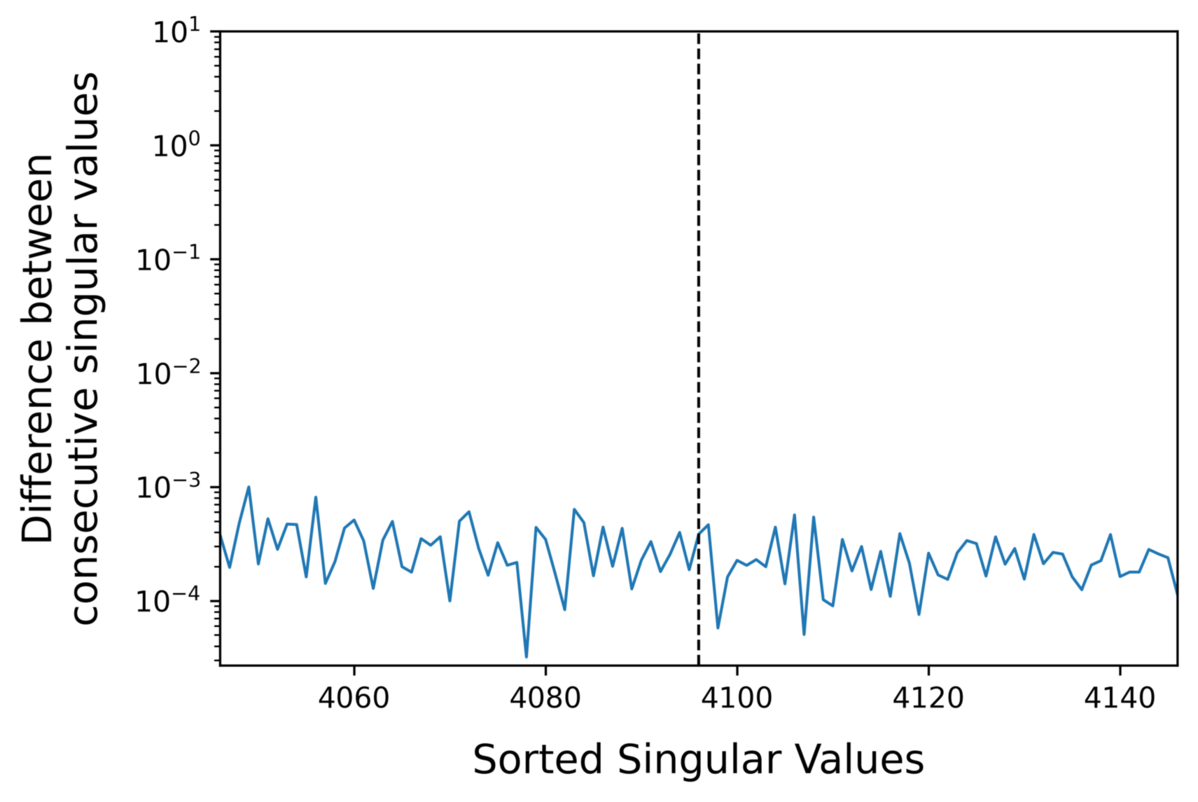}
        \caption{Noise 1.0, Sparsity 0.1}
    \end{subfigure}
    \begin{subfigure}{0.24\textwidth}
        \centering
        \includegraphics[width=\textwidth]{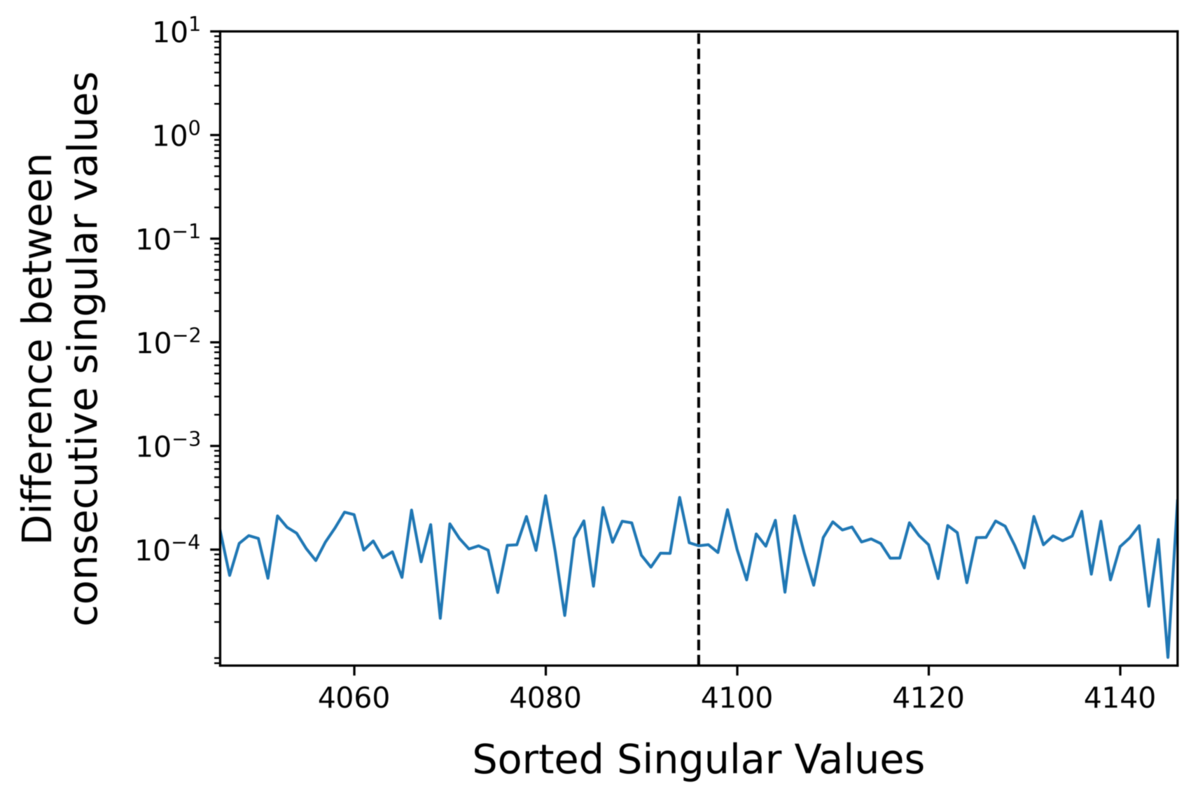}
        \caption{Noise 2.0, Sparsity 0.1}
    \end{subfigure}


\vspace{.2in}

    \begin{subfigure}{0.24\textwidth} 
        \centering
        \includegraphics[width=\textwidth]{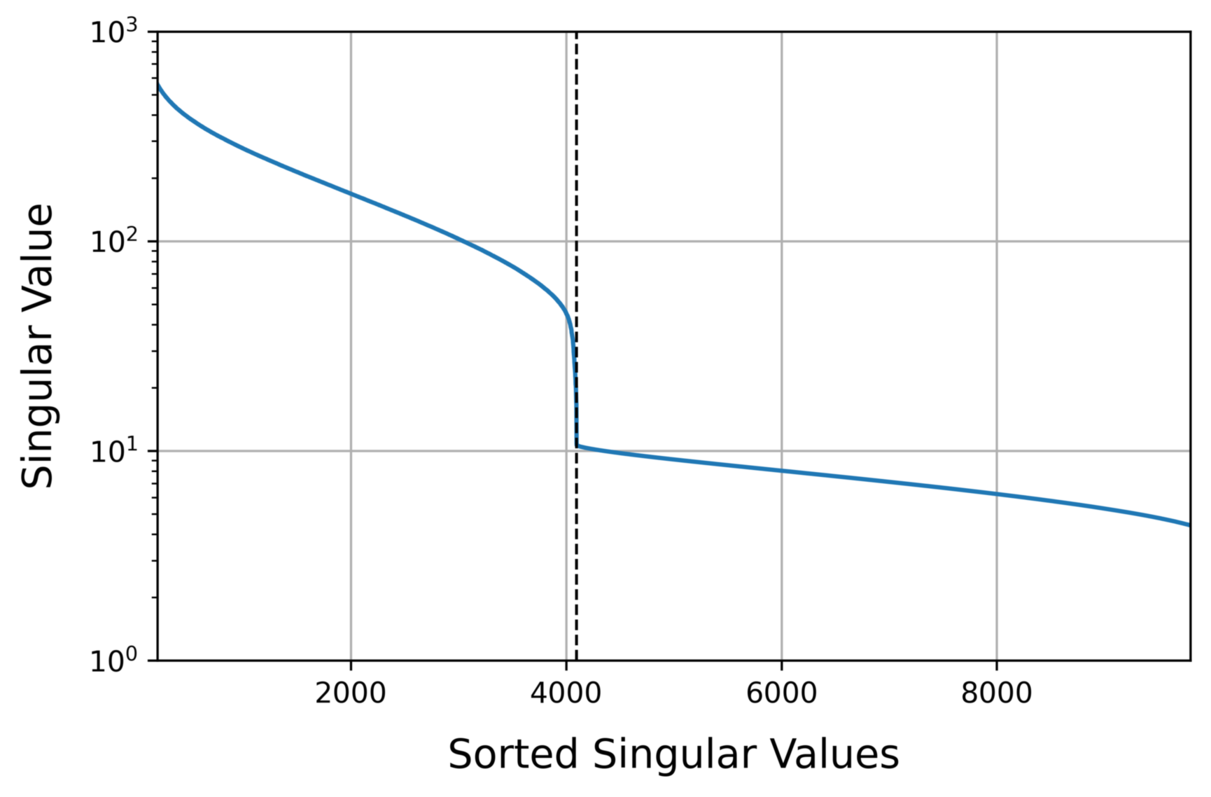}
    \end{subfigure}
    \begin{subfigure}{0.24\textwidth}
        \centering
        \includegraphics[width=\textwidth]{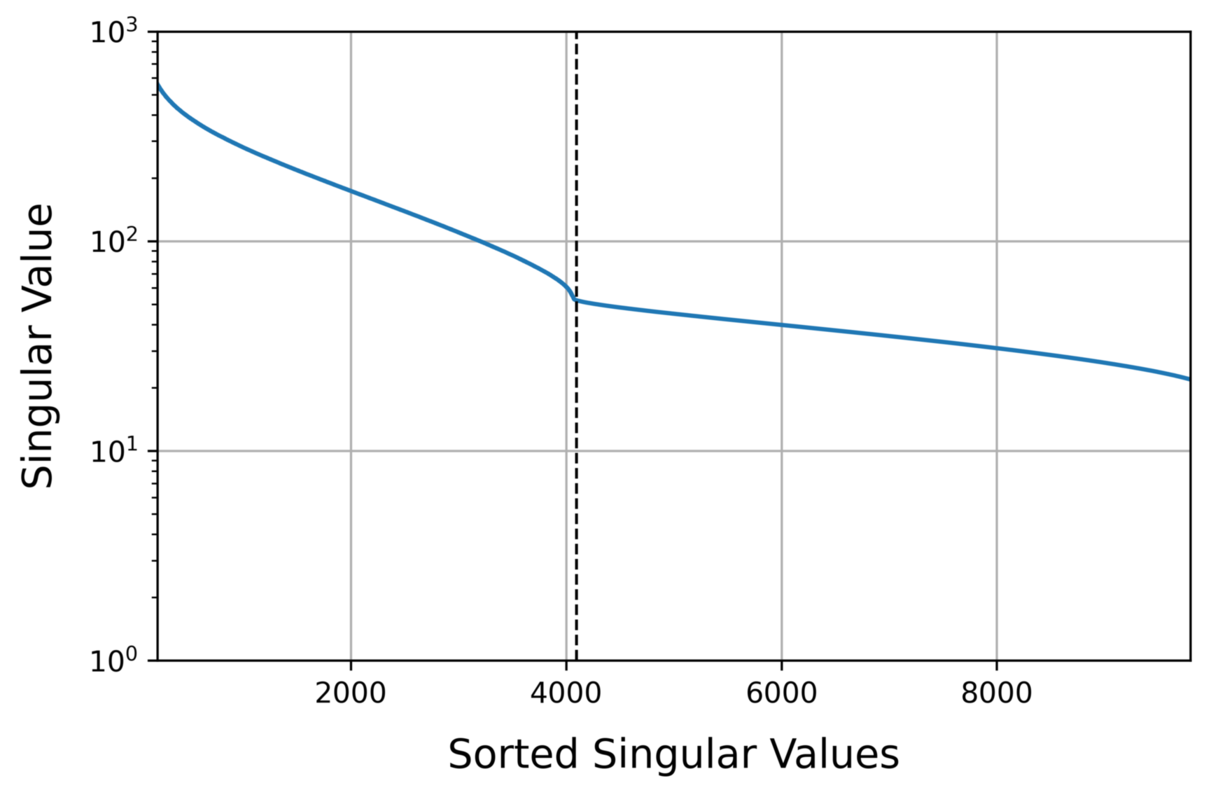}
    \end{subfigure}
    \begin{subfigure}{0.24\textwidth}
        \centering
        \includegraphics[width=\textwidth]{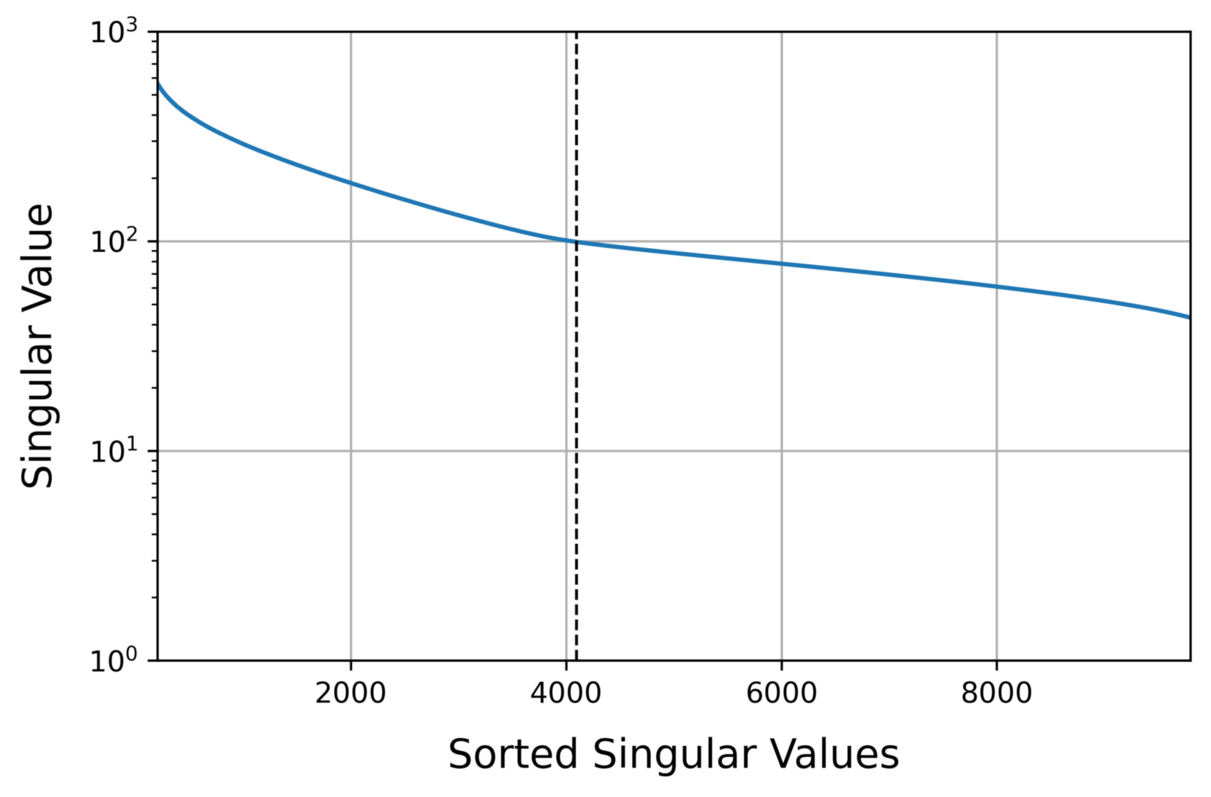}
    \end{subfigure}
    \begin{subfigure}{0.24\textwidth}
        \centering
        \includegraphics[width=\textwidth]{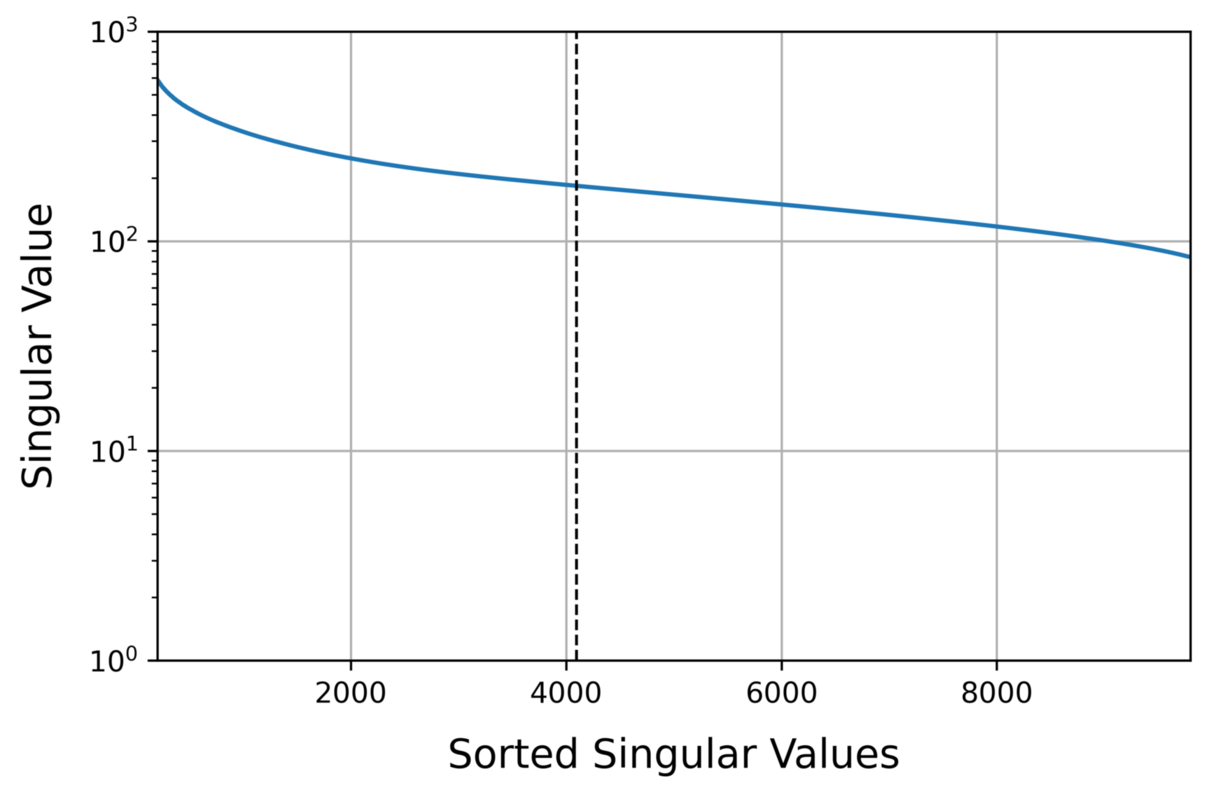}
    \end{subfigure}


    \begin{subfigure}{0.24\textwidth} 
        \centering
        \includegraphics[width=\textwidth]{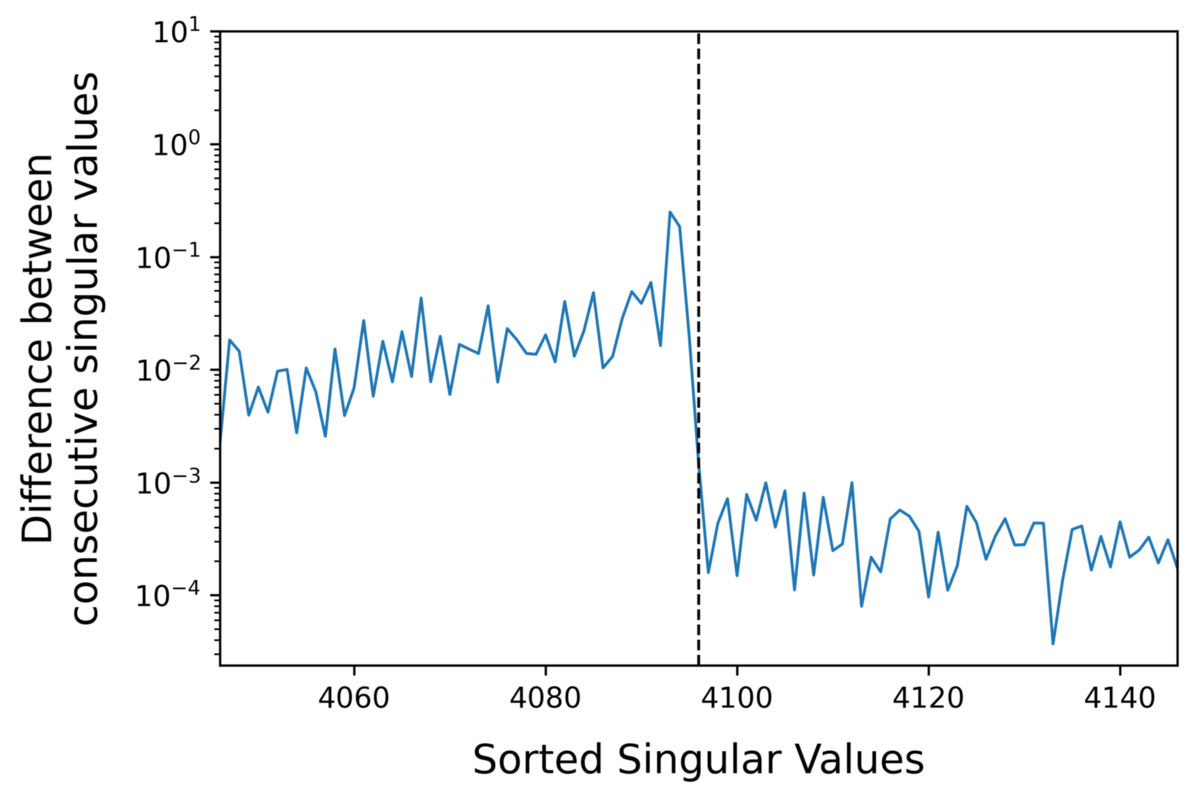}
        \caption{Noise 0.1, Sparsity 0.25}
    \end{subfigure}
    \begin{subfigure}{0.24\textwidth}
        \centering
        \includegraphics[width=\textwidth]{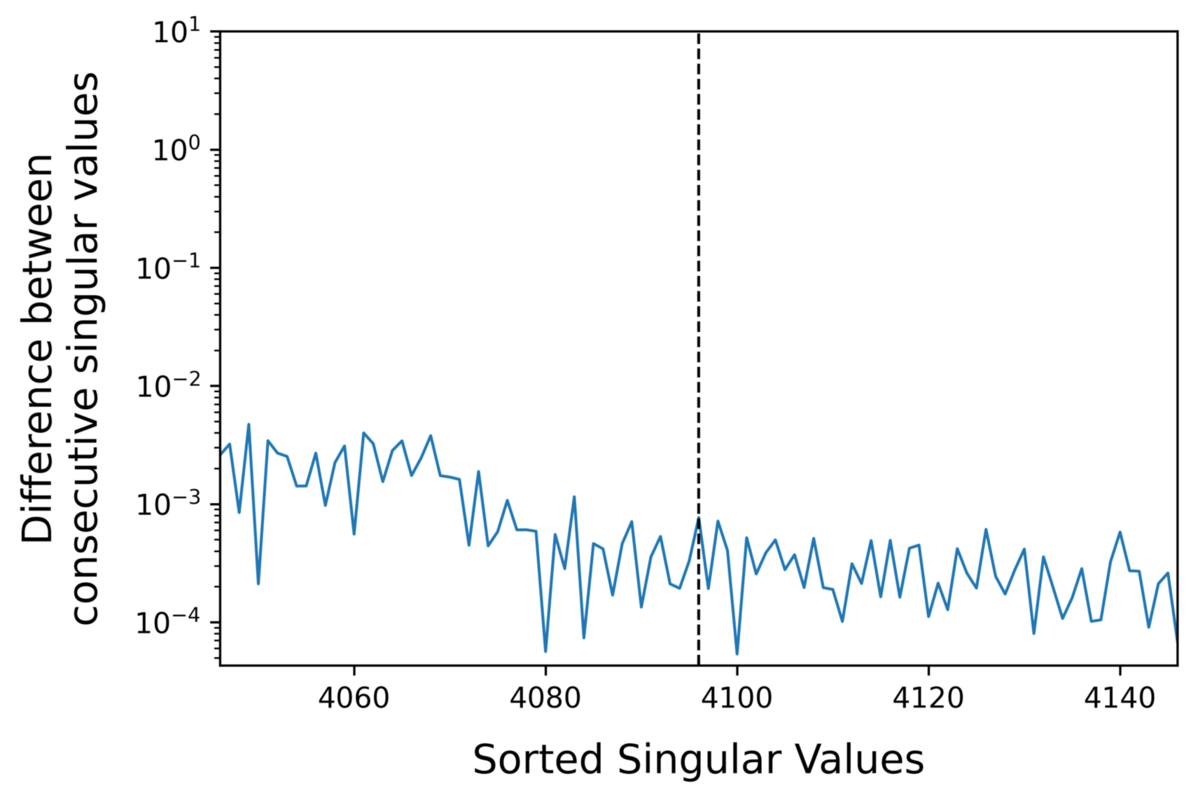}
        \caption{Noise 0.5, Sparsity 0.25}
    \end{subfigure}
    \begin{subfigure}{0.24\textwidth}
        \centering
        \includegraphics[width=\textwidth]{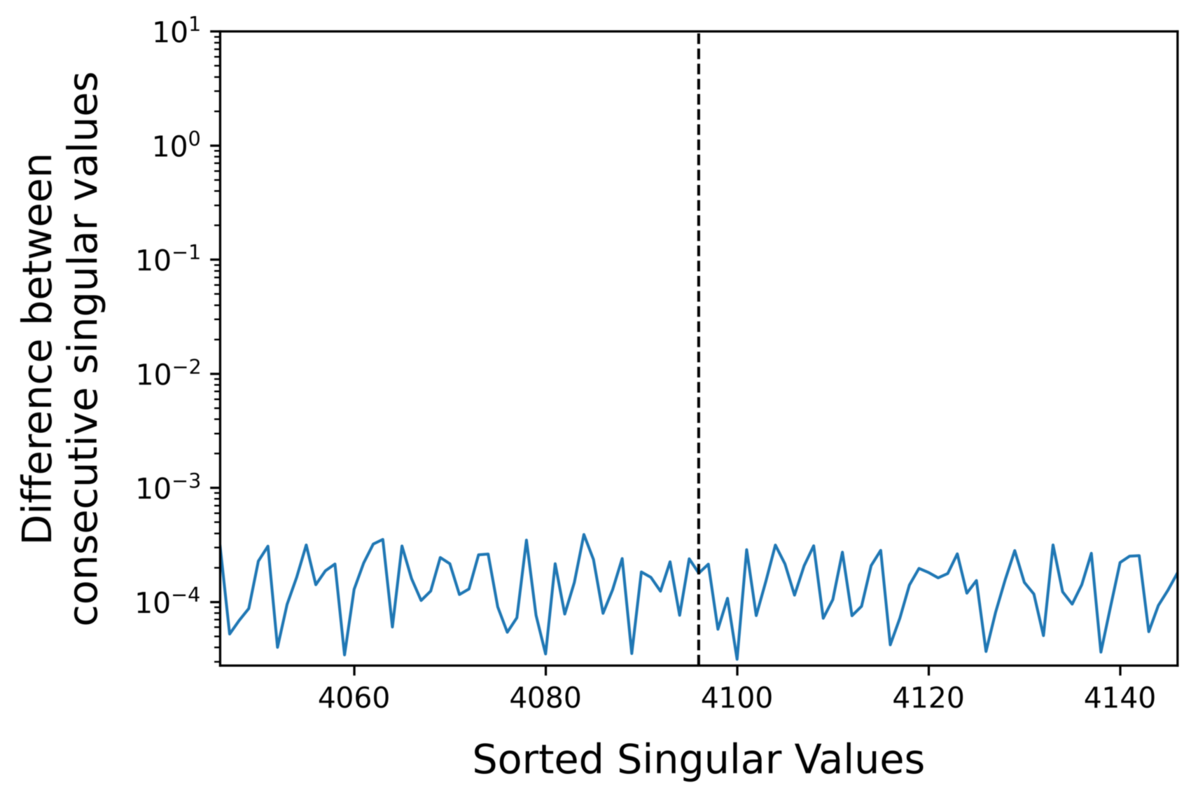}
        \caption{Noise 1.0, Sparsity 0.25}
    \end{subfigure}
    \begin{subfigure}{0.24\textwidth}
        \centering
        \includegraphics[width=\textwidth]{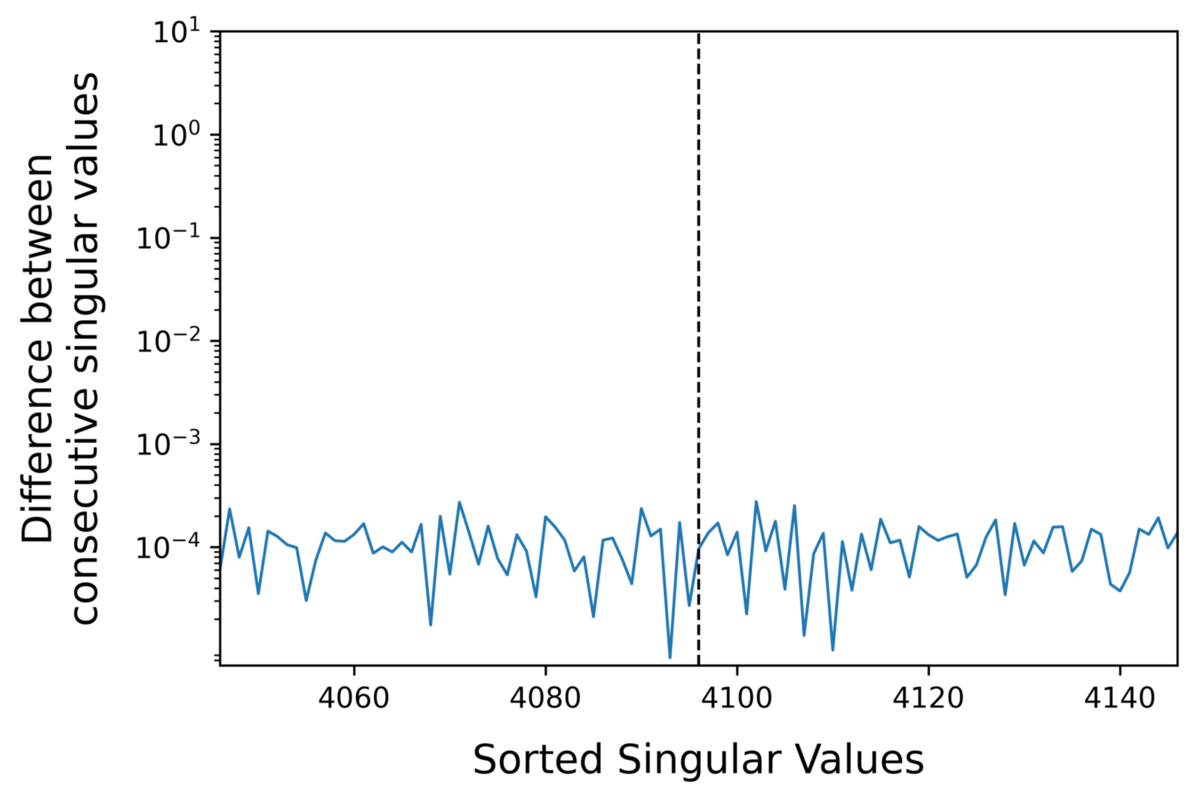}
        \caption{Noise 2.0, Sparsity 0.25}
    \end{subfigure}


\vspace{.2in}

    \begin{subfigure}{0.24\textwidth} 
        \centering
        \includegraphics[width=\textwidth]{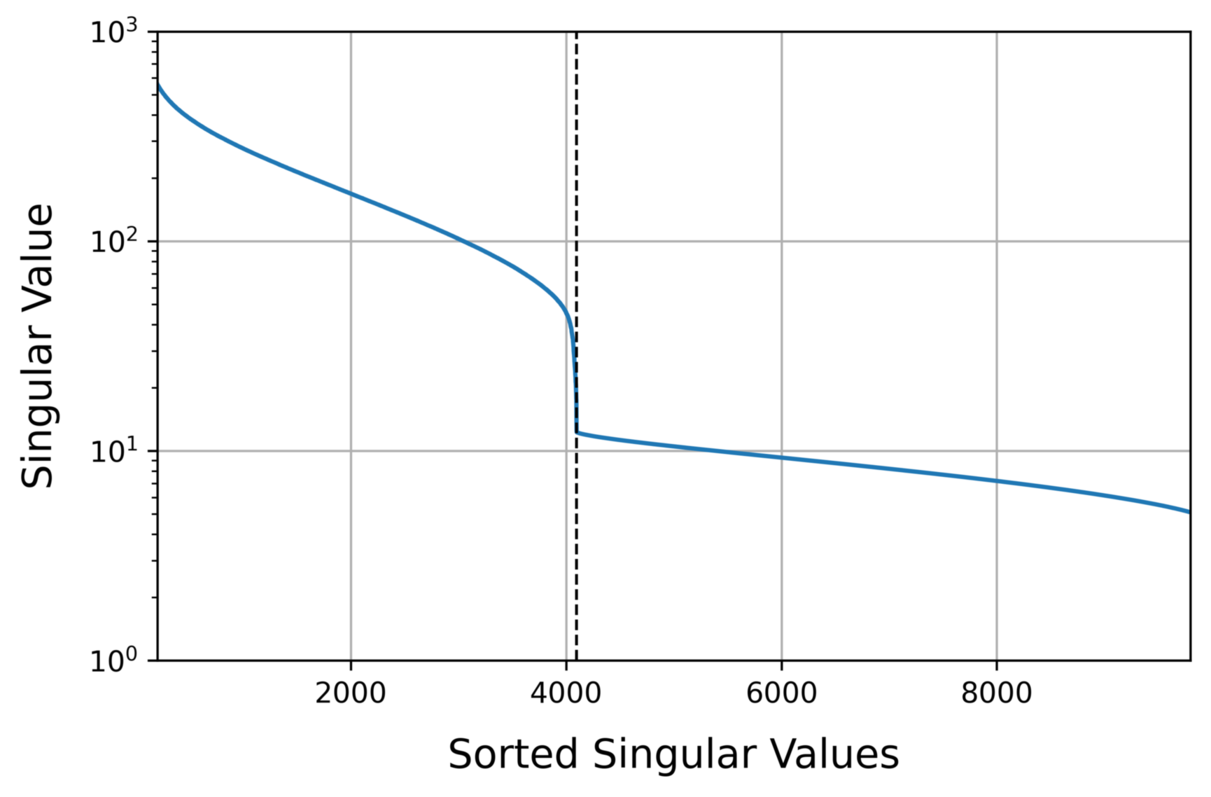}
    \end{subfigure}
    \begin{subfigure}{0.24\textwidth}
        \centering
        \includegraphics[width=\textwidth]{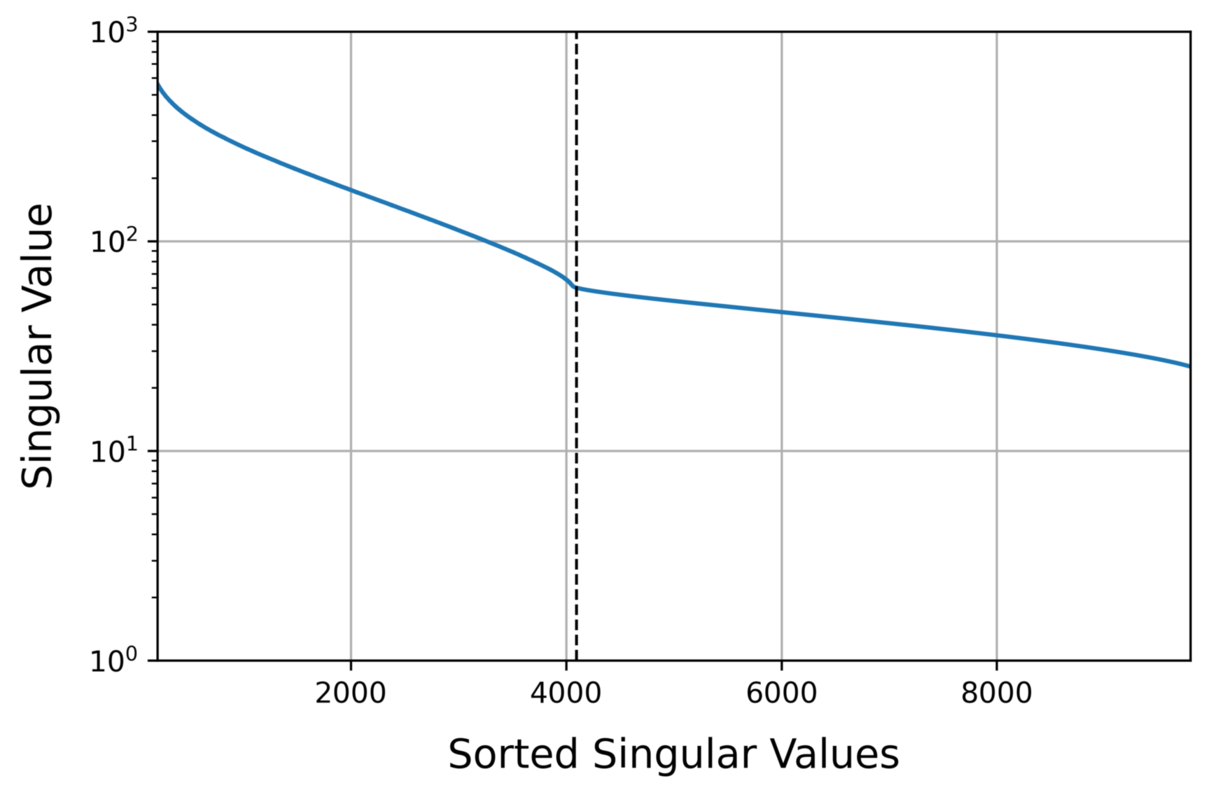}
    \end{subfigure}
    \begin{subfigure}{0.24\textwidth}
        \centering
        \includegraphics[width=\textwidth]{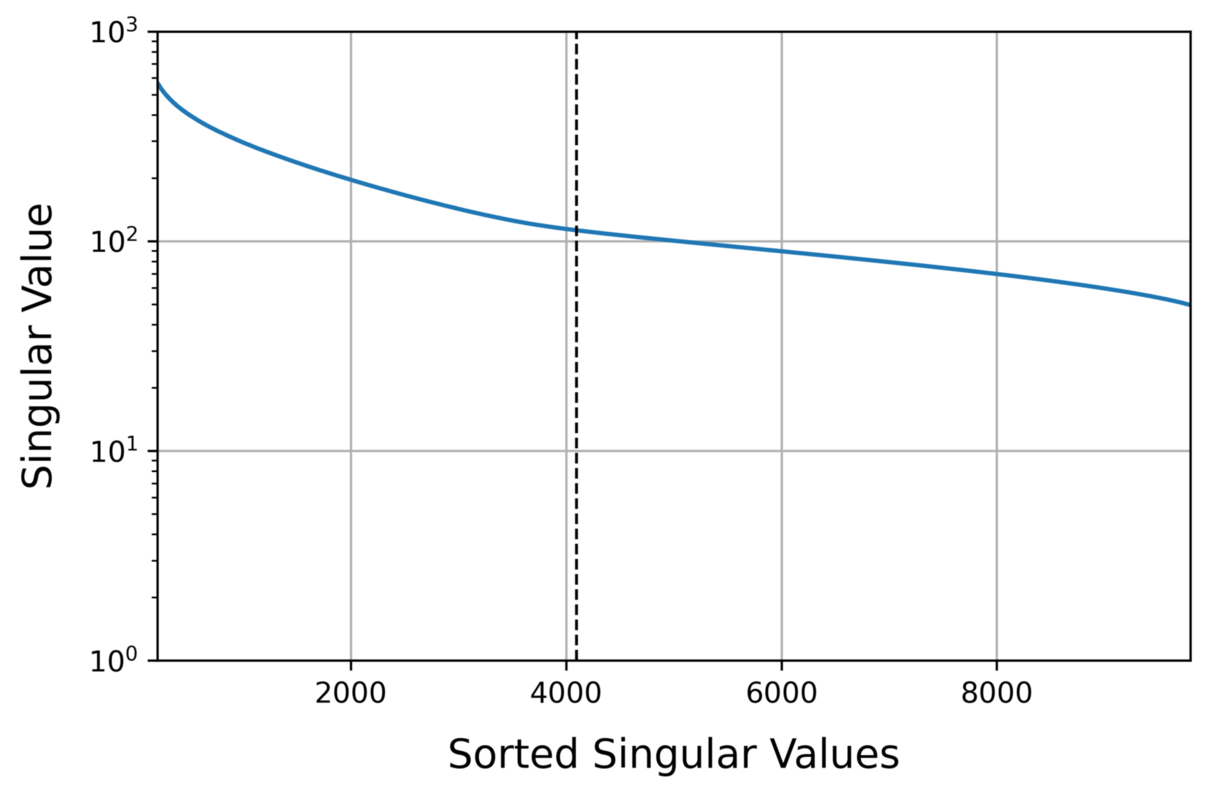}
    \end{subfigure}
    \begin{subfigure}{0.24\textwidth}
        \centering
        \includegraphics[width=\textwidth]{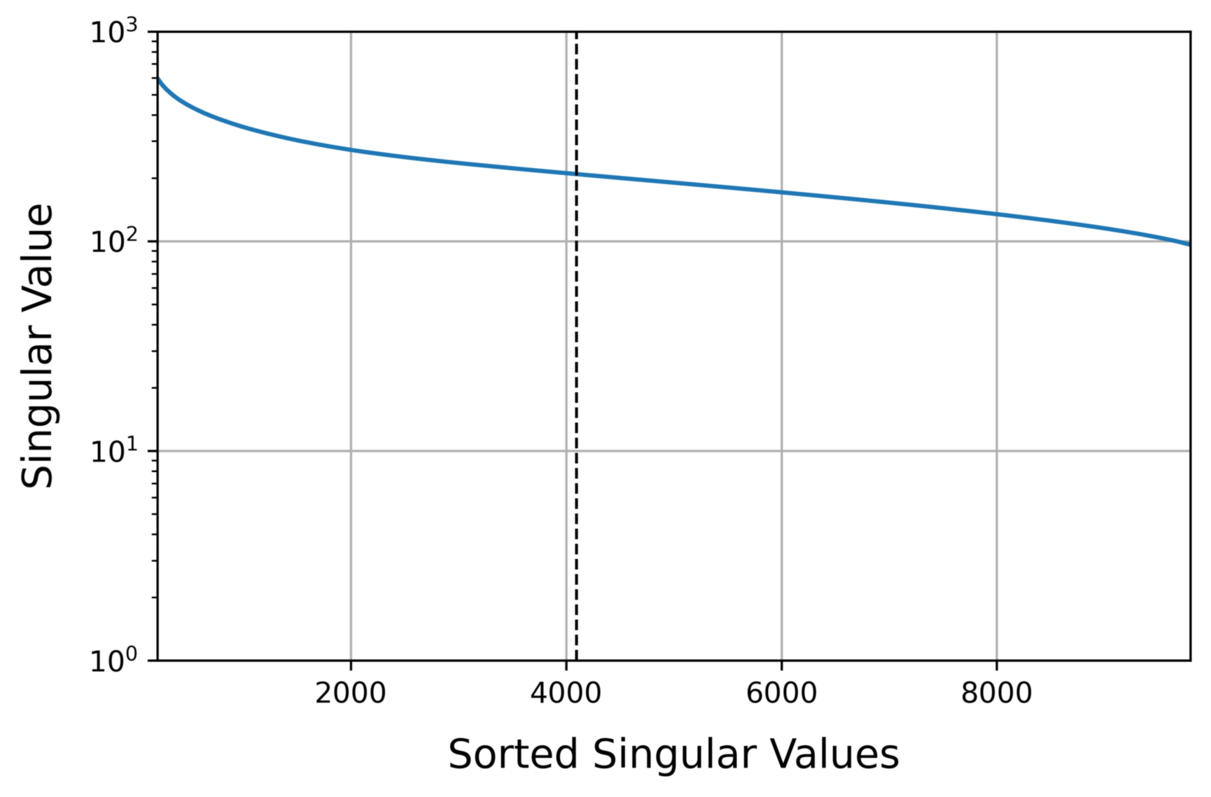}
    \end{subfigure}


    \begin{subfigure}{0.24\textwidth} 
        \centering
        \includegraphics[width=\textwidth]{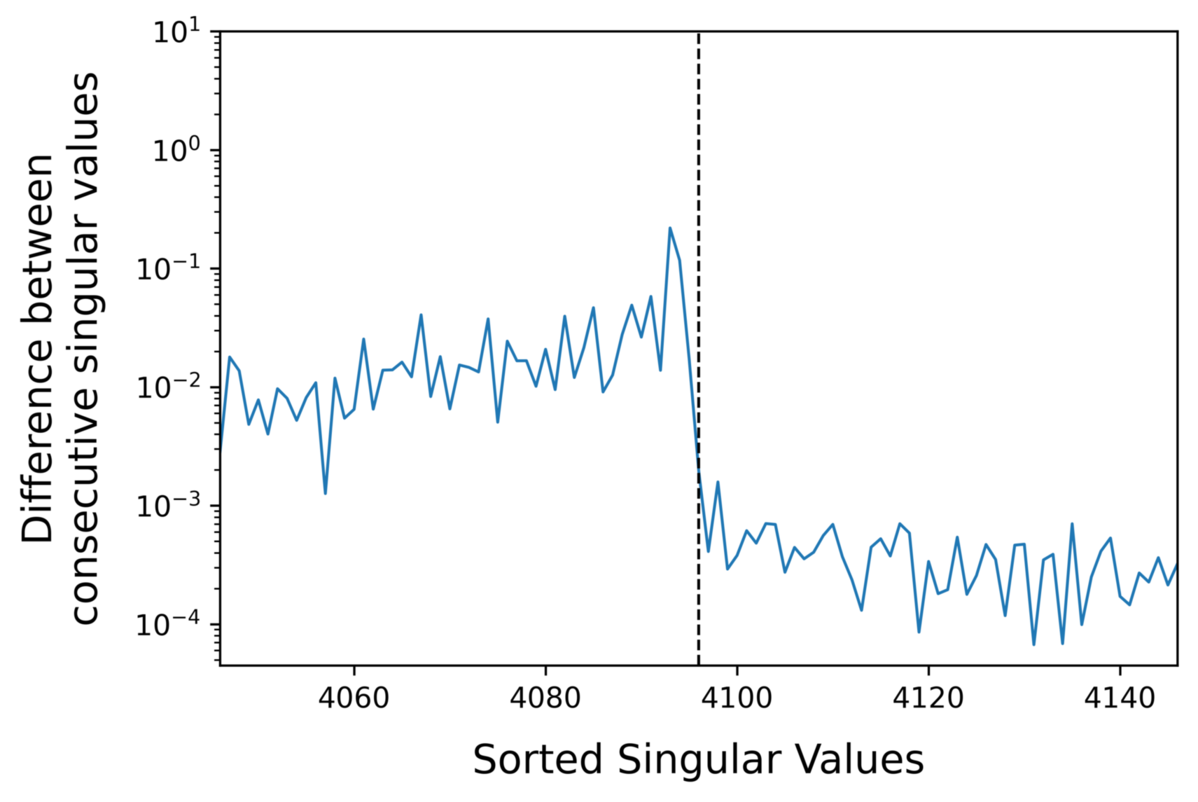}
        \caption{Noise 0.1, Sparsity 0.5}
    \end{subfigure}
    \begin{subfigure}{0.24\textwidth}
        \centering
        \includegraphics[width=\textwidth]{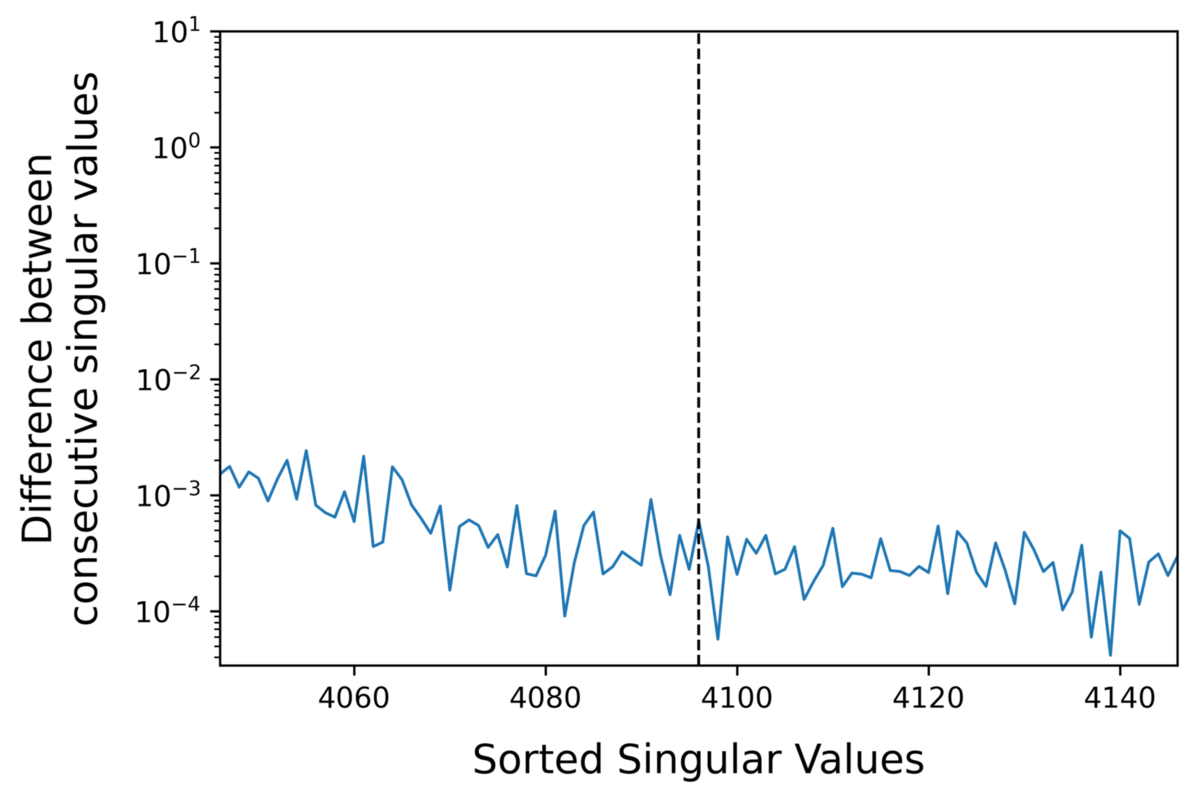}
        \caption{Noise 0.5, Sparsity 0.5}
    \end{subfigure}
    \begin{subfigure}{0.24\textwidth}
        \centering
        \includegraphics[width=\textwidth]{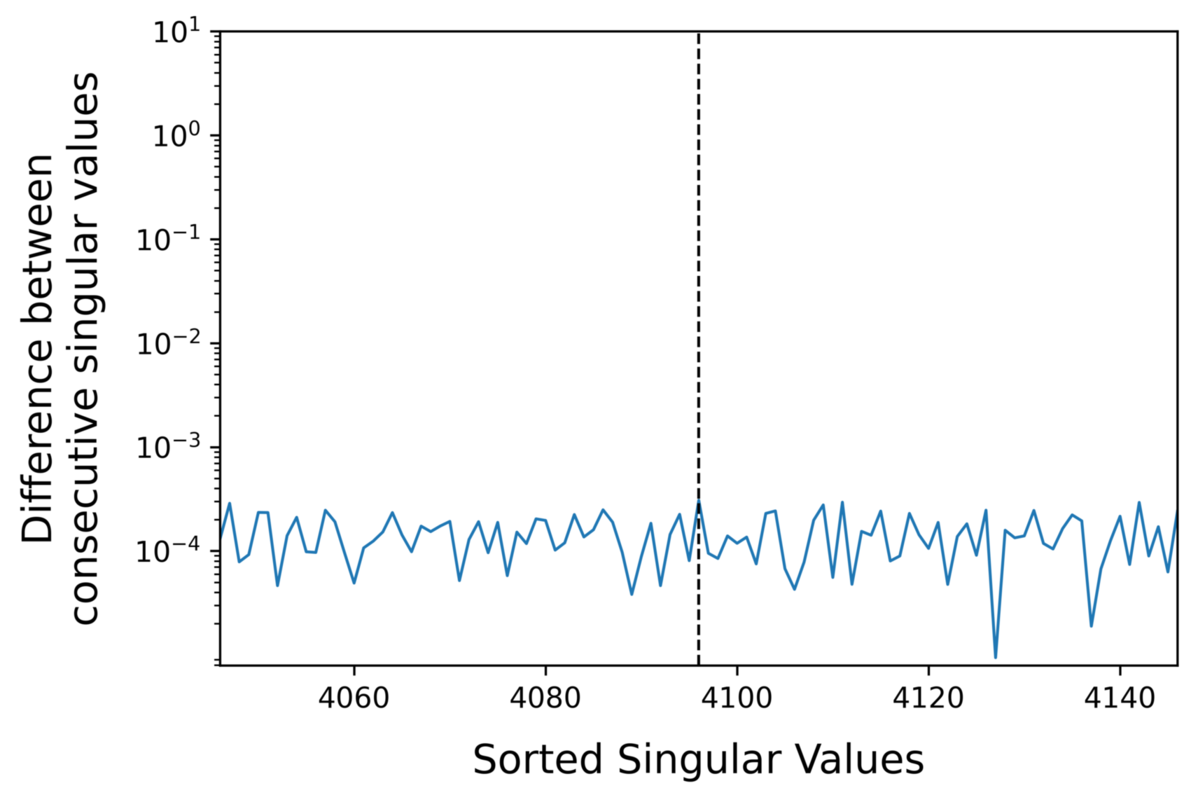}
        \caption{Noise 1.0, Sparsity 0.5}
    \end{subfigure}
    \begin{subfigure}{0.24\textwidth}
        \centering
        \includegraphics[width=\textwidth]{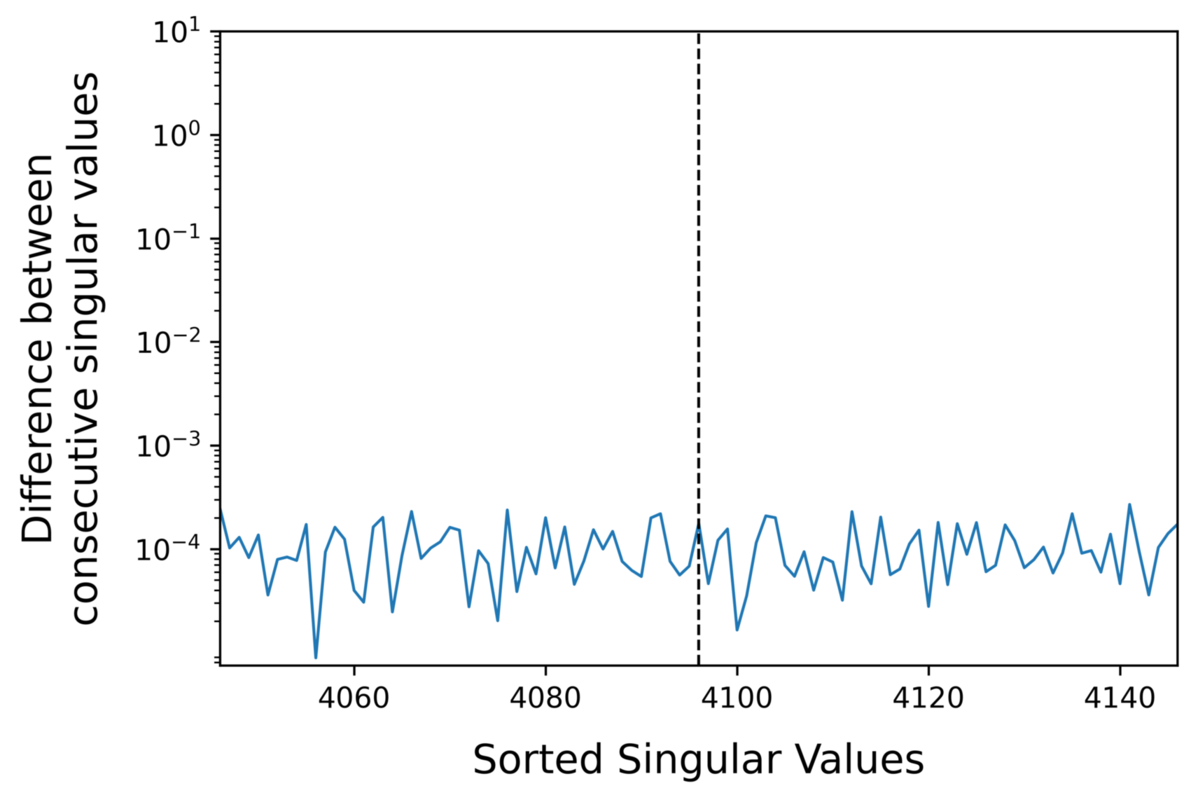}
        \caption{Noise 2.0, Sparsity 0.5}
    \end{subfigure}

    \caption{(Mistral-7B Discrete Noise). Dimension extraction attack on Mistral-7B by SVD singular values for discrete noise addition with varying sparsity (0.1, 0.25, 0.5) and magnitude (0.1, 0.5, 1.0, 2.0). Each top plot depicts the sorted singular values of the corresponding noisy logits matrix, where a jump at correct position 4,096 yields a successful dimension extraction. Each bottom plot provides a zoomed in picture of the difference between consecutive singular values; the attack outputs the index for which this difference is maximized.}
    \label{fig:Mistral-many-noise-discrete}
\end{figure*}

\begin{figure*}[h!] 
    \centering
    

    \begin{subfigure}{0.24\textwidth} 
        \centering
        \includegraphics[width=\textwidth]{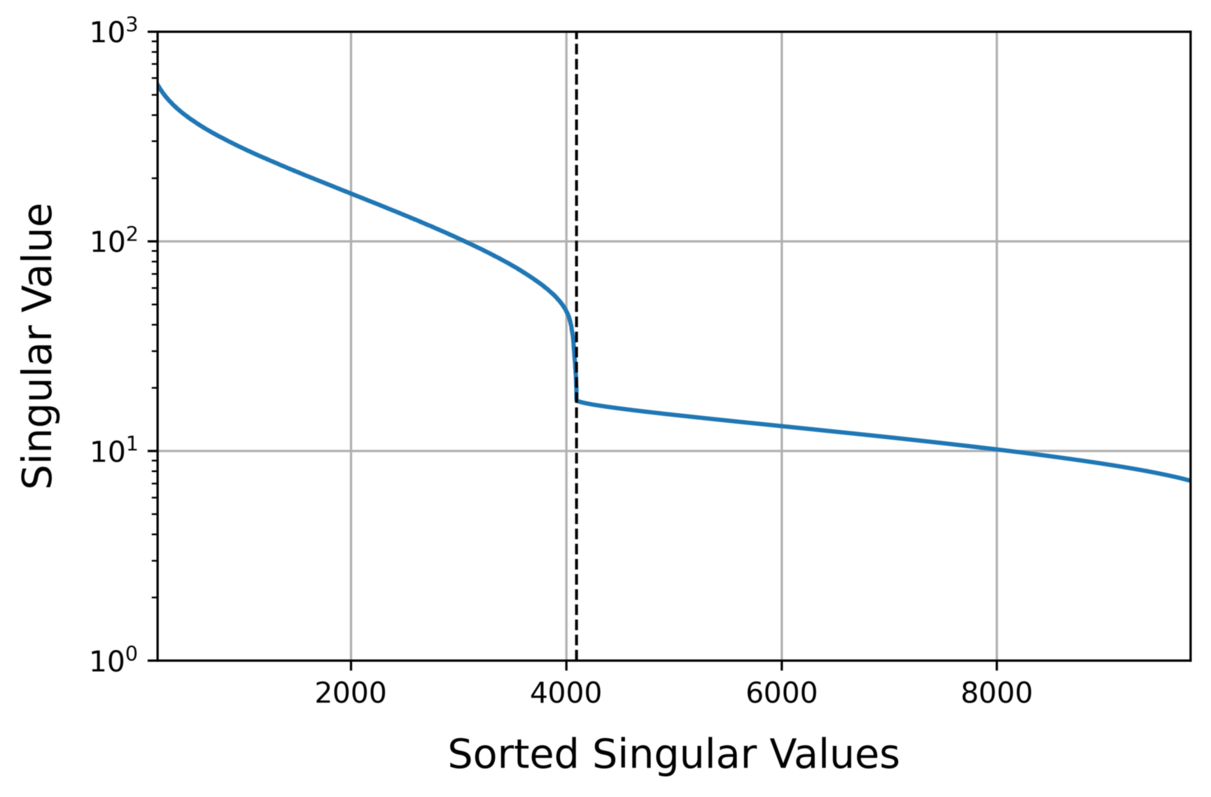}
    \end{subfigure}
    \begin{subfigure}{0.24\textwidth}
        \centering
        \includegraphics[width=\textwidth]{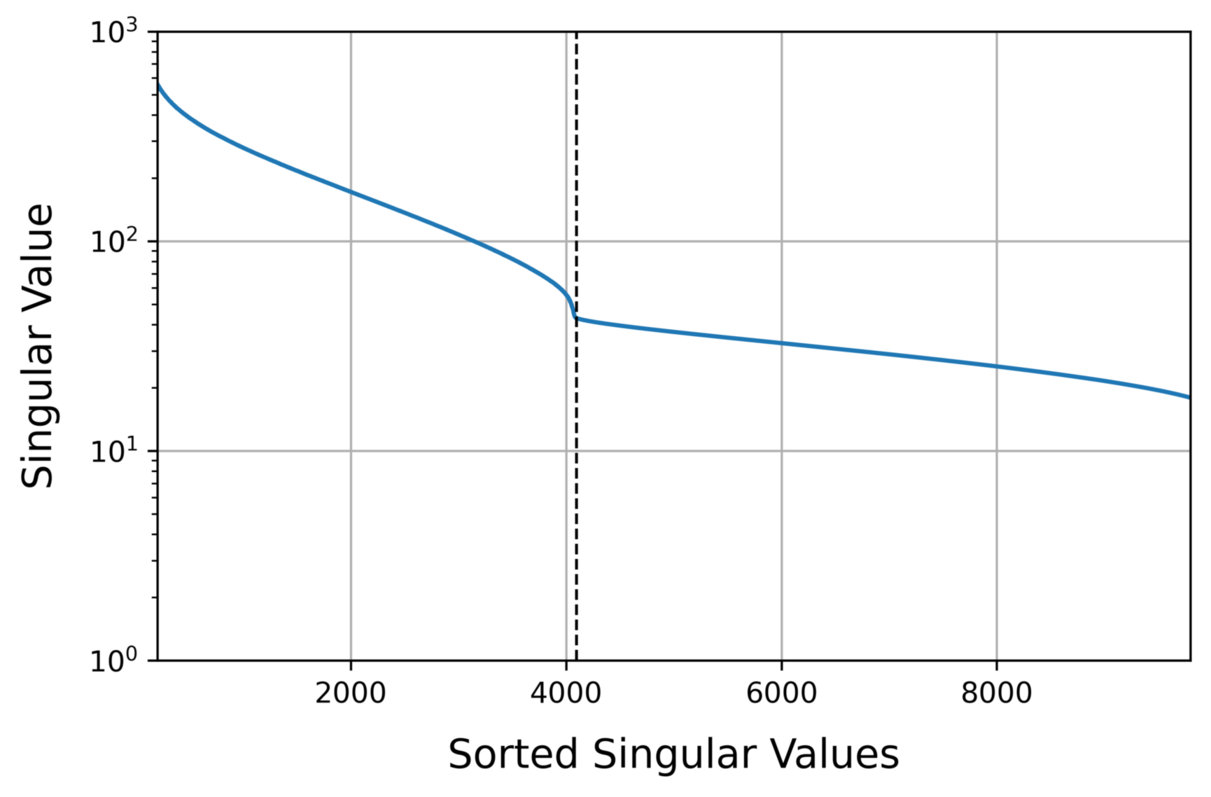}
    \end{subfigure}
    \begin{subfigure}{0.24\textwidth}
        \centering
        \includegraphics[width=\textwidth]{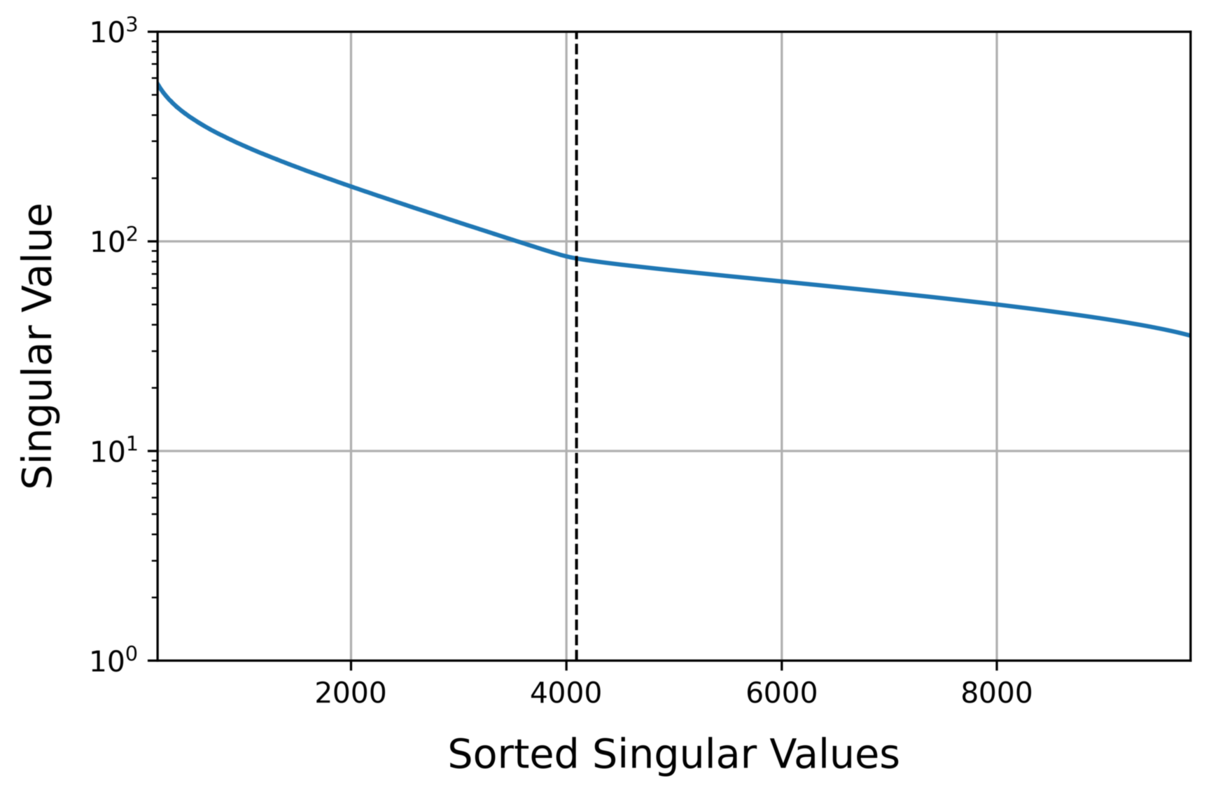}
    \end{subfigure}
    \begin{subfigure}{0.24\textwidth}
        \centering
        \includegraphics[width=\textwidth]{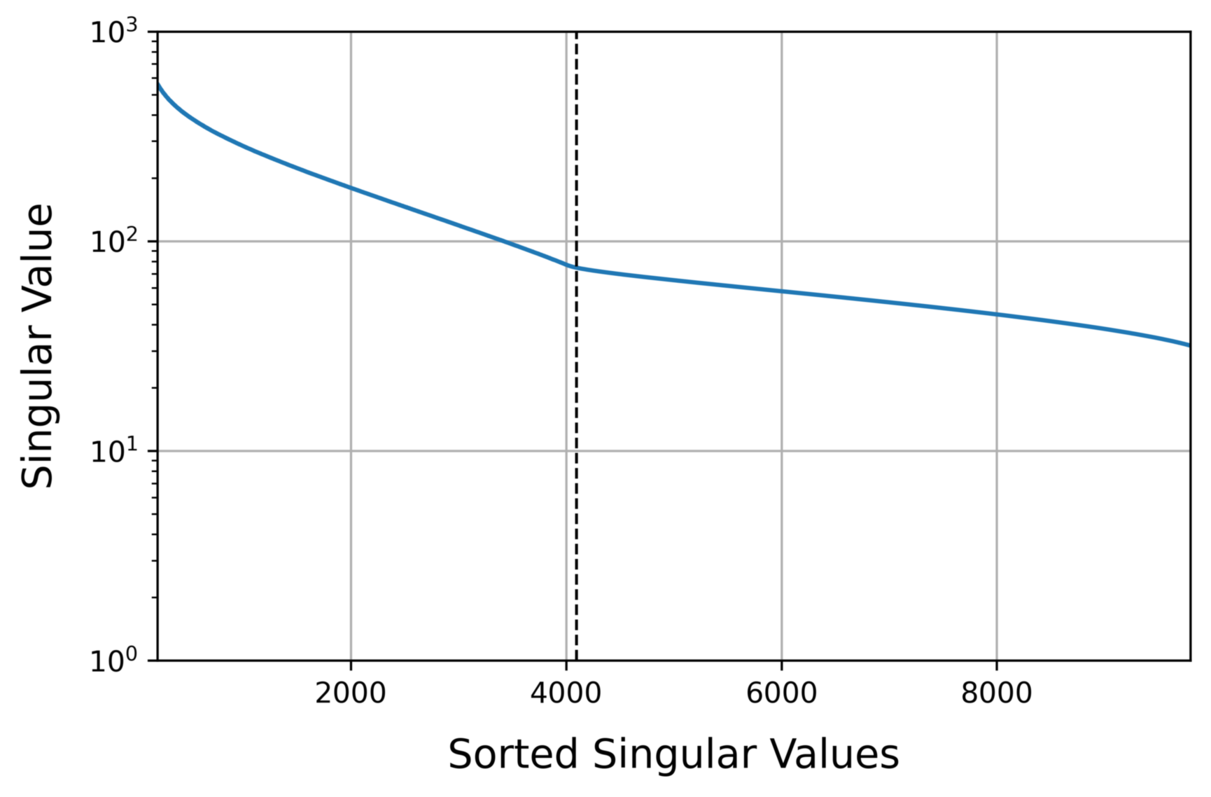}
    \end{subfigure}


    \begin{subfigure}{0.24\textwidth} 
        \centering
        \includegraphics[width=\textwidth]{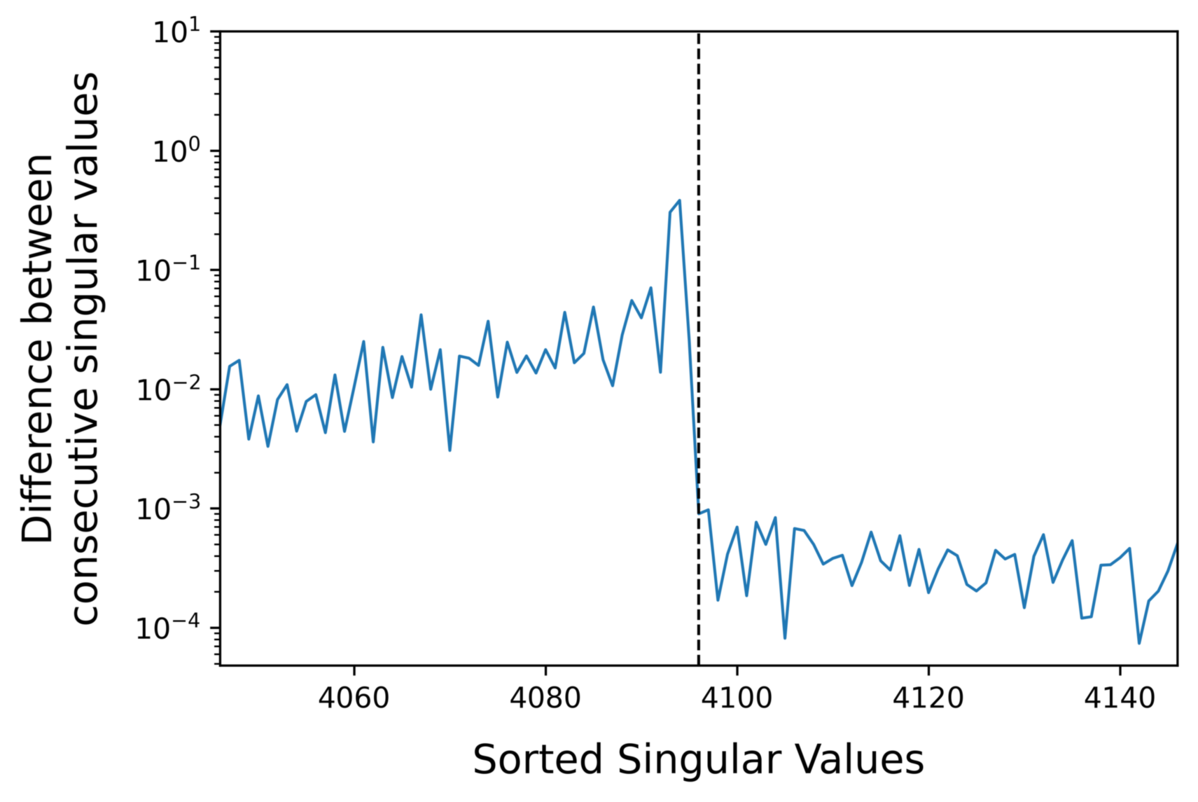}
        \caption{$\text{std}=0.1$, $\gamma=0.1$}
    \end{subfigure}
    \begin{subfigure}{0.24\textwidth}
        \centering
        \includegraphics[width=\textwidth]{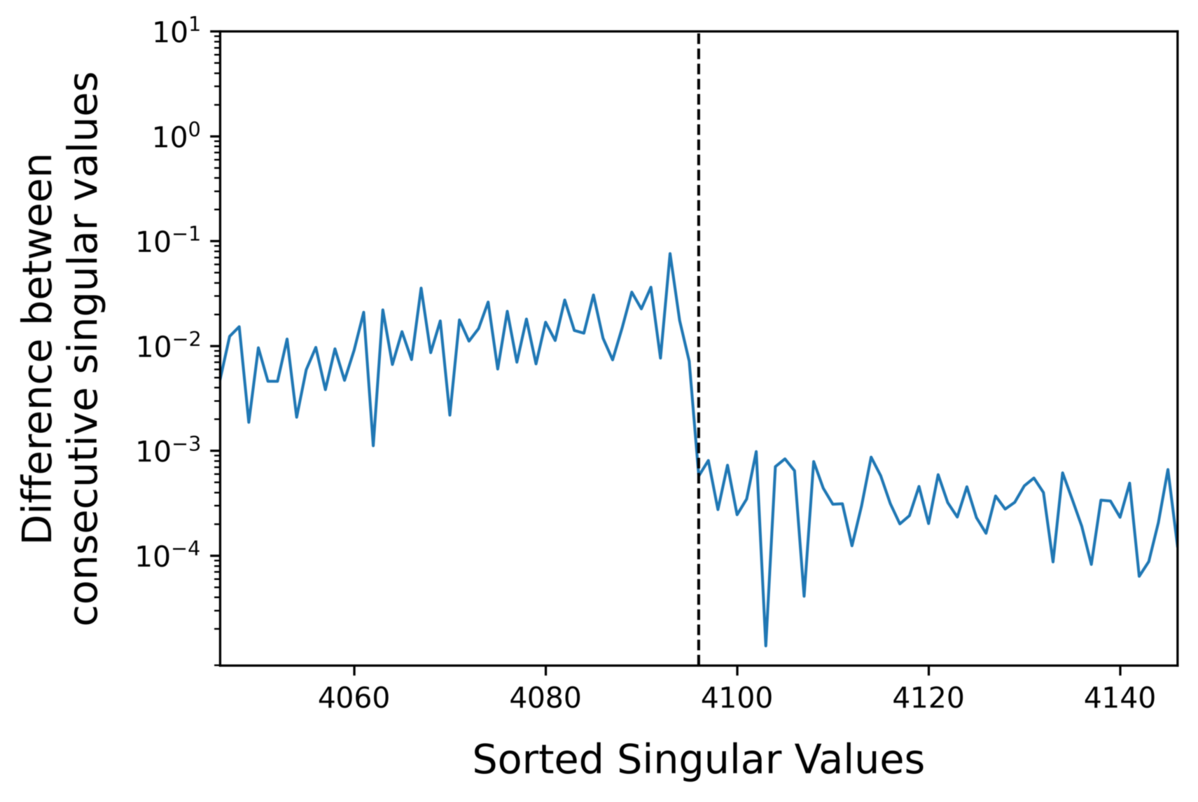}
        \caption{$\text{std}=0.25$, $\gamma=0.1$}
    \end{subfigure}
    \begin{subfigure}{0.24\textwidth}
        \centering
        \includegraphics[width=\textwidth]{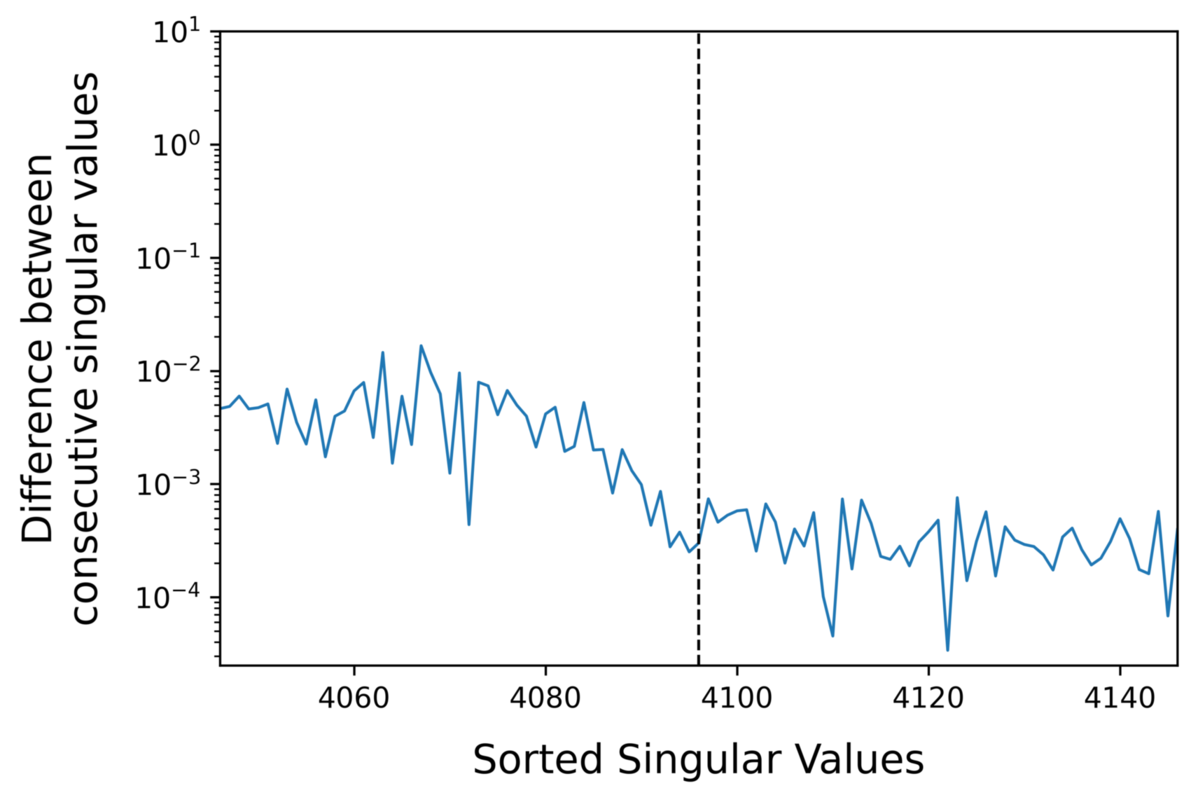}
        \caption{Standard Deviation}
        \caption{$\text{std}=0.5$, $\gamma=0.1$}
    \end{subfigure}
    \begin{subfigure}{0.24\textwidth}
        \centering
        \includegraphics[width=\textwidth]{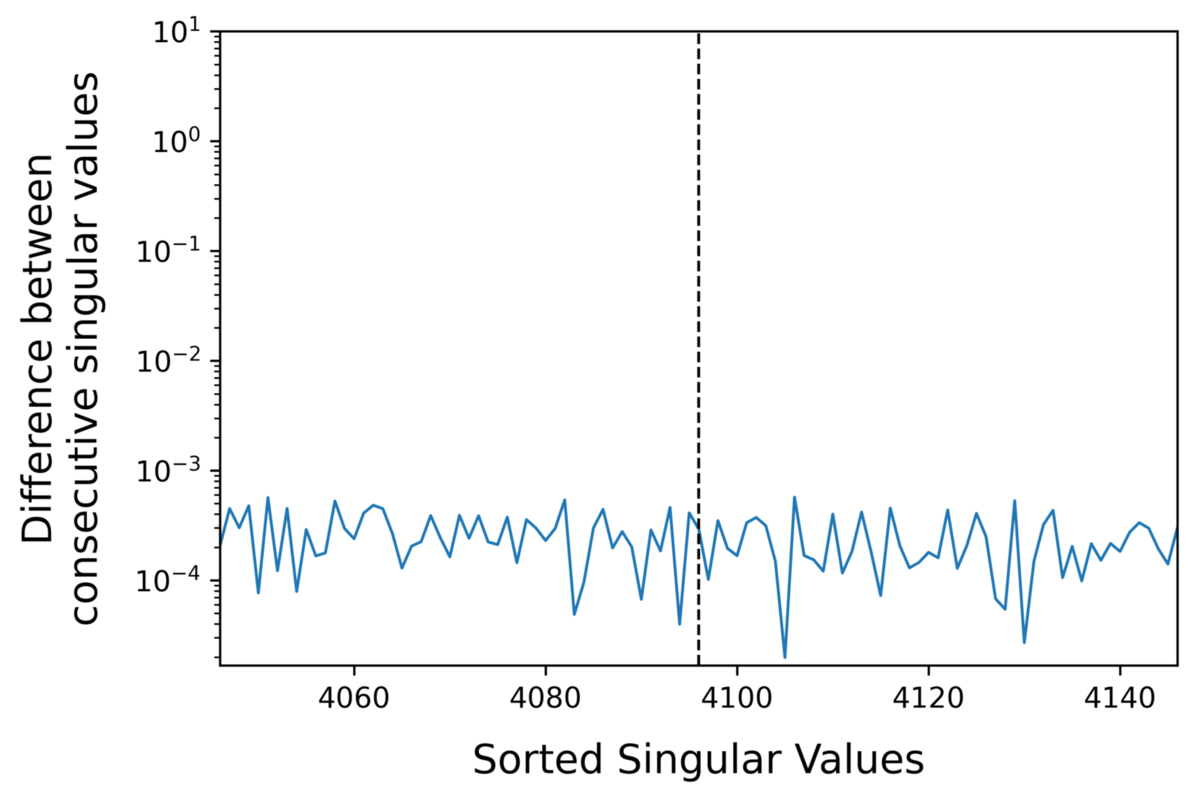}
        \caption{$\text{std}=1.0$, $\gamma=0.1$}
    \end{subfigure}


\vspace{.2in}

    \begin{subfigure}{0.24\textwidth} 
        \centering
        \includegraphics[width=\textwidth]{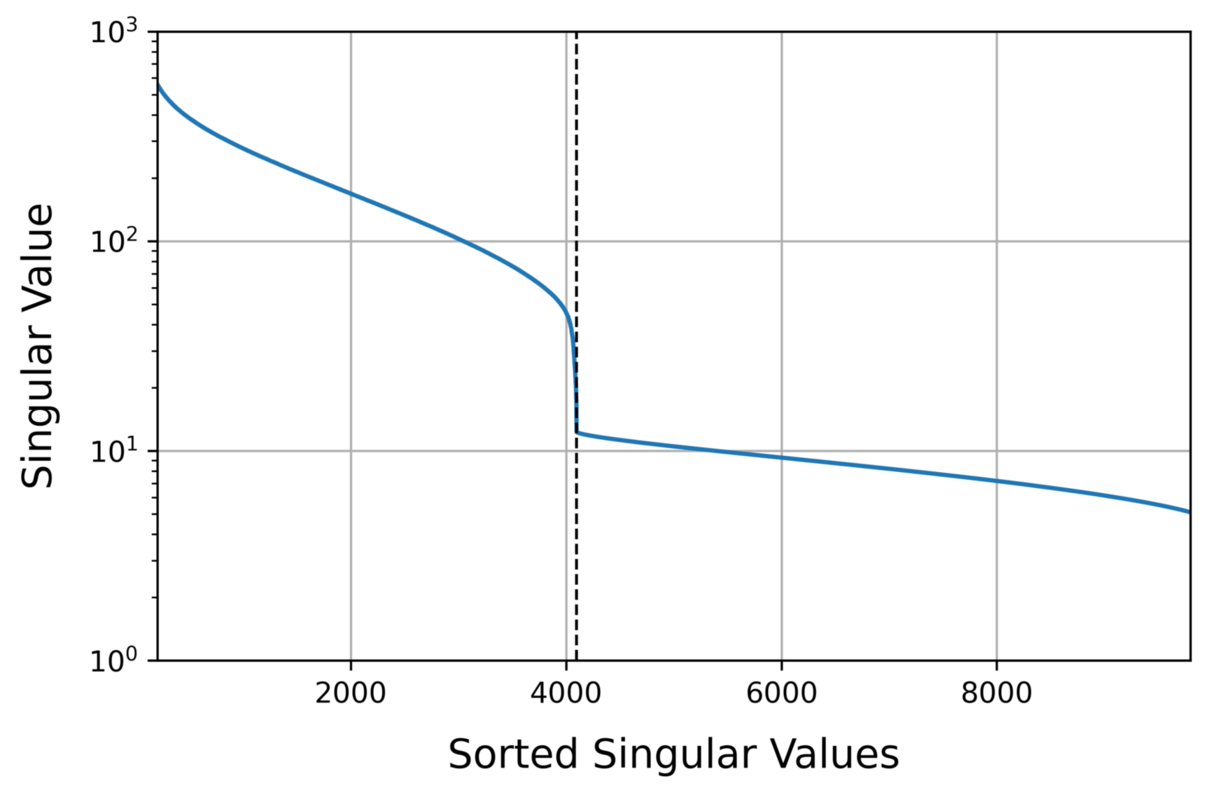}
    \end{subfigure}
    \begin{subfigure}{0.24\textwidth}
        \centering
        \includegraphics[width=\textwidth]{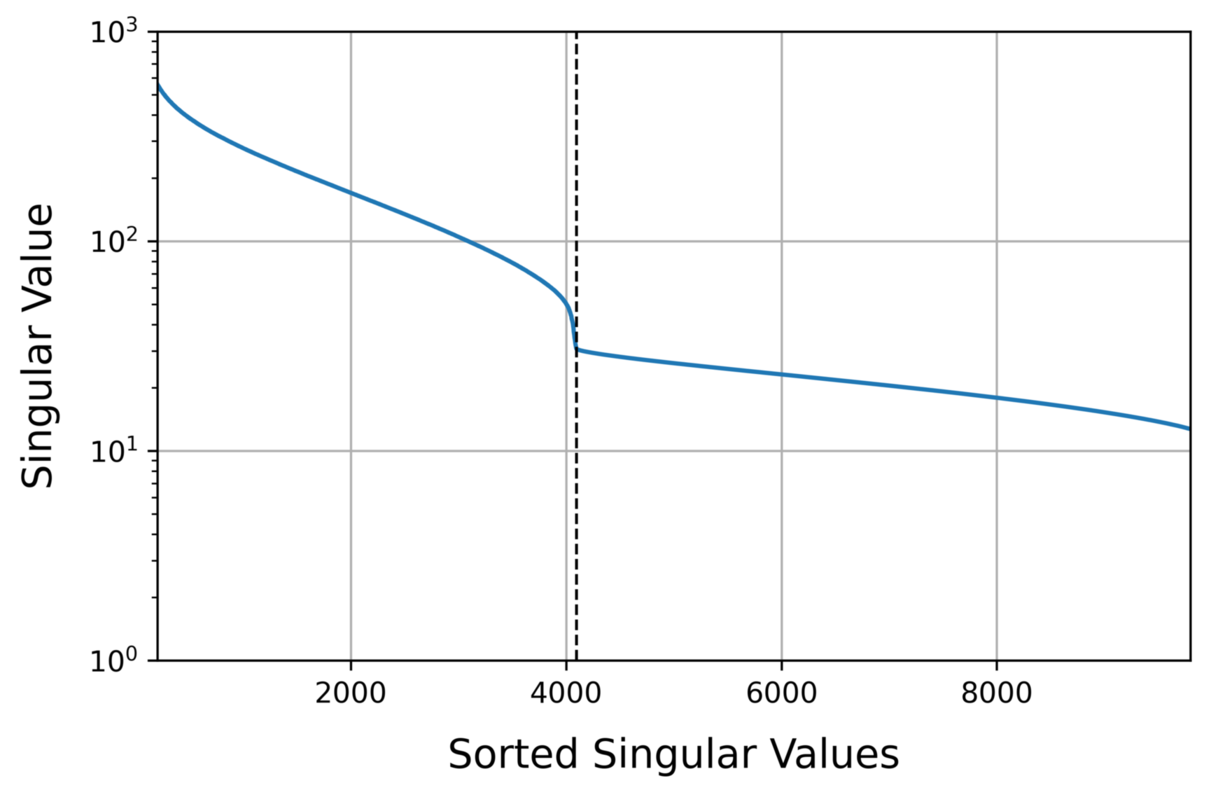}
    \end{subfigure}
    \begin{subfigure}{0.24\textwidth}
        \centering
        \includegraphics[width=\textwidth]{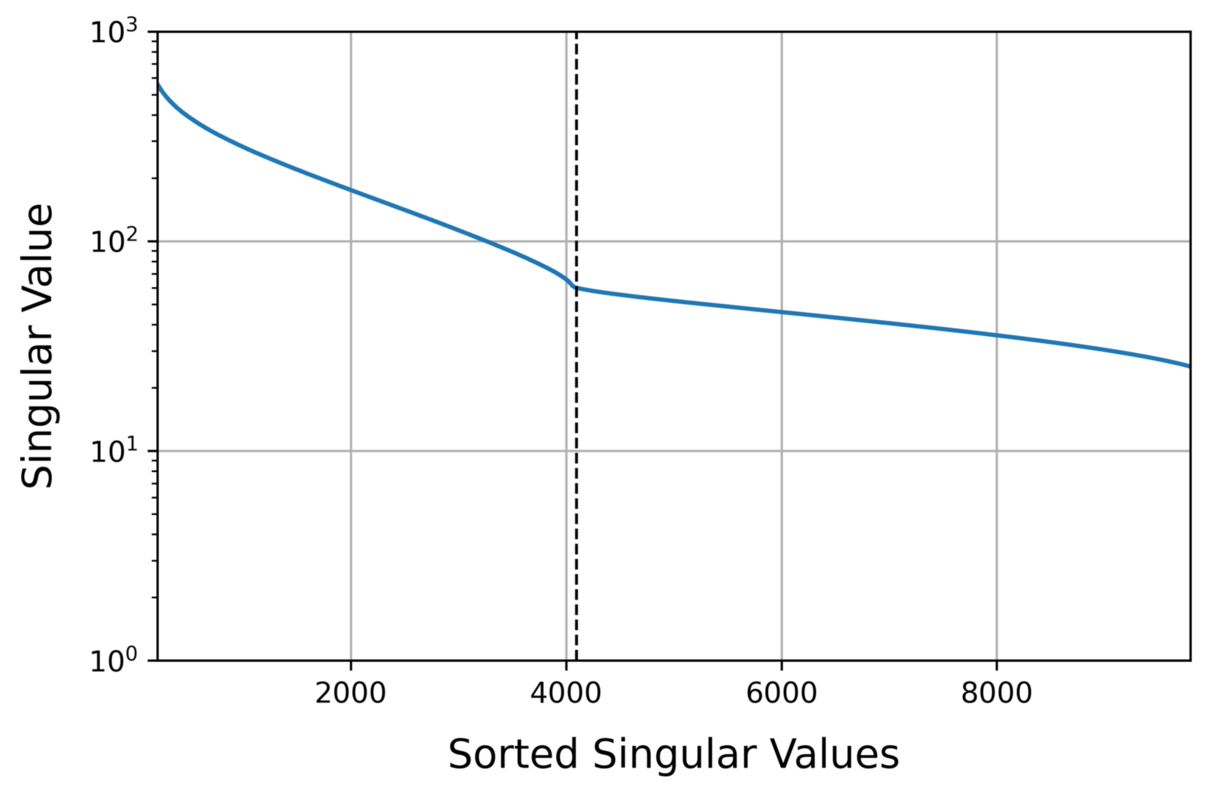}
    \end{subfigure}
    \begin{subfigure}{0.24\textwidth}
        \centering
        \includegraphics[width=\textwidth]{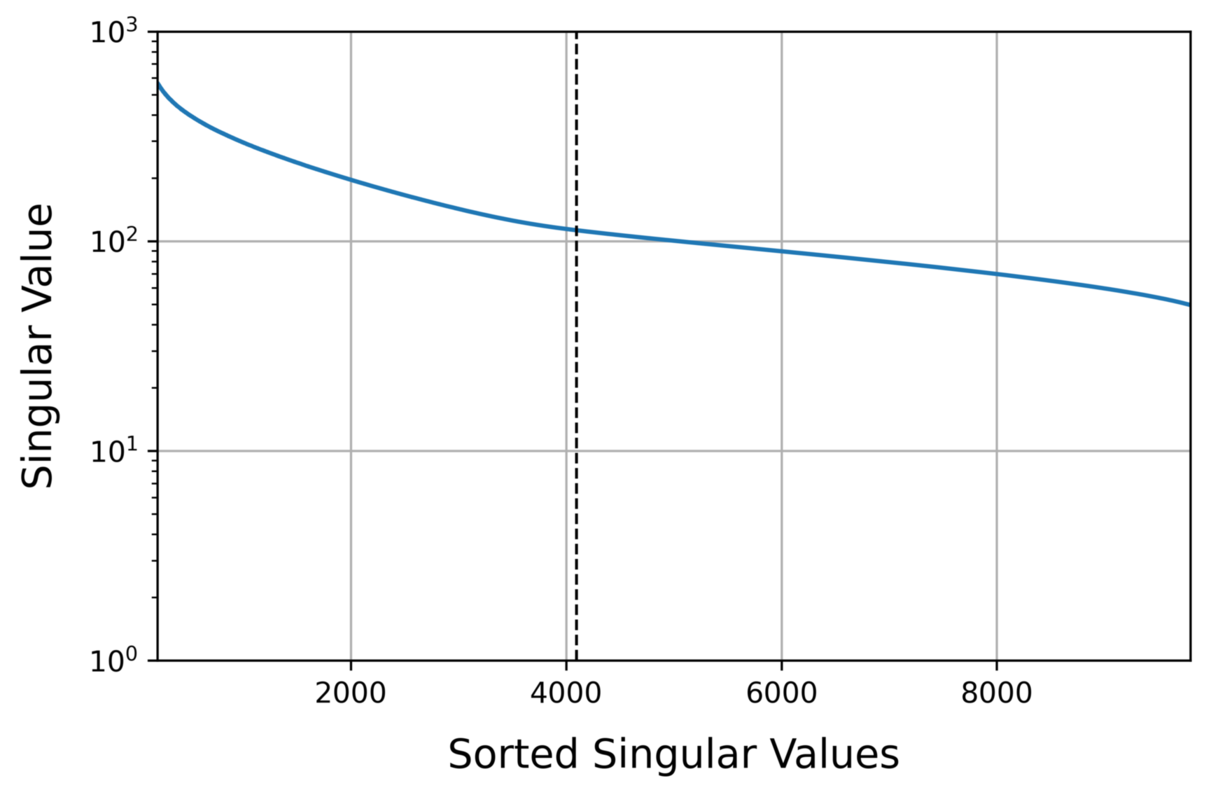}
    \end{subfigure}


    \begin{subfigure}{0.24\textwidth} 
        \centering
        \includegraphics[width=\textwidth]{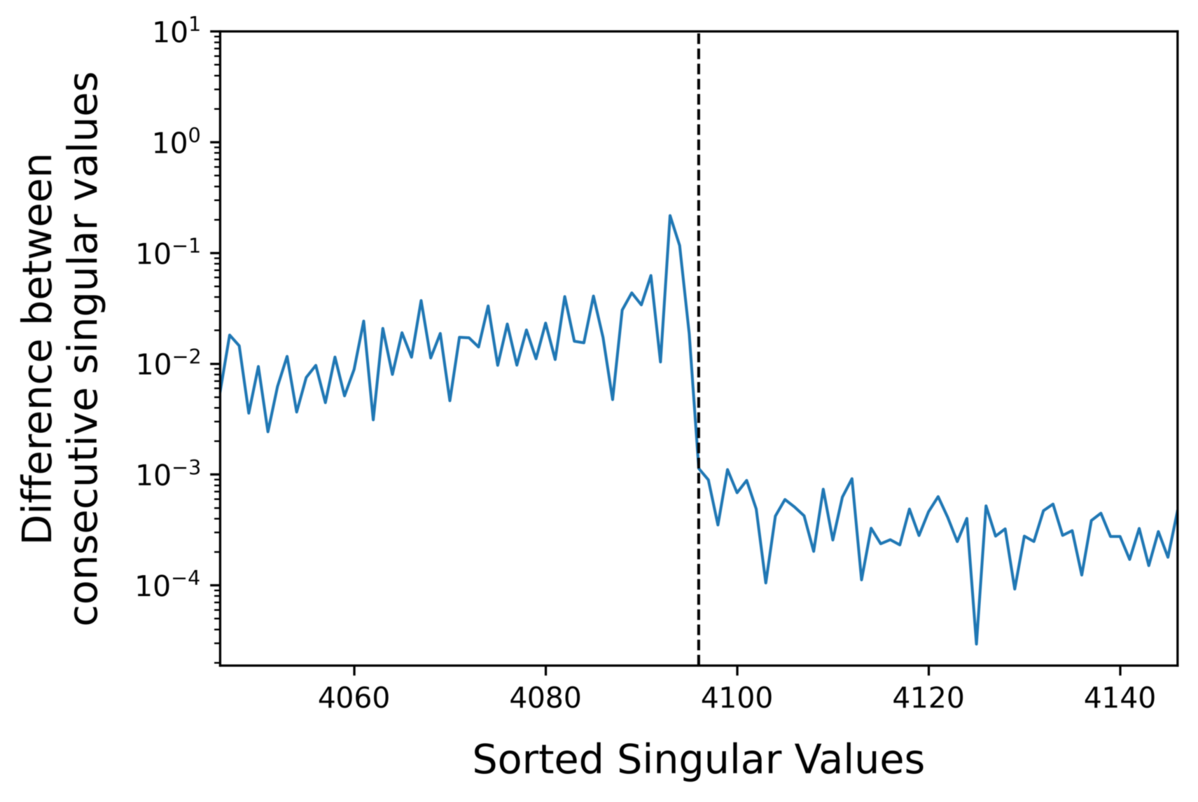}
        \caption{$\text{std}=0.1$, $\gamma=0.25$}
    \end{subfigure}
    \begin{subfigure}{0.24\textwidth}
        \centering
        \includegraphics[width=\textwidth]{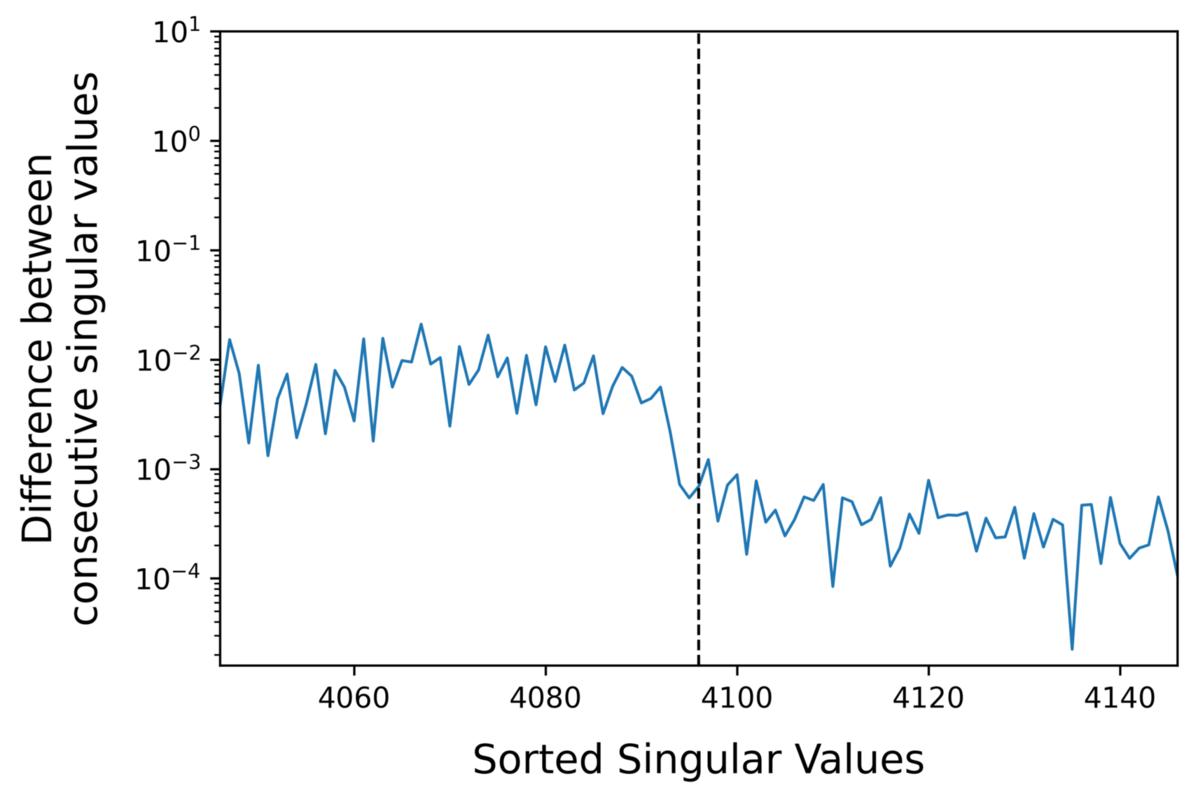}
        \caption{$\text{std}=0.25$, $\gamma=0.25$}
    \end{subfigure}
    \begin{subfigure}{0.24\textwidth}
        \centering
        \includegraphics[width=\textwidth]{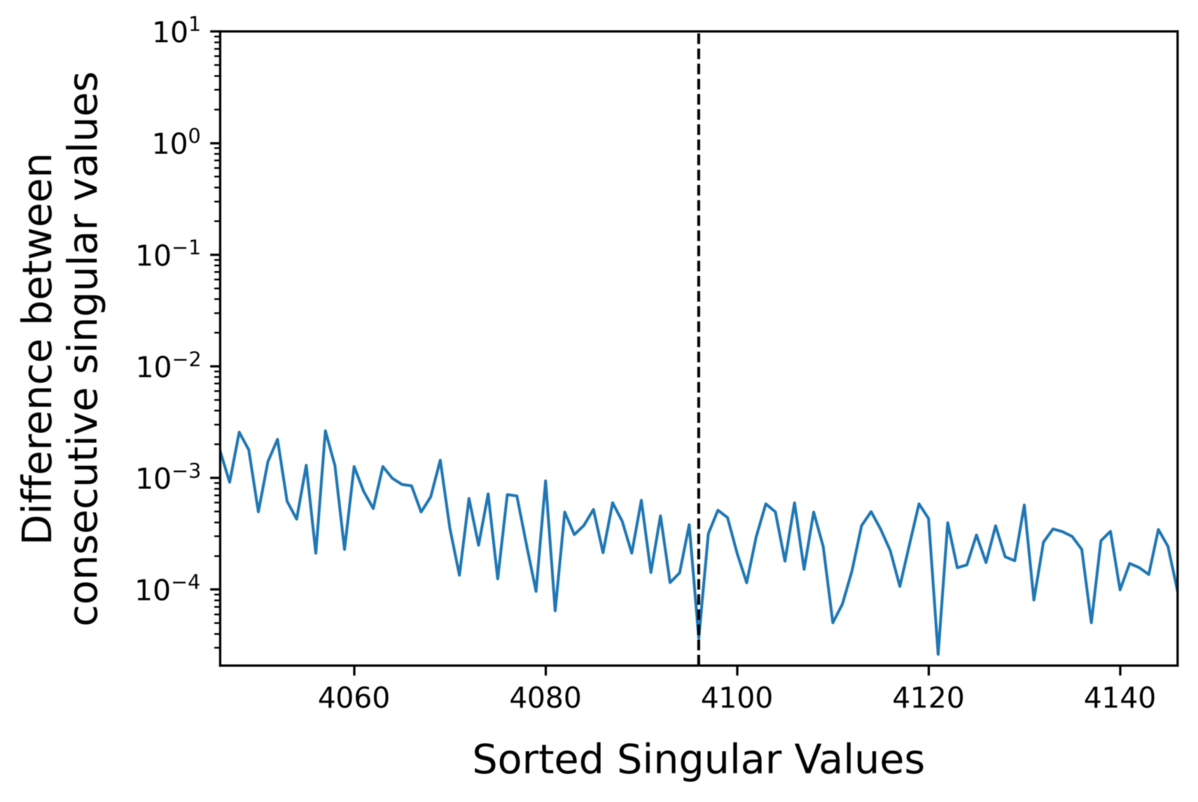}
        \caption{$\text{std}=0.5$, $\gamma=0.25$}
    \end{subfigure}
    \begin{subfigure}{0.24\textwidth}
        \centering
        \includegraphics[width=\textwidth]{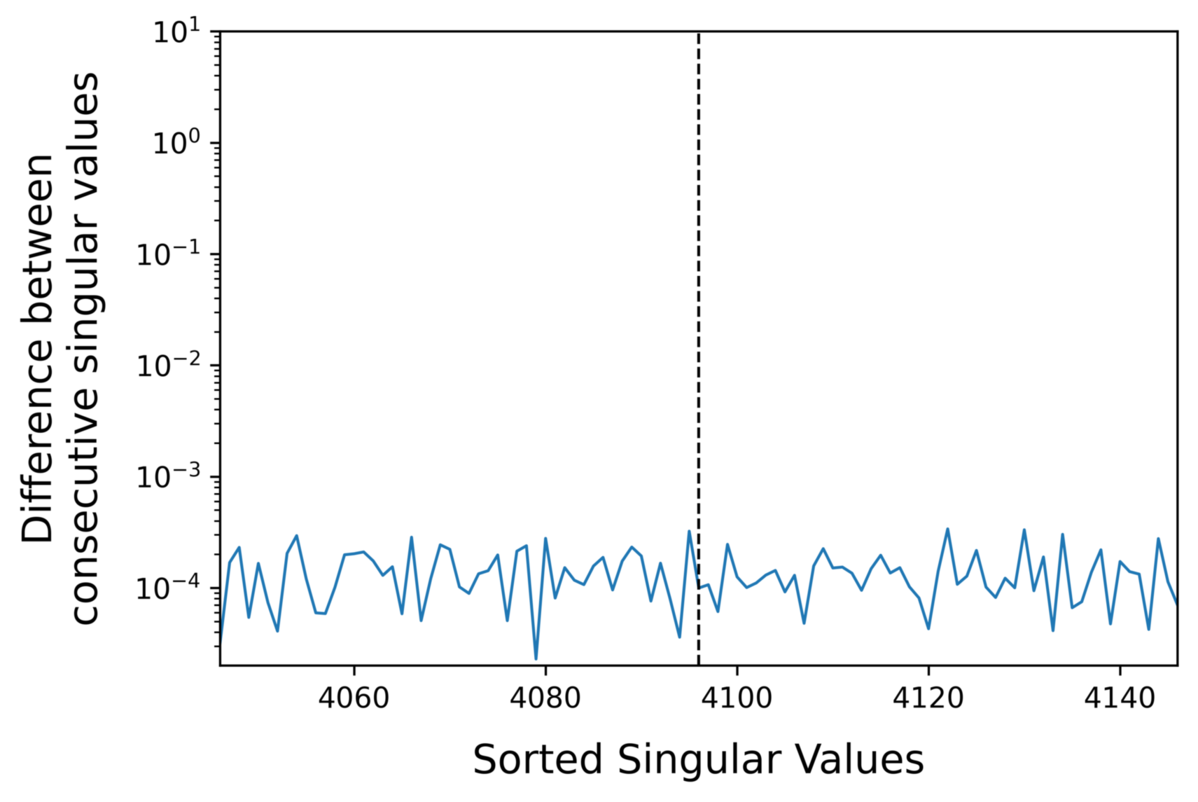}
        \caption{$\text{std}=1.0$, $\gamma=0.25$}
    \end{subfigure}


\vspace{.2in}

    \begin{subfigure}{0.24\textwidth} 
        \centering
        \includegraphics[width=\textwidth]{images/mistral_sparsity_noise/mistral_hidden_dim_gamma_0.5_std_0.1_trimmed_values.png}
    \end{subfigure}
    \begin{subfigure}{0.24\textwidth}
        \centering
        \includegraphics[width=\textwidth]{images/mistral_sparsity_noise/mistral_hidden_dim_gamma_0.5_std_0.25_trimmed_values.png}
    \end{subfigure}
    \begin{subfigure}{0.24\textwidth}
        \centering
        \includegraphics[width=\textwidth]{images/mistral_sparsity_noise/mistral_hidden_dim_gamma_0.5_std_0.5_trimmed_values.png}
    \end{subfigure}
    \begin{subfigure}{0.24\textwidth}
        \centering
        \includegraphics[width=\textwidth]{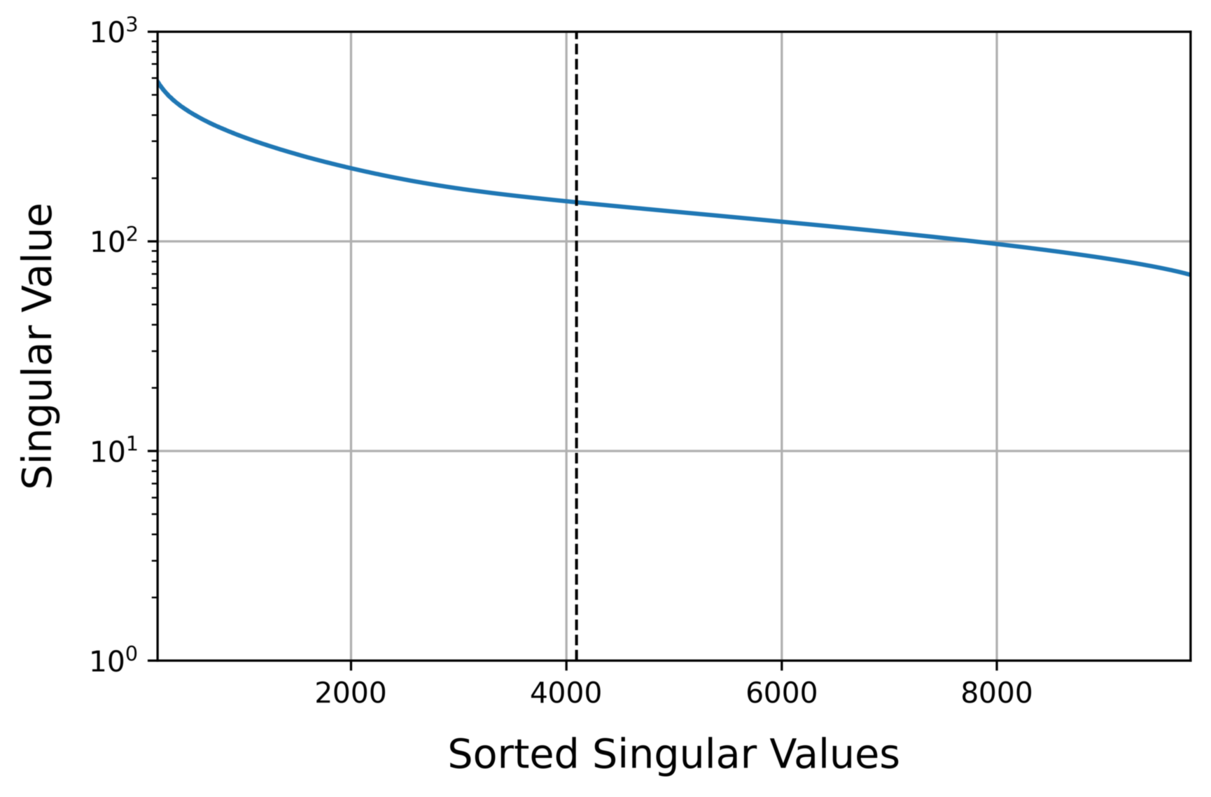}
    \end{subfigure}


    \begin{subfigure}{0.24\textwidth} 
        \centering
        \includegraphics[width=\textwidth]{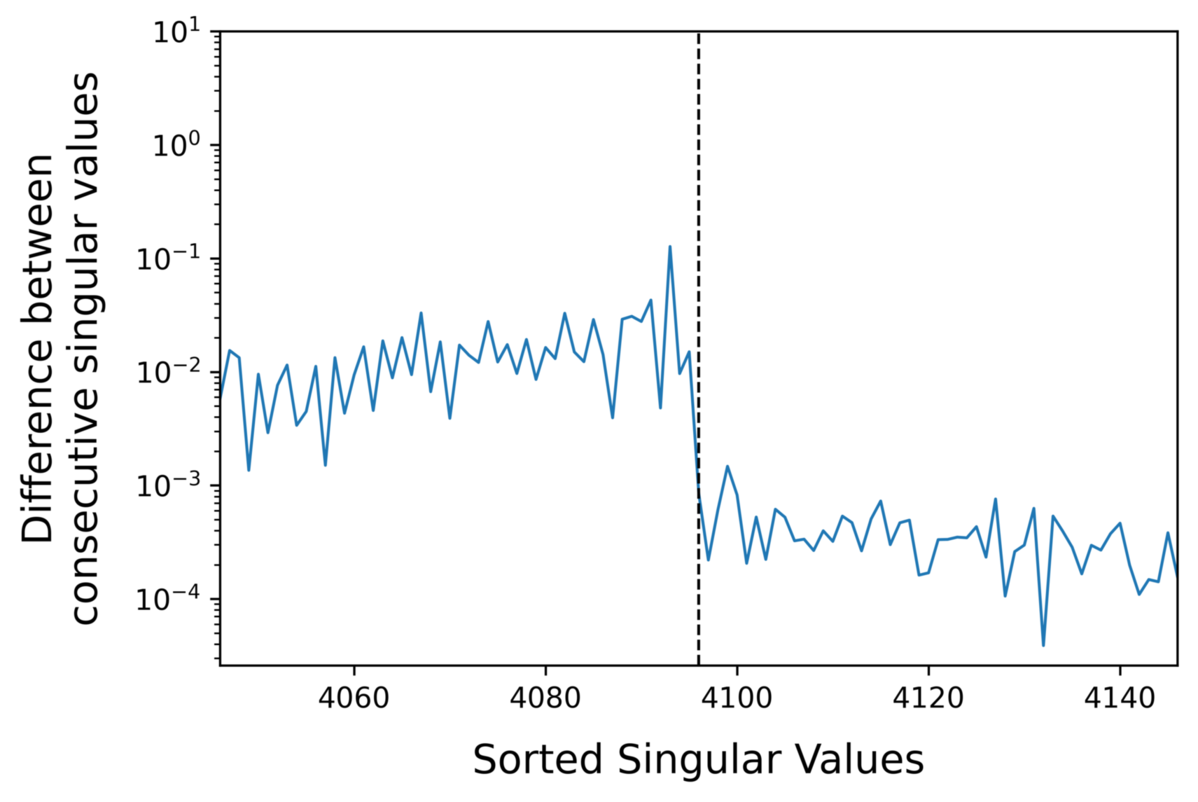}
        \caption{$\text{std}=0.1$, $\gamma=0.5$}
    \end{subfigure}
    \begin{subfigure}{0.24\textwidth}
        \centering
        \includegraphics[width=\textwidth]{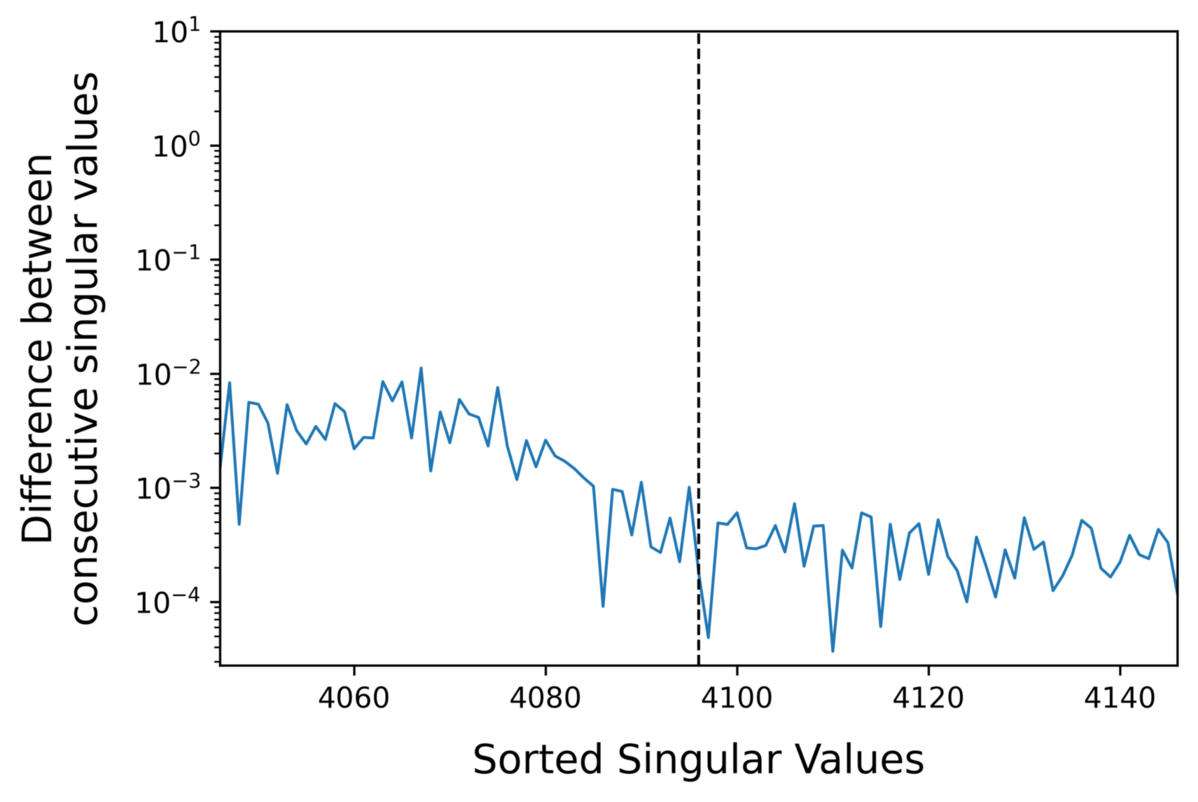}
        \caption{$\text{std}=0.25$, $\gamma=0.5$}
    \end{subfigure}
    \begin{subfigure}{0.24\textwidth}
        \centering
        \includegraphics[width=\textwidth]{images/mistral_sparsity_noise/mistral_hidden_dim_gamma_0.5_noise_1.0_windowed.png}
        \caption{$\text{std}=0.5$, $\gamma=0.5$}
    \end{subfigure}
    \begin{subfigure}{0.24\textwidth}
        \centering
        \includegraphics[width=\textwidth]{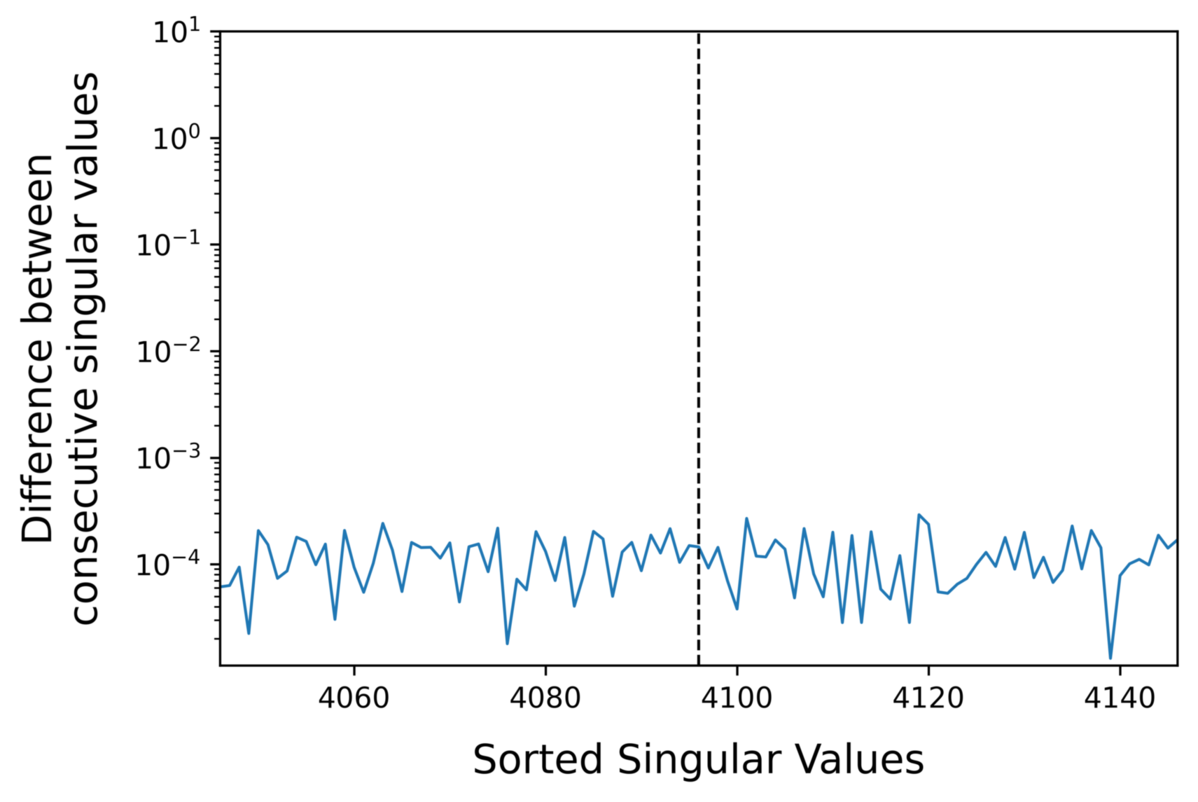}
        \caption{$\text{std}=1.0$, $\gamma=0.5$}
    \end{subfigure}

    \caption{(Mistral-7B Gaussian Noise). Dimension extraction attack on Mistral-7B by SVD singular values for {\em Gaussian} noise addition with varying sparsity $\gamma \in \{0.1, 0.25, 0.5\}$ and standard deviation $\text{std} \in \{0.1, 0.25, 0.5, 1.0\}$. Each top plot depicts the sorted singular values of the corresponding noisy logits matrix, where a jump at correct position 4,096 yields a successful dimension extraction. Each bottom plot provides a zoomed in picture of the difference between consecutive singular values; the attack outputs the index for which this difference is maximized.}
    \label{fig:Mistral-many-noise-Gaussian}
\end{figure*}

We present experimental results for Mistral-7B with a range of noise types and parameters.  In Figure~\ref{fig:Mistral-many-noise-discrete} we give our findings for applying discrete noise with varying sparsity $\gamma \in \{0.1, 0.25, 0.5\}$ and magnitude $\in \{0.1, 0.5, 1.0, 2.0\}$.  In Figure~\ref{fig:Mistral-many-noise-Gaussian}, we provide the comparable plots for Mistral-7B with Gaussian noise.

We observe a clear trend where increased noise parameters lead to decreased success of the attack, in a gradual fashion. In particular, we see that noise magnitude (or standard deviation, in the case of Gaussian noise) plays a seemingly stronger role than sparsity.  

\subsection{Results for GPT-2} \label{appendix:gpt2}

For completeness, we performed comparable measurements on GPT-2, with vocabulary size $T=50,257$ and hidden dimension $768$.  Experiments were run with $N=2,300$ prompts for each $N \times T$ logits matrix.

\begin{figure*}[h!] 
    \centering
    

    \begin{subfigure}{0.24\textwidth} 
        \centering
        \includegraphics[width=\textwidth]{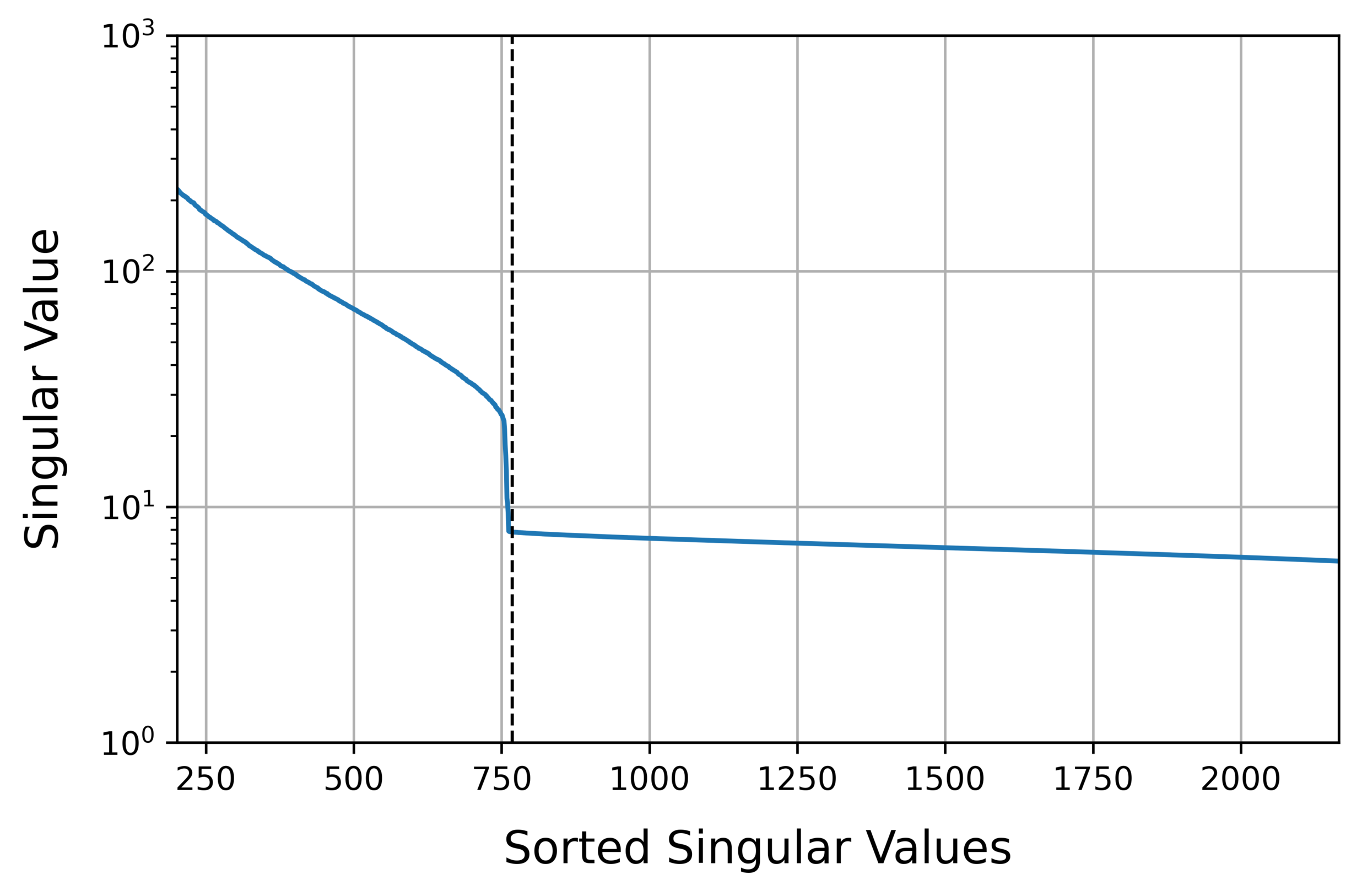}
    \end{subfigure}
    \begin{subfigure}{0.24\textwidth}
        \centering
        \includegraphics[width=\textwidth]{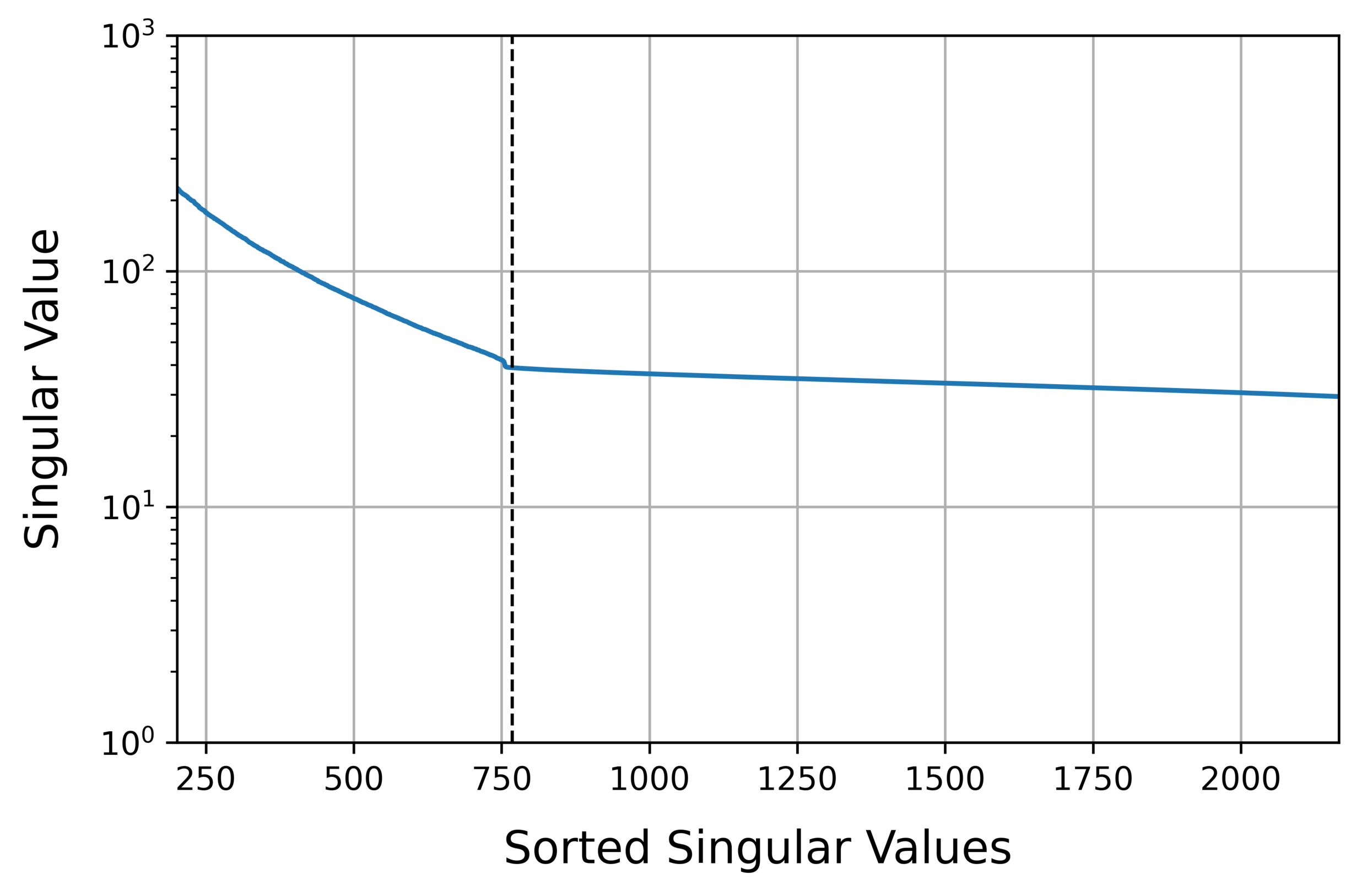}
    \end{subfigure}
    \begin{subfigure}{0.24\textwidth}
        \centering
        \includegraphics[width=\textwidth]{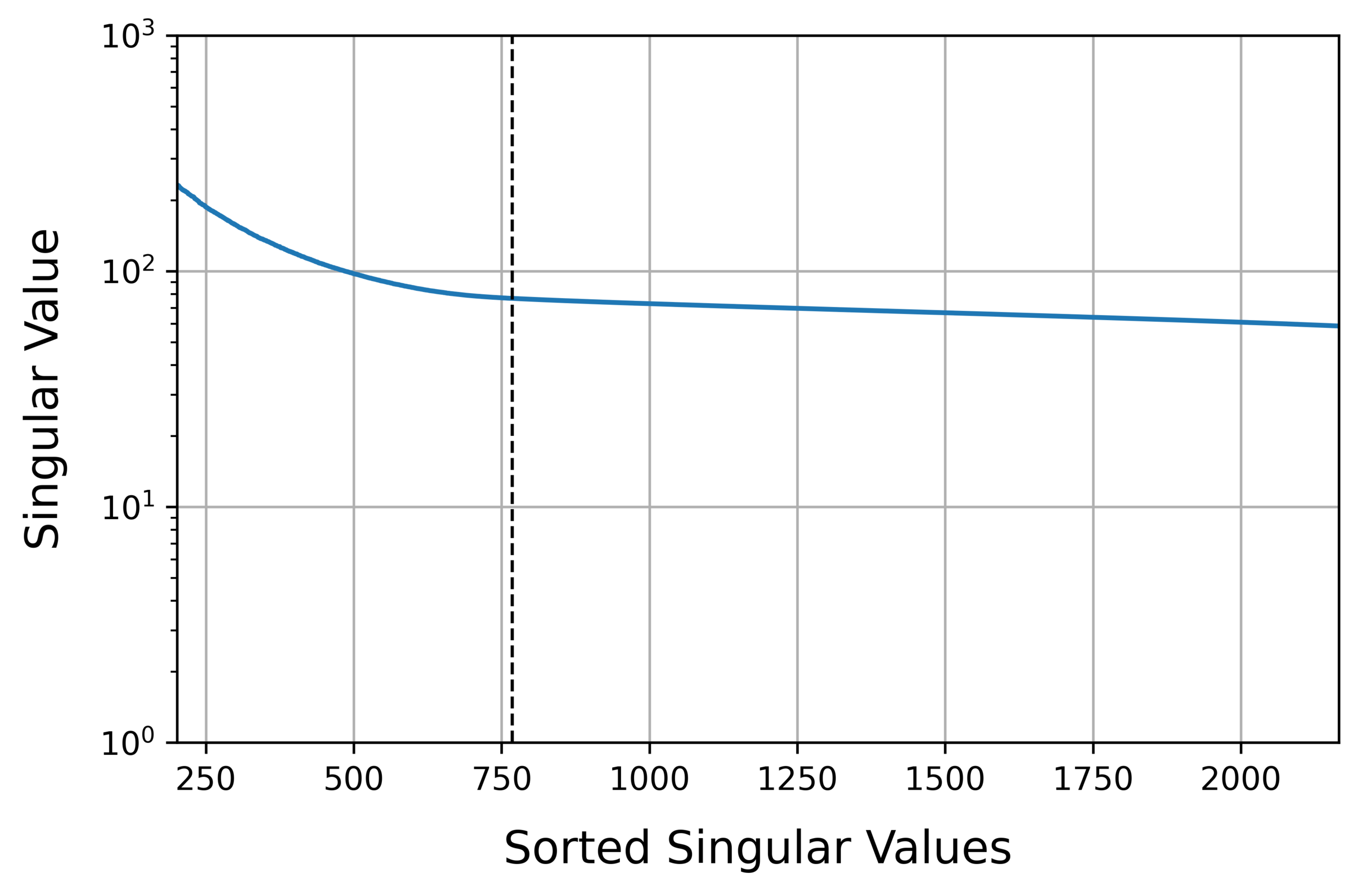}
    \end{subfigure}
    \begin{subfigure}{0.24\textwidth}
        \centering
        \includegraphics[width=\textwidth]{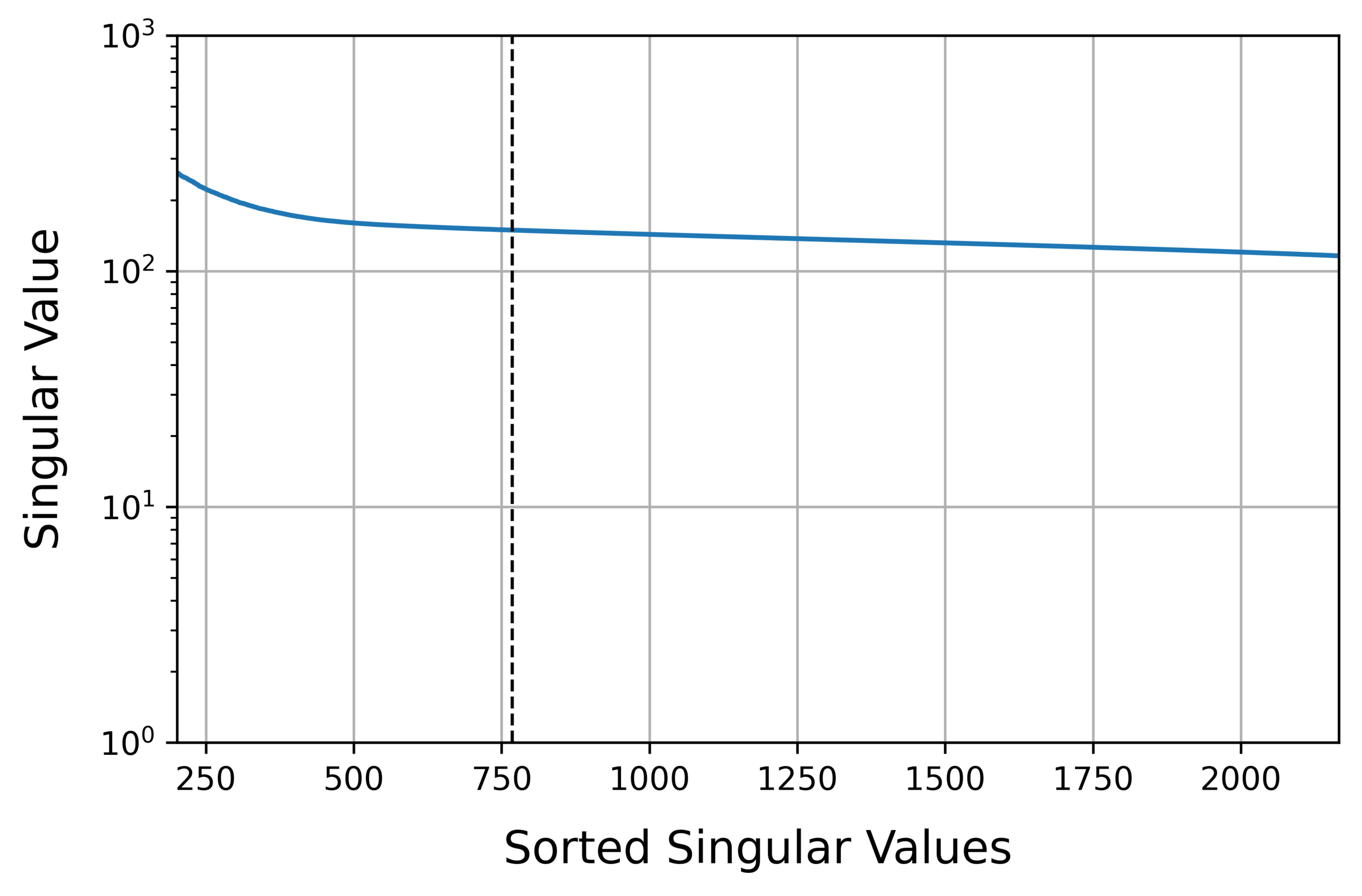}
    \end{subfigure}


    \begin{subfigure}{0.24\textwidth} 
        \centering
        \includegraphics[width=\textwidth]{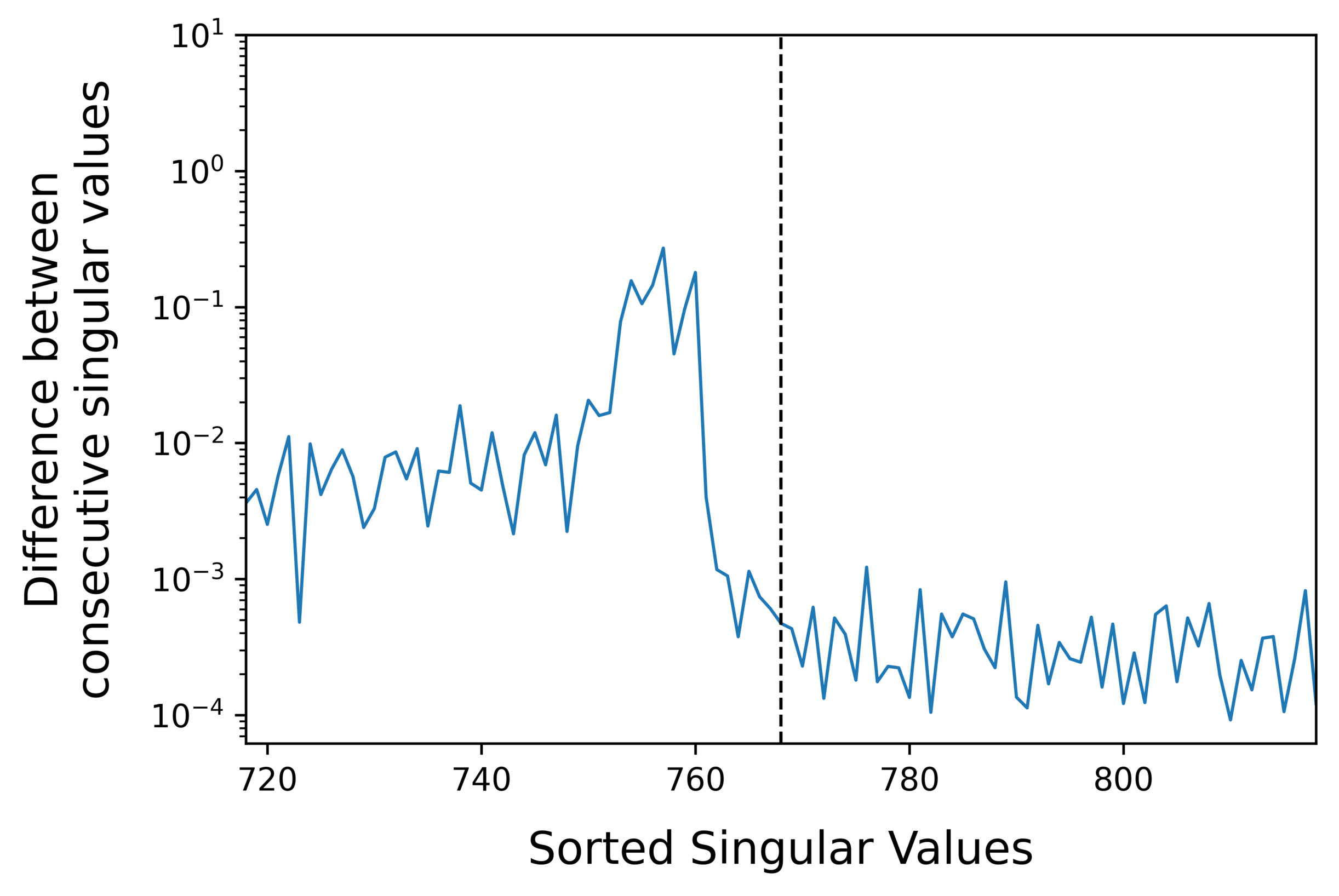}
        \caption{Noise 0.1, Sparsity 0.1}
    \end{subfigure}
    \begin{subfigure}{0.24\textwidth}
        \centering
        \includegraphics[width=\textwidth]{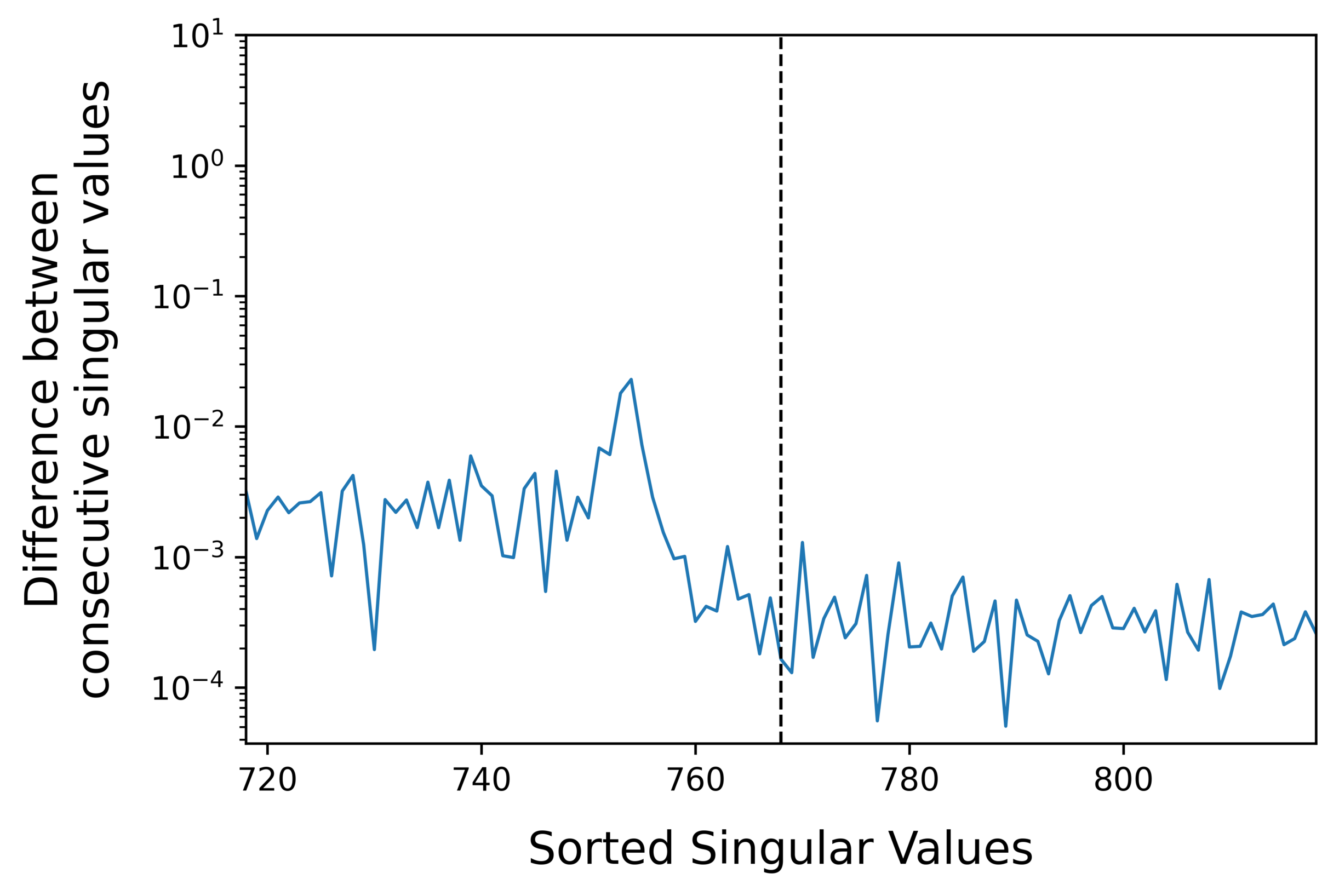}
        \caption{Noise 0.5, Sparsity 0.1}
    \end{subfigure}
    \begin{subfigure}{0.24\textwidth}
        \centering
        \includegraphics[width=\textwidth]{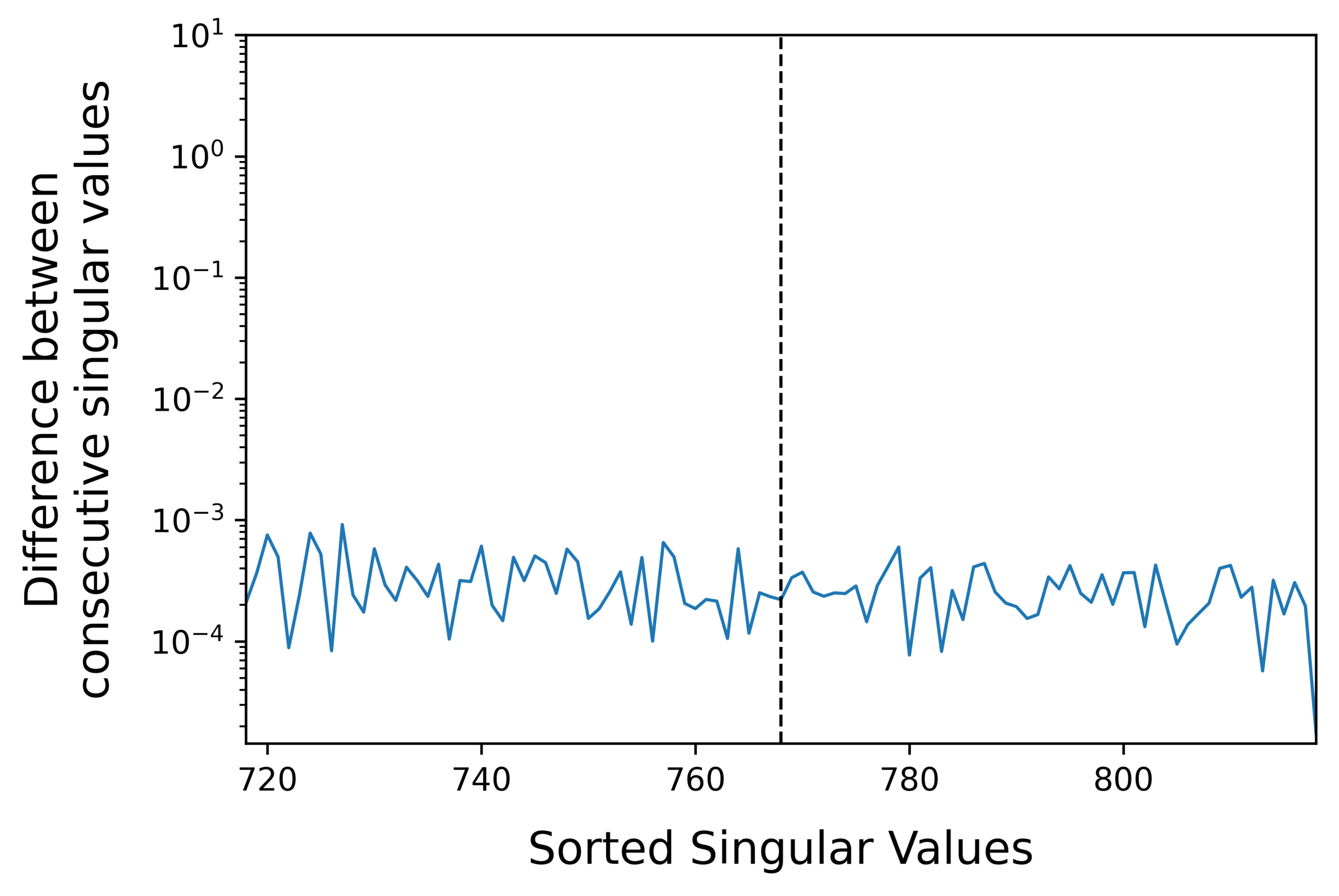}
        \caption{Noise 1.0, Sparsity 0.1}
    \end{subfigure}
    \begin{subfigure}{0.24\textwidth}
        \centering
        \includegraphics[width=\textwidth]{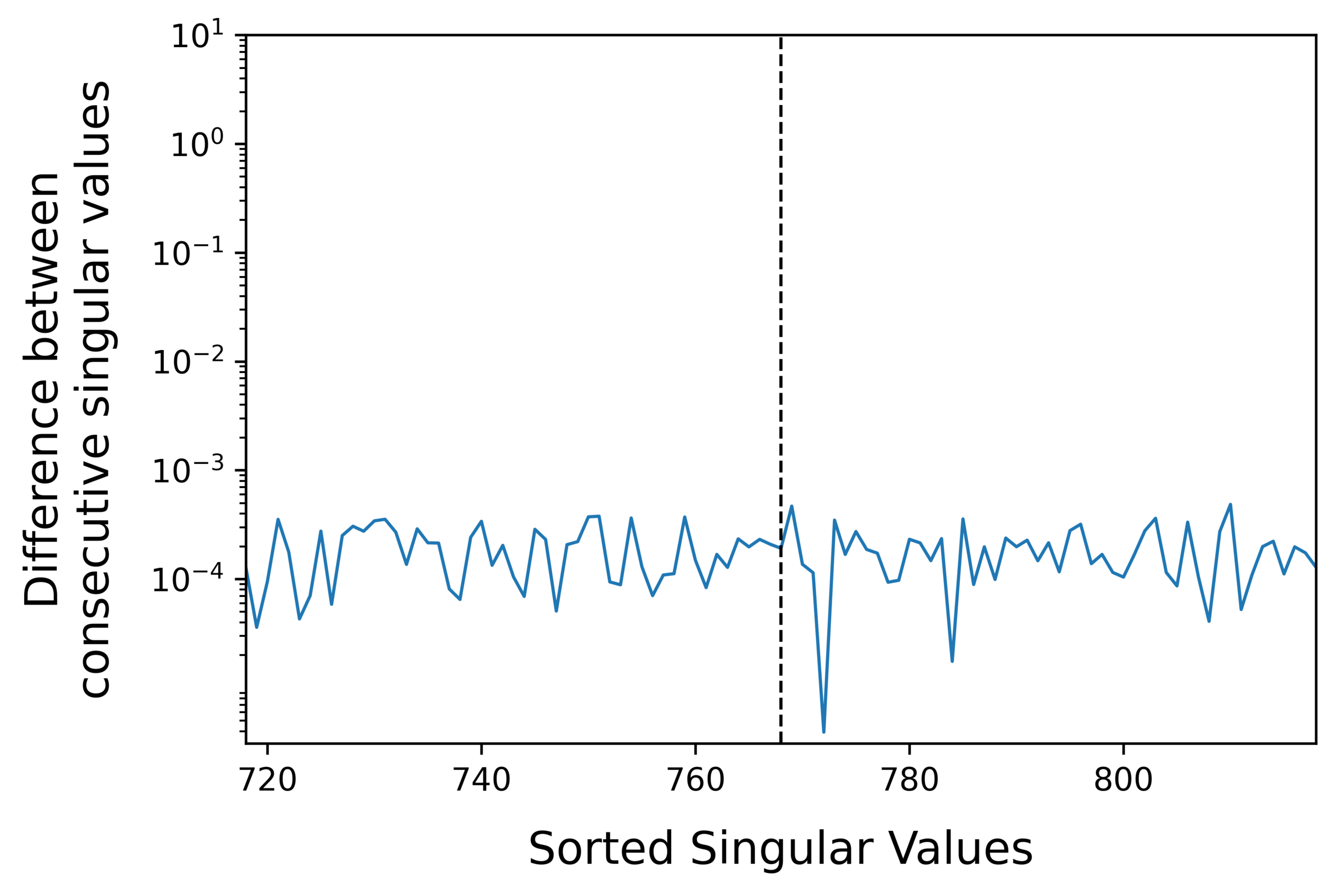}
        \caption{Noise 2.0, Sparsity 0.1}
    \end{subfigure}


\vspace{.2in}

    \begin{subfigure}{0.24\textwidth} 
        \centering
        \includegraphics[width=\textwidth]{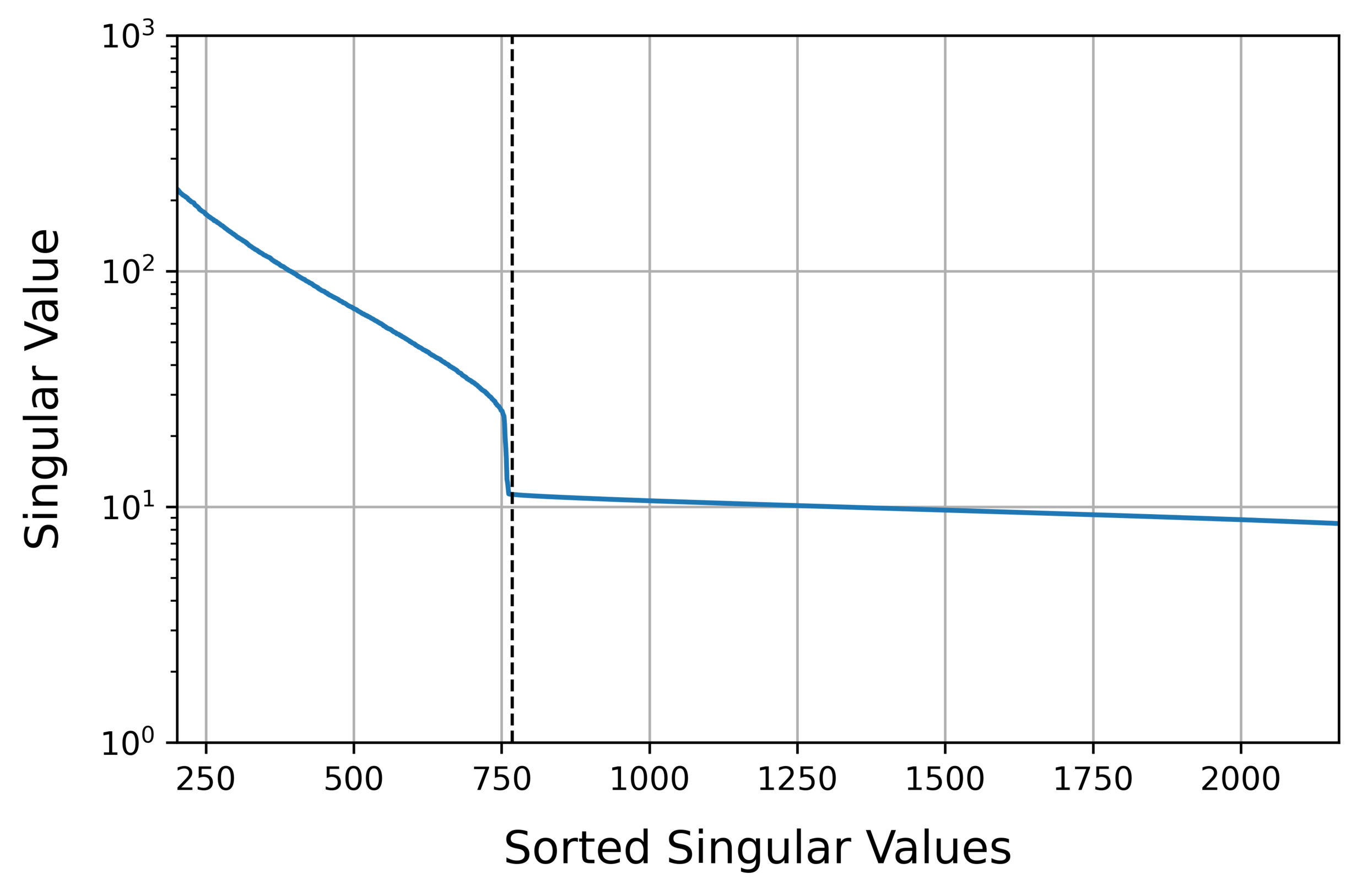}
    \end{subfigure}
    \begin{subfigure}{0.24\textwidth}
        \centering
        \includegraphics[width=\textwidth]{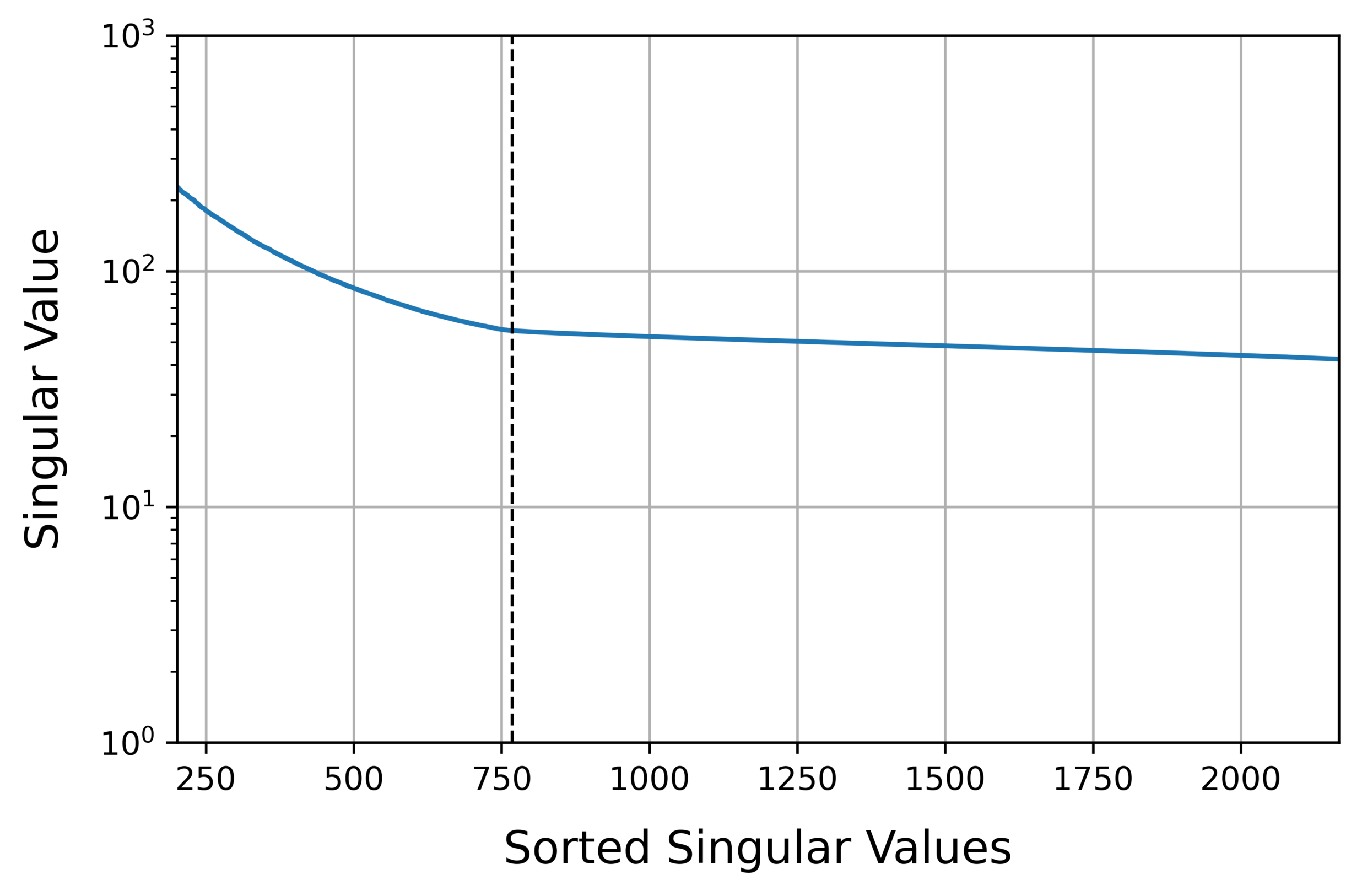}
    \end{subfigure}
    \begin{subfigure}{0.24\textwidth}
        \centering
        \includegraphics[width=\textwidth]{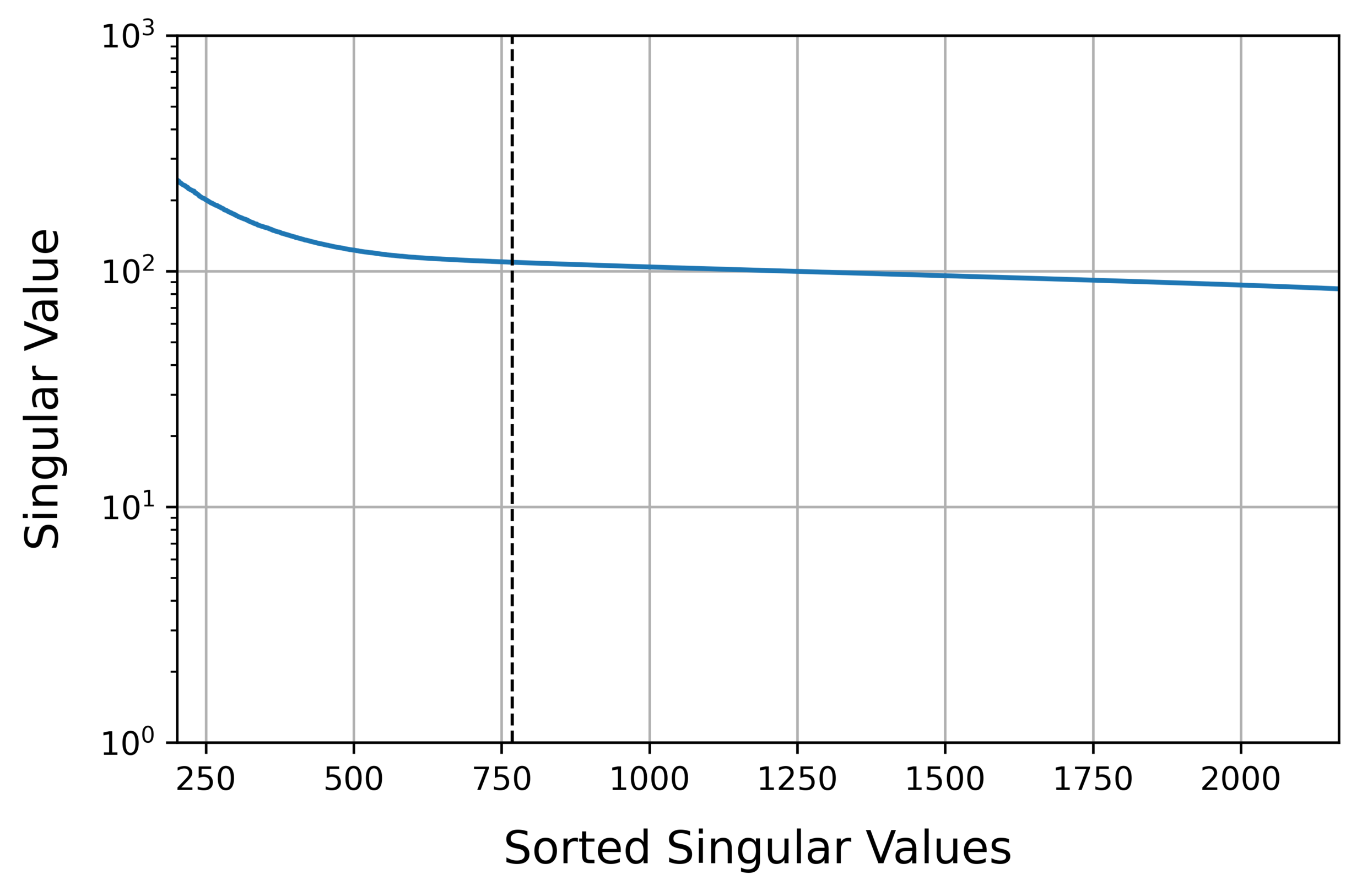}
    \end{subfigure}
    \begin{subfigure}{0.24\textwidth}
        \centering
        \includegraphics[width=\textwidth]{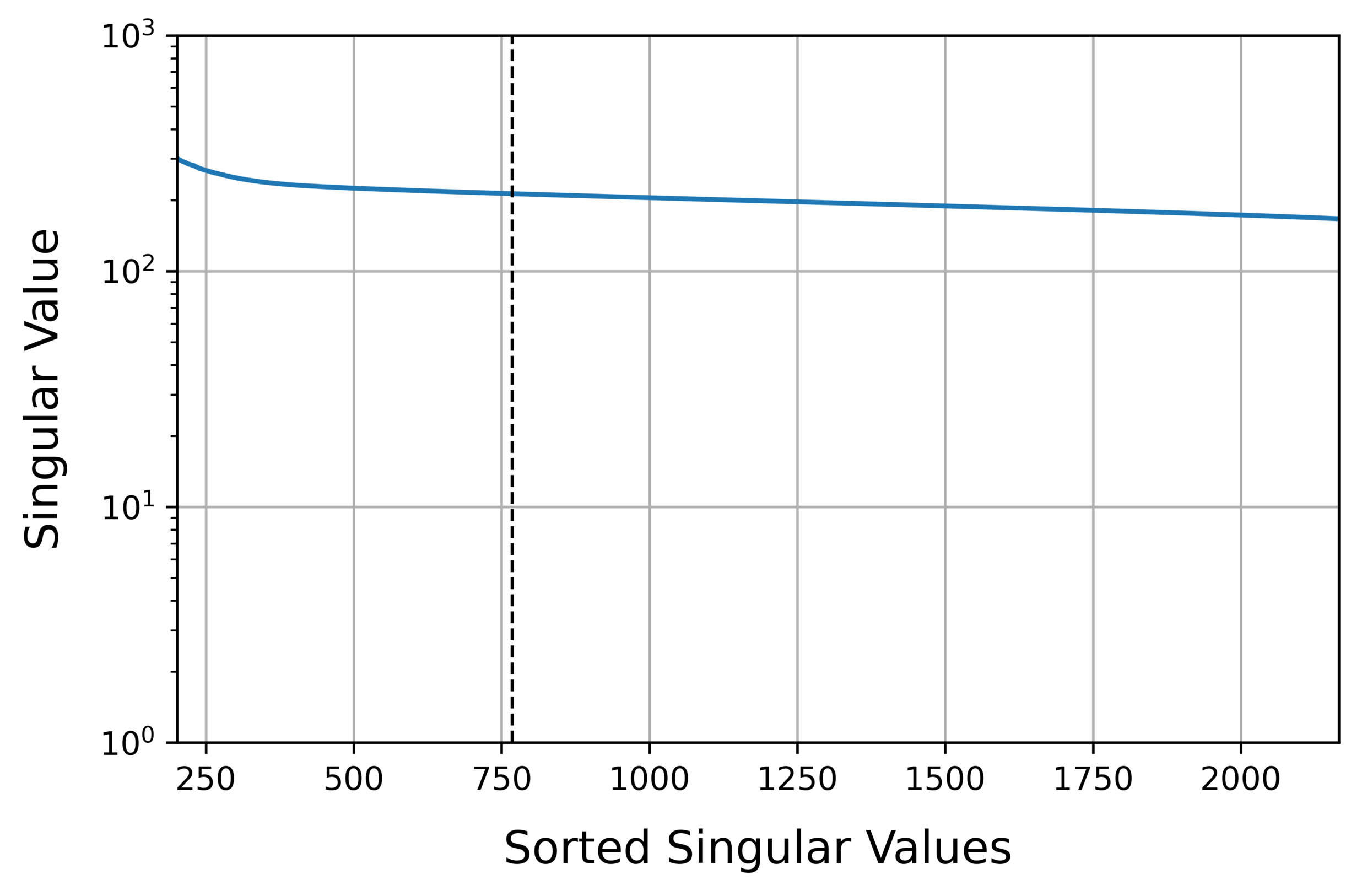}
    \end{subfigure}


    \begin{subfigure}{0.24\textwidth} 
        \centering
        \includegraphics[width=\textwidth]{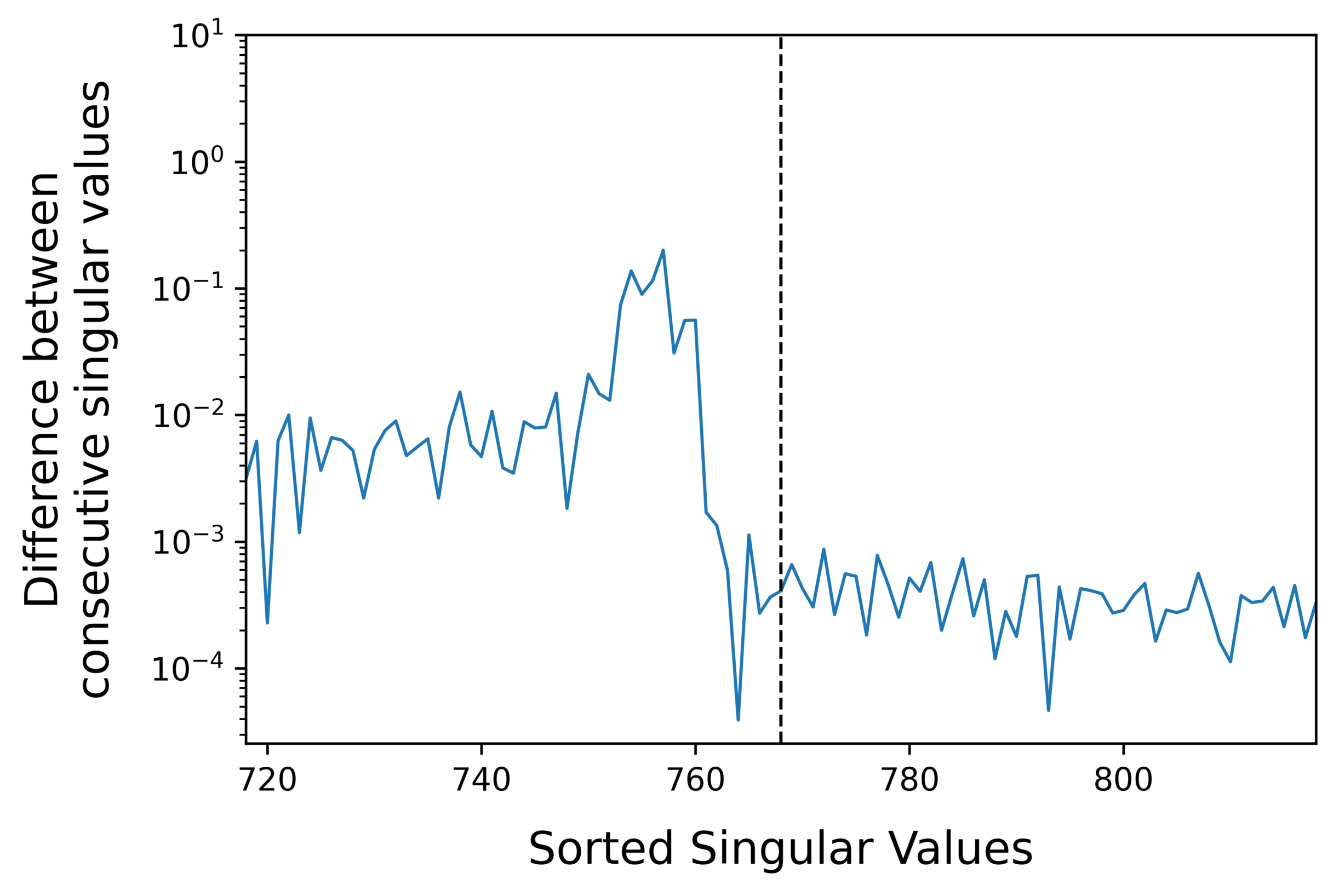}
        \caption{Noise 0.1, Sparsity 0.25}
    \end{subfigure}
    \begin{subfigure}{0.24\textwidth}
        \centering
        \includegraphics[width=\textwidth]{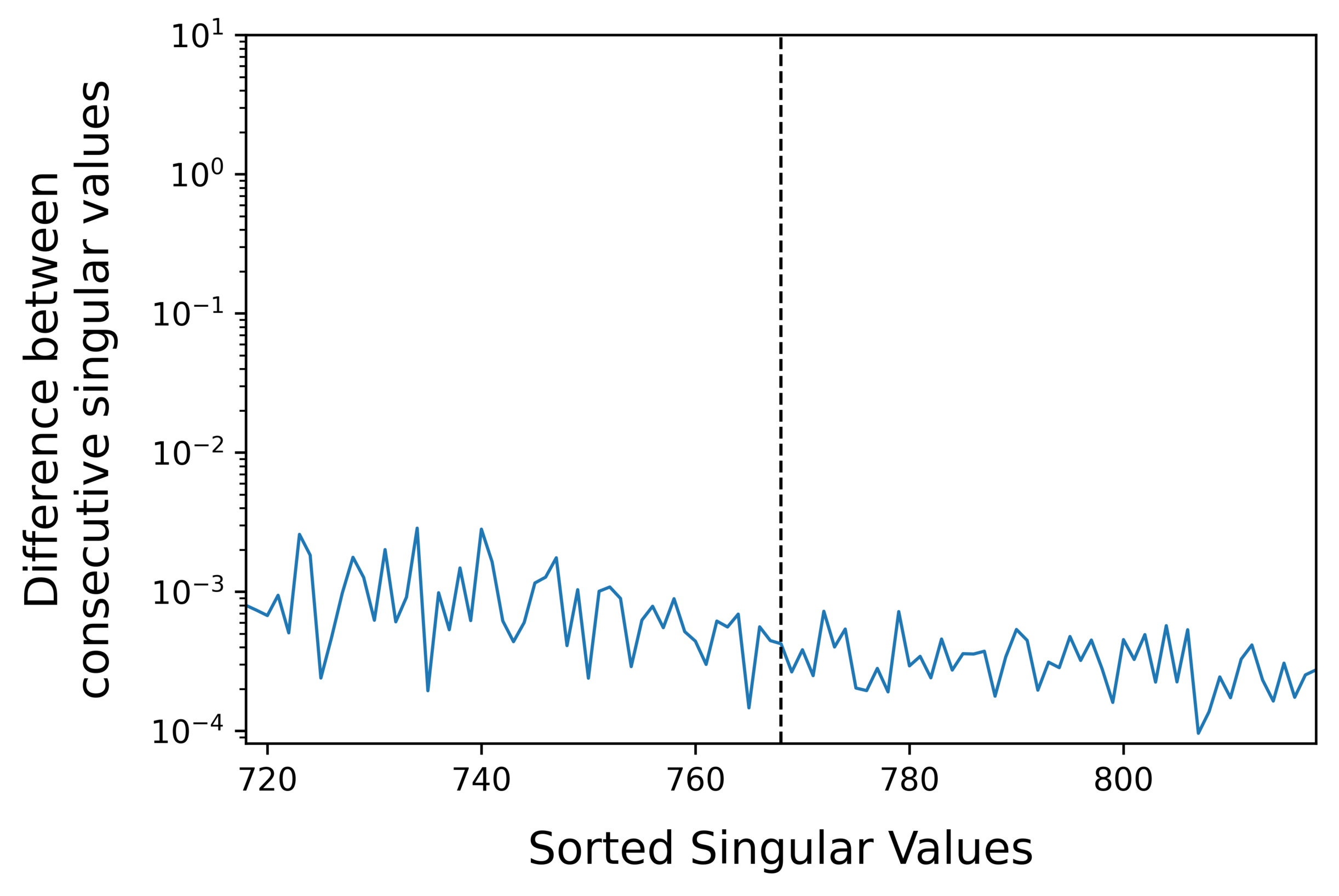}
        \caption{Noise 0.5, Sparsity 0.25}
    \end{subfigure}
    \begin{subfigure}{0.24\textwidth}
        \centering
        \includegraphics[width=\textwidth]{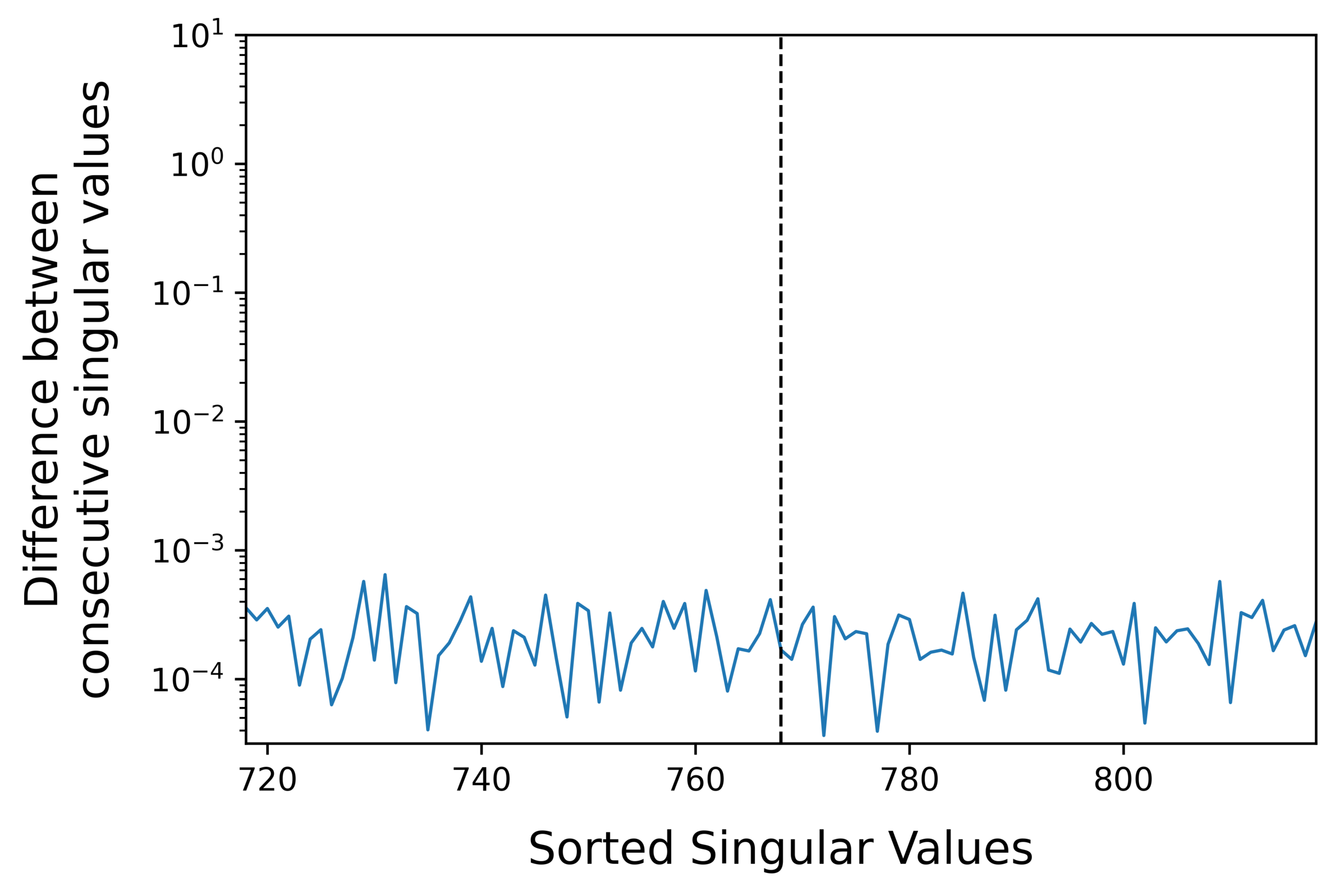}
        \caption{Noise 1.0, Sparsity 0.25}
    \end{subfigure}
    \begin{subfigure}{0.24\textwidth}
        \centering
        \includegraphics[width=\textwidth]{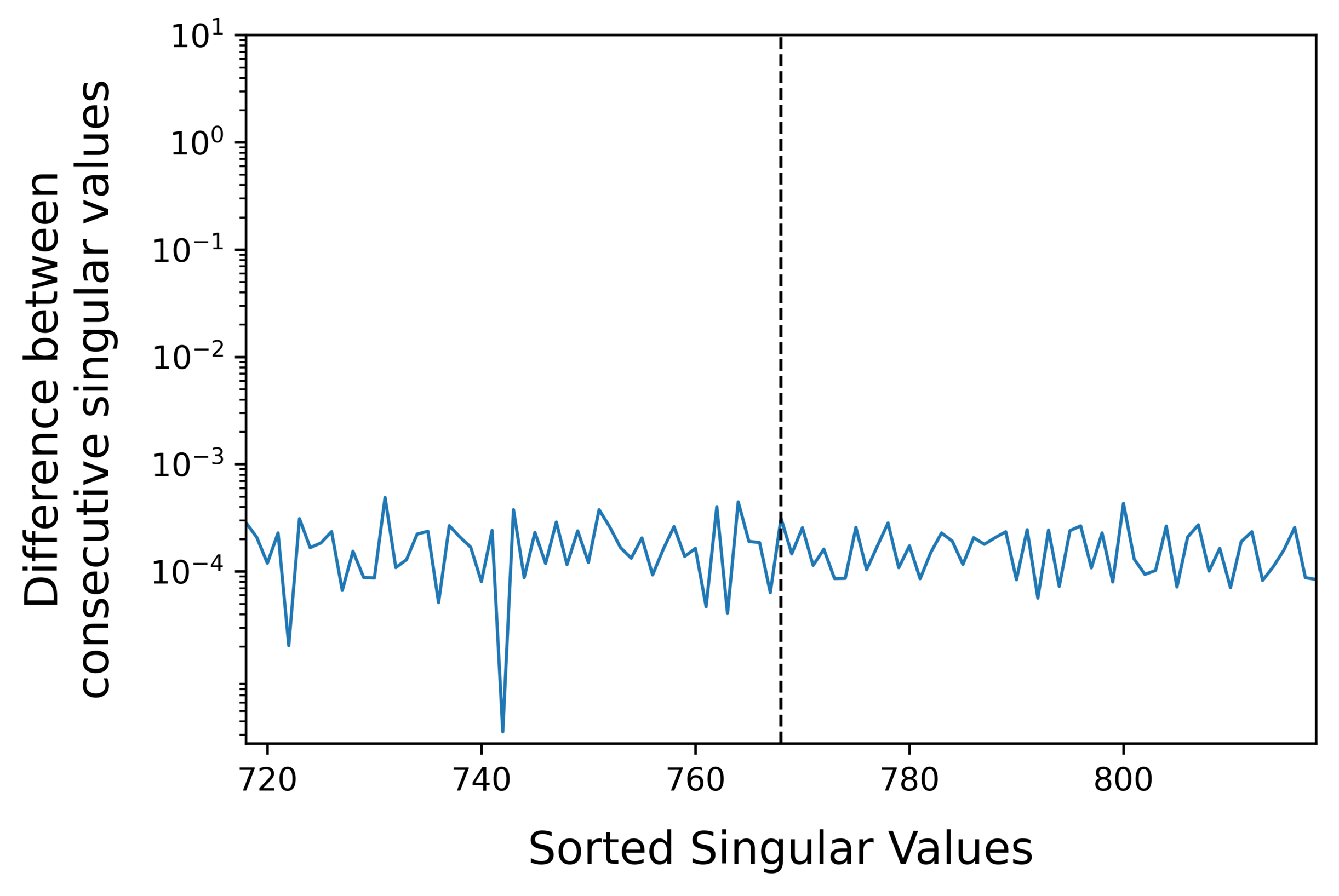}
        \caption{Noise 2.0, Sparsity 0.25}
    \end{subfigure}


\vspace{.2in}

    \begin{subfigure}{0.24\textwidth} 
        \centering
        \includegraphics[width=\textwidth]{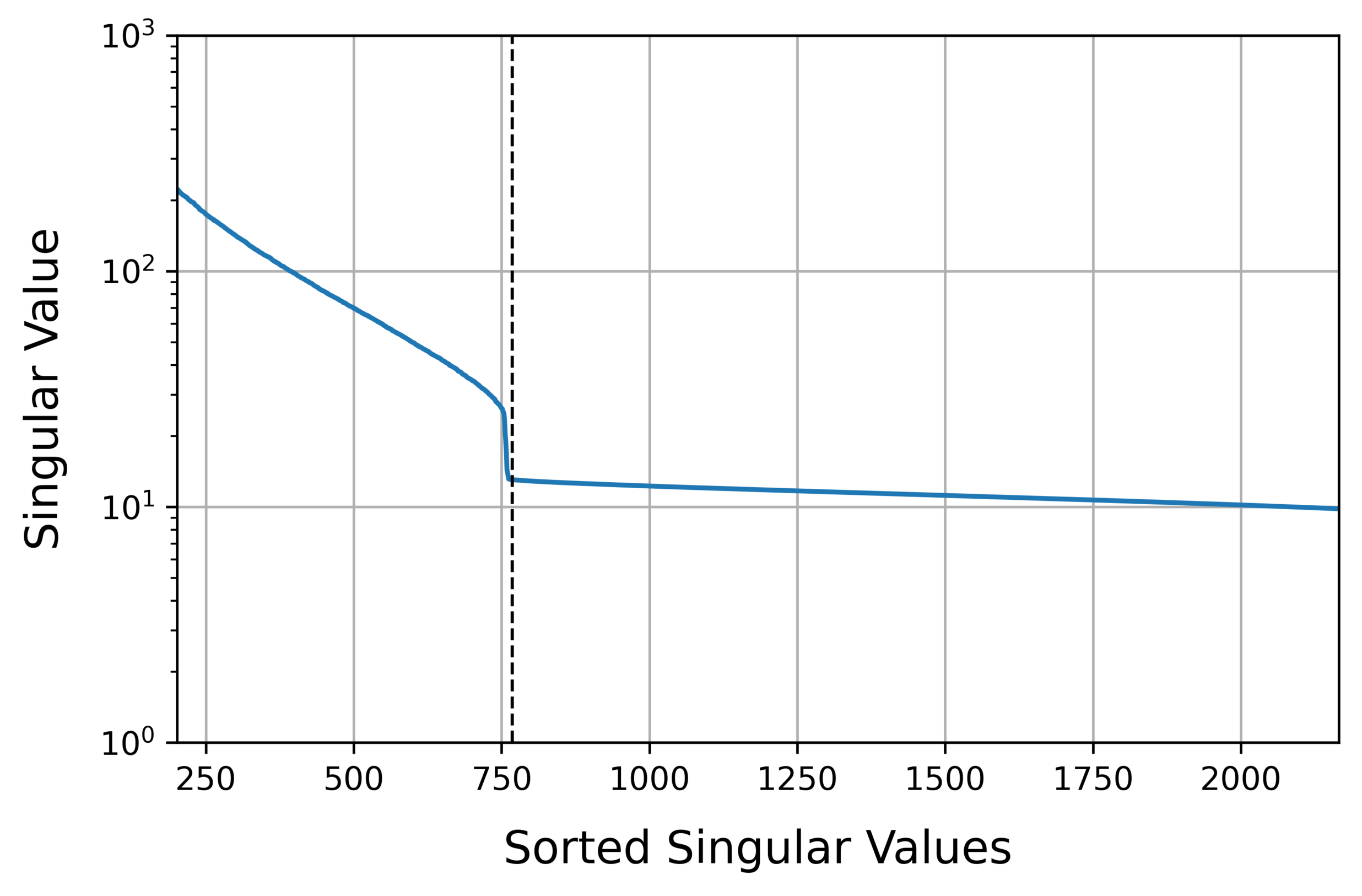}
    \end{subfigure}
    \begin{subfigure}{0.24\textwidth}
        \centering
        \includegraphics[width=\textwidth]{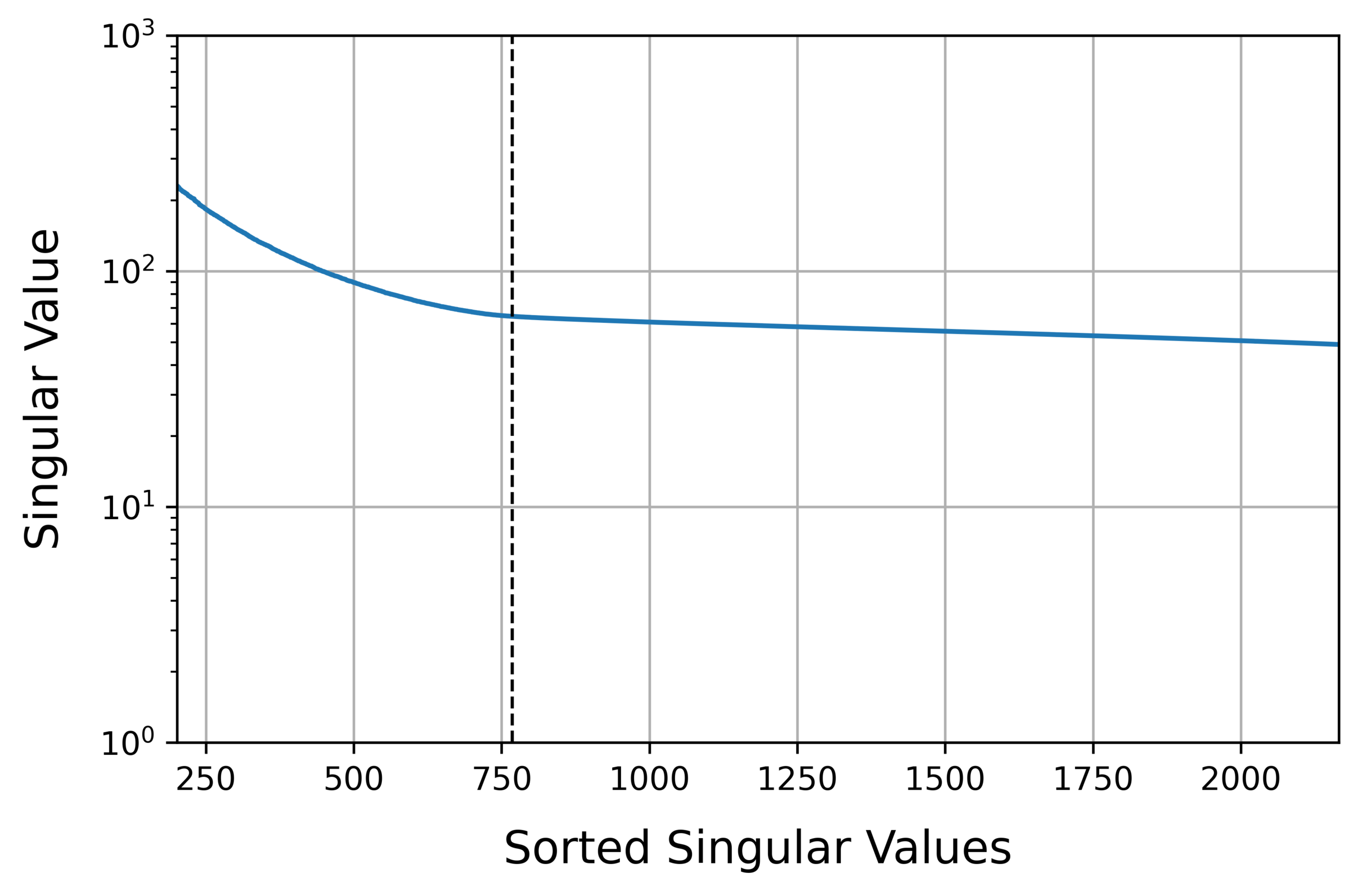}
    \end{subfigure}
    \begin{subfigure}{0.24\textwidth}
        \centering
        \includegraphics[width=\textwidth]{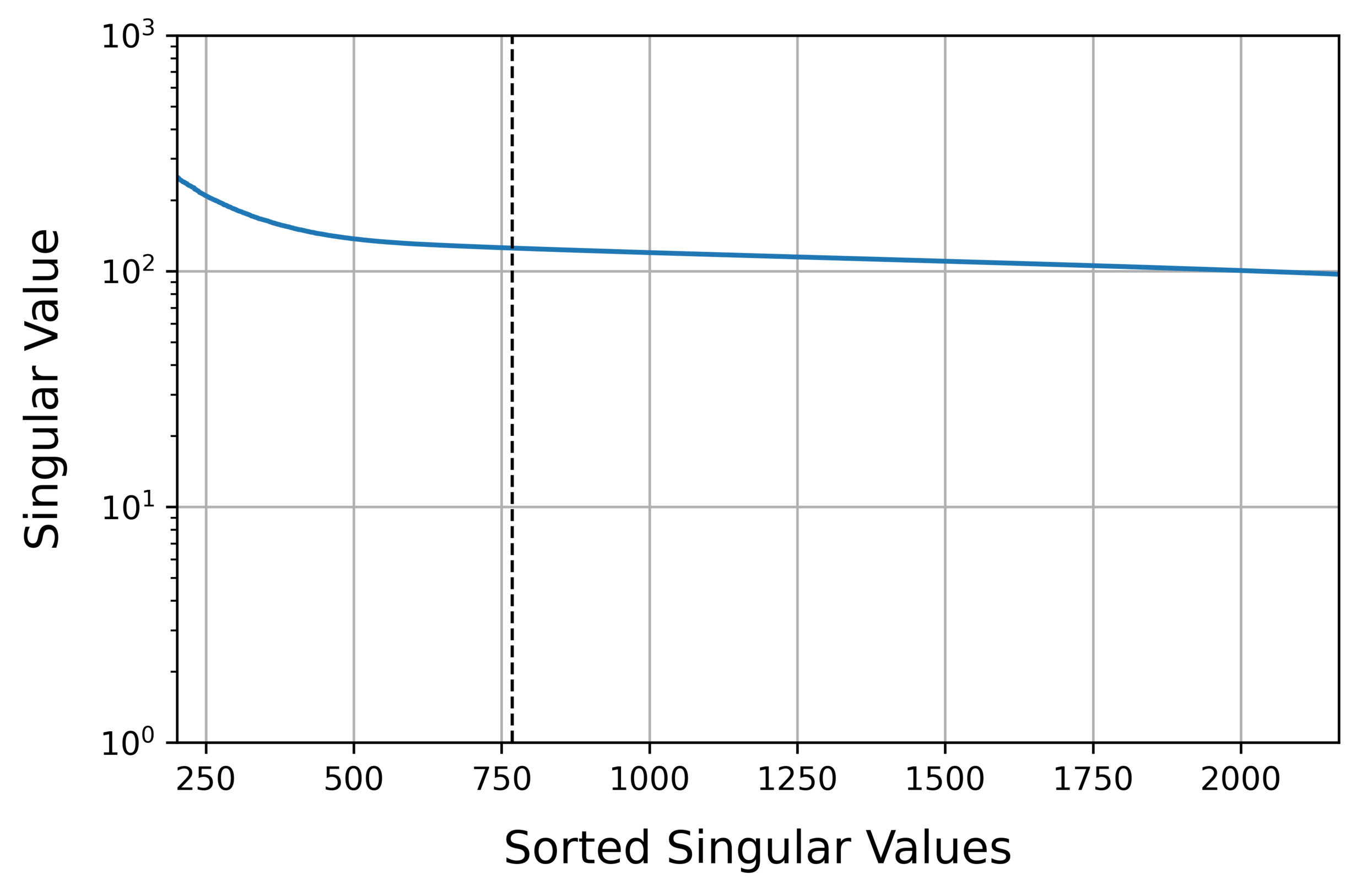}
    \end{subfigure}
    \begin{subfigure}{0.24\textwidth}
        \centering
        \includegraphics[width=\textwidth]{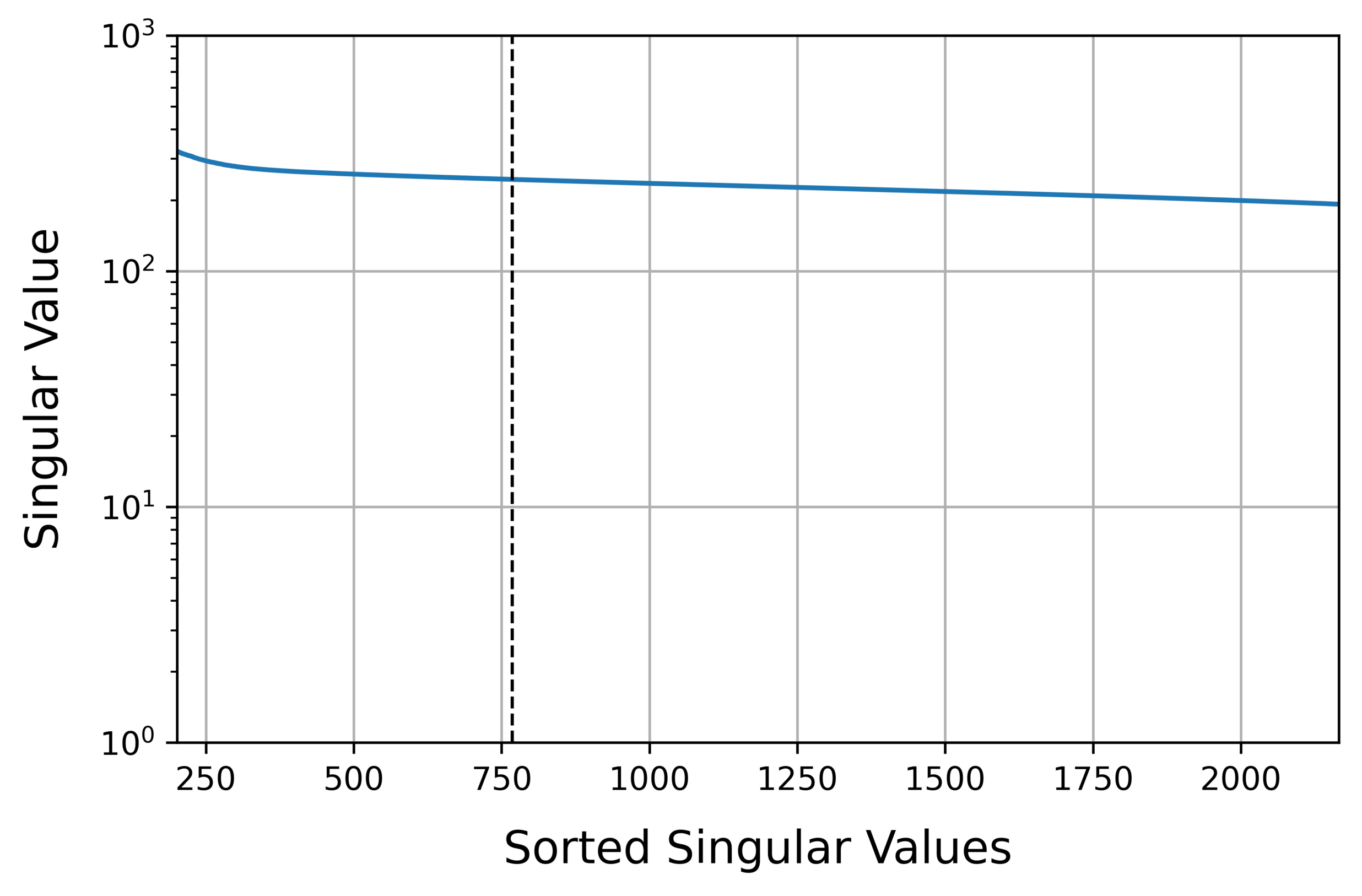}
    \end{subfigure}


    \begin{subfigure}{0.24\textwidth} 
        \centering
        \includegraphics[width=\textwidth]{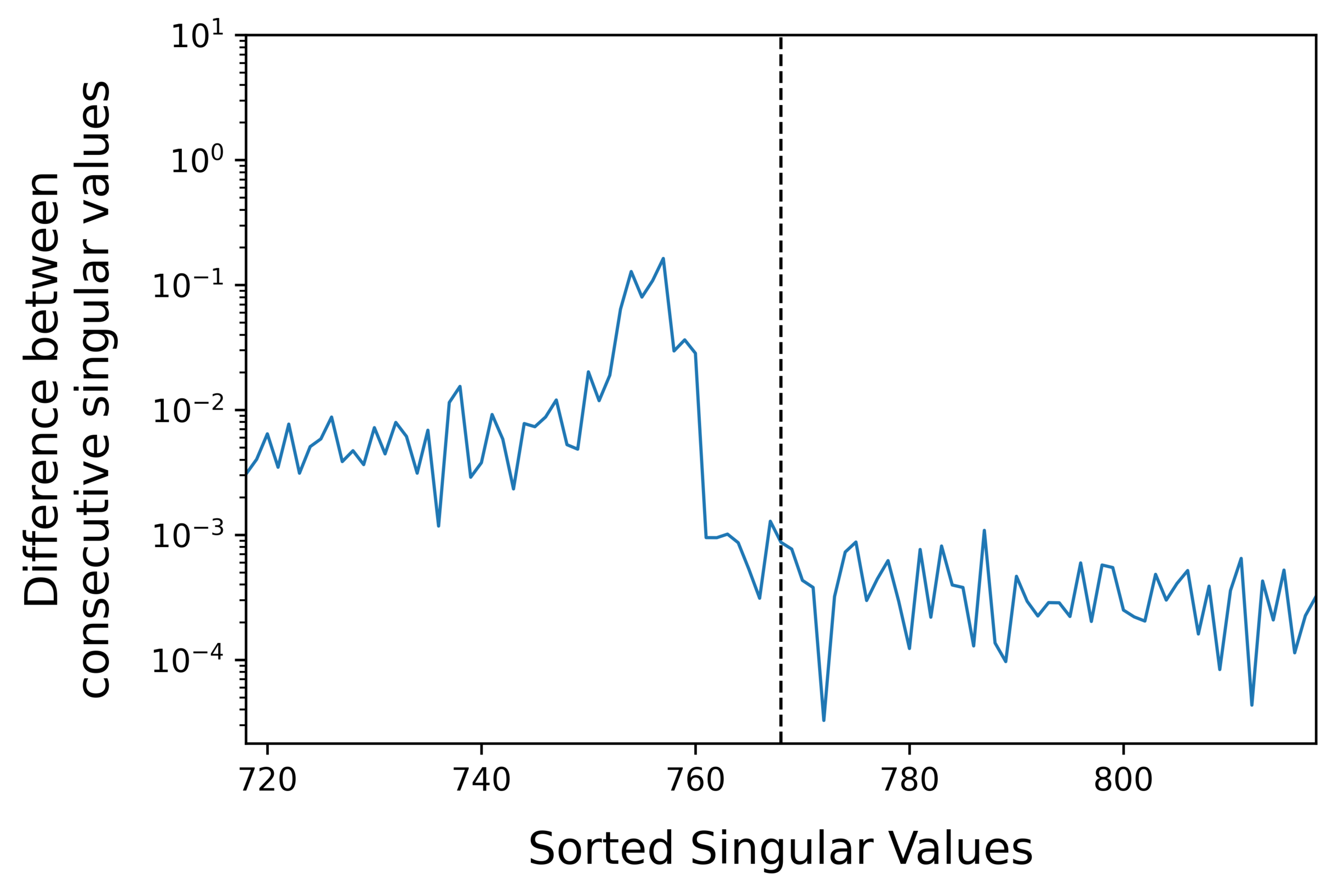}
        \caption{Noise 0.1, Sparsity 0.5}
    \end{subfigure}
    \begin{subfigure}{0.24\textwidth}
        \centering
        \includegraphics[width=\textwidth]{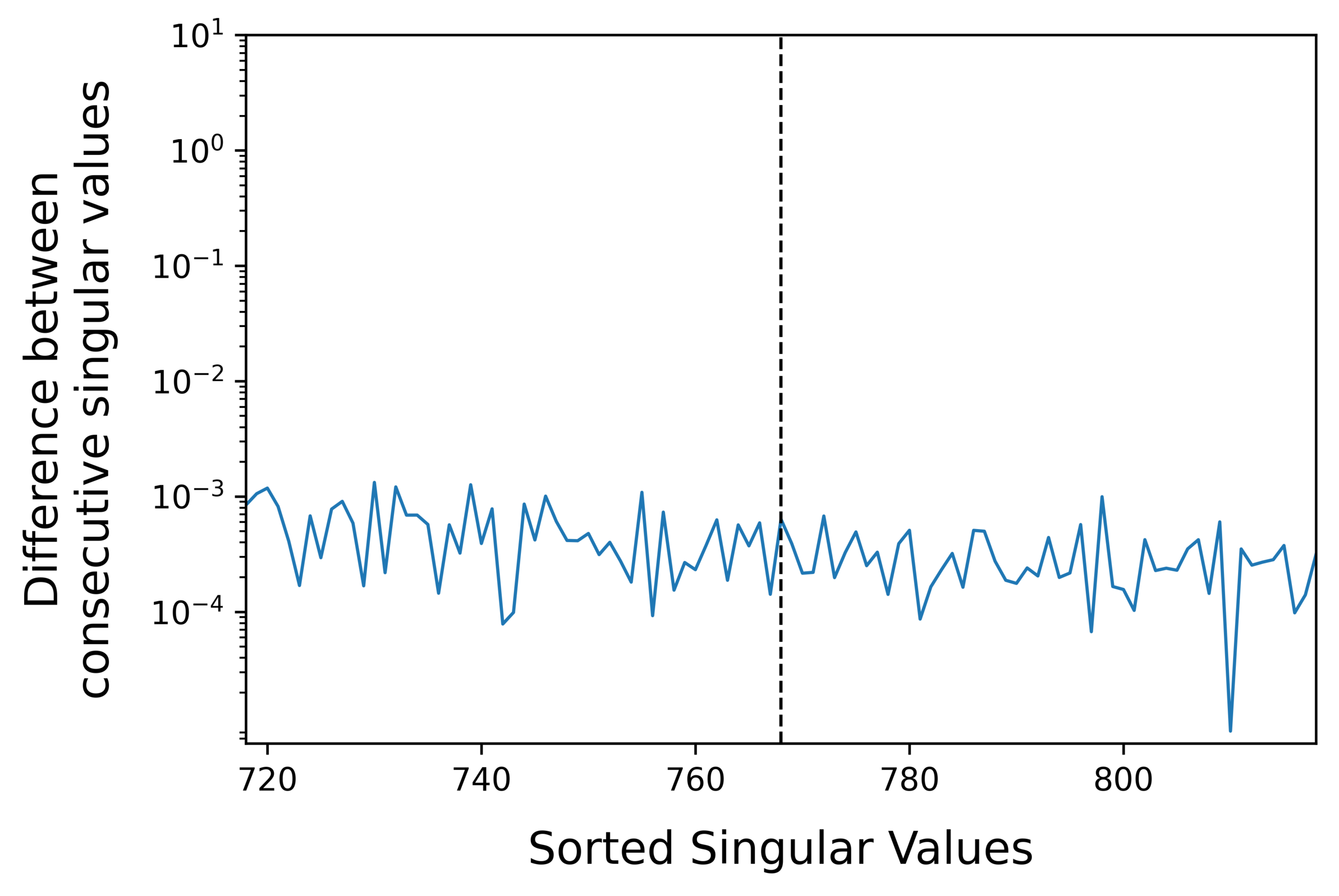}
        \caption{Noise 0.5, Sparsity 0.5}
    \end{subfigure}
    \begin{subfigure}{0.24\textwidth}
        \centering
        \includegraphics[width=\textwidth]{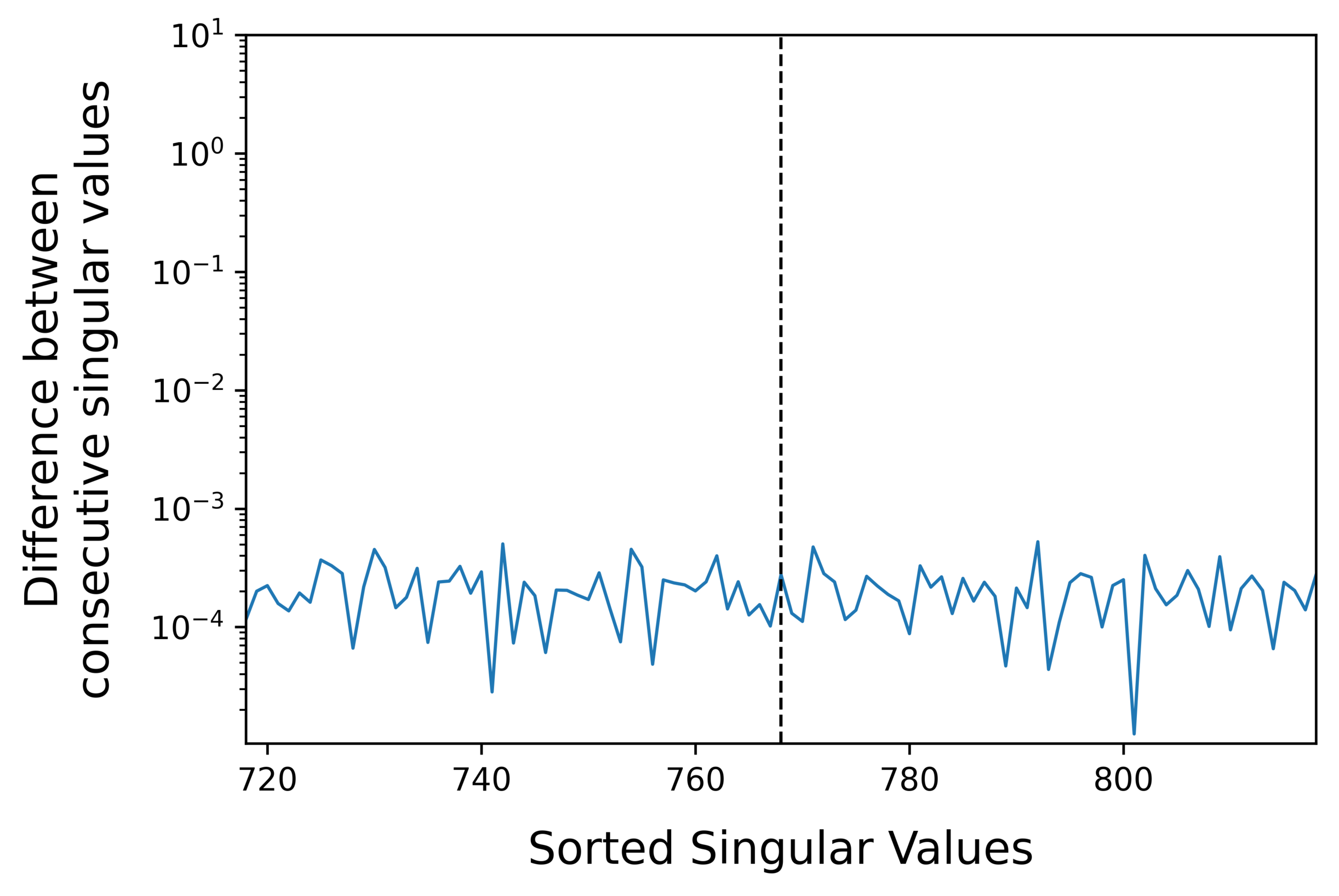}
        \caption{Noise 1.0, Sparsity 0.5}
    \end{subfigure}
    \begin{subfigure}{0.24\textwidth}
        \centering
        \includegraphics[width=\textwidth]{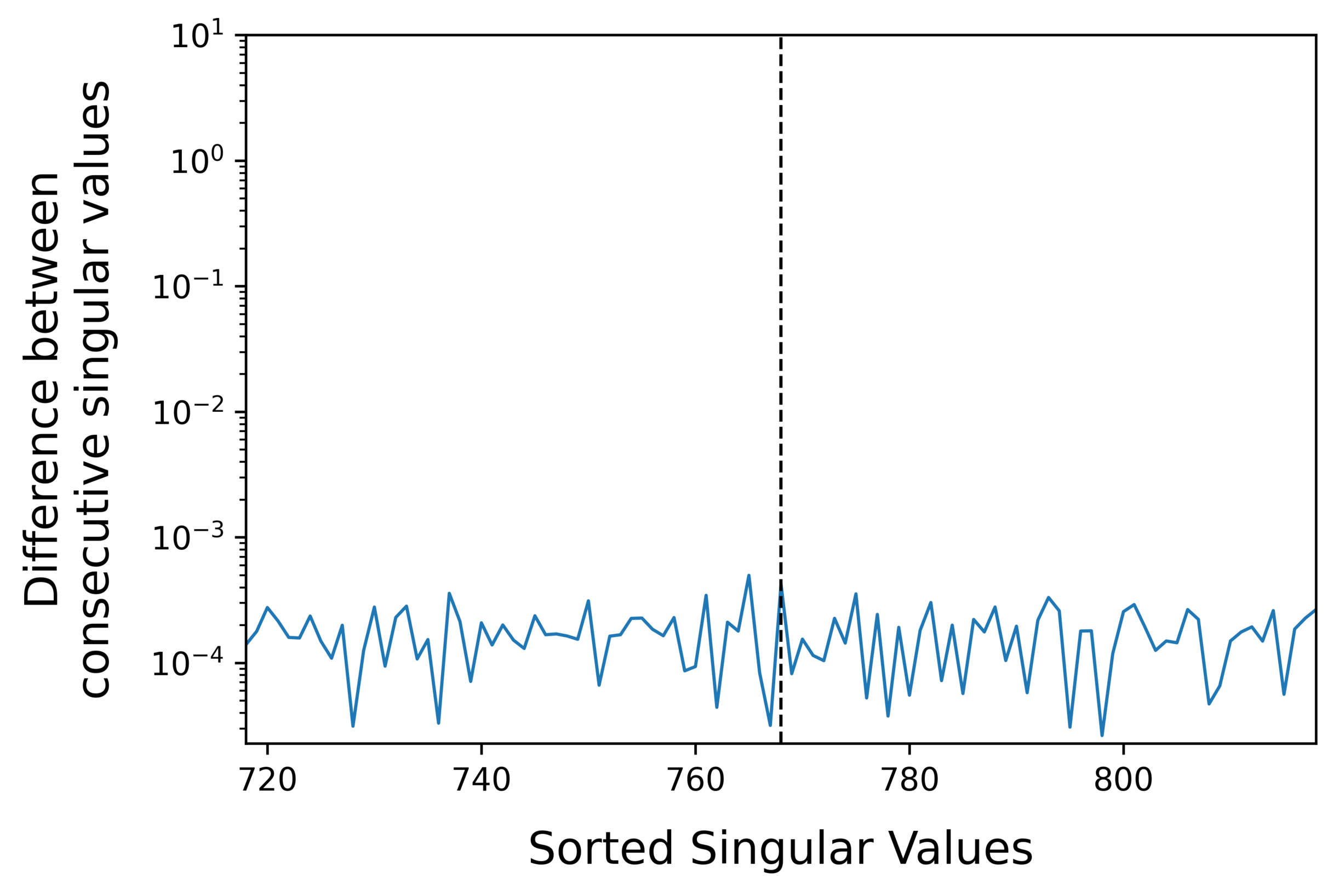}
        \caption{Noise 2.0, Sparsity 0.5}
    \end{subfigure}

    \caption{(GPT-2 Discrete Noise). Dimension extraction attack on GPT-2 by SVD singular values for discrete noise addition with varying sparsity (0.1, 0.25, 0.5) and magnitude (0.1, 0.5, 1.0, 2.0). Each top plot depicts the sorted singular values of the corresponding noisy logits matrix, where a jump at correct position 768 yields a successful dimension extraction. Each bottom plot provides a zoomed in picture of the difference between consecutive singular values; the attack outputs the index for which this difference is maximized.}
    \label{fig:GPT2-many-noise-discrete}
\end{figure*}

\begin{figure*}[h!] 
    \centering
    

    \begin{subfigure}{0.24\textwidth} 
        \centering
        \includegraphics[width=\textwidth]{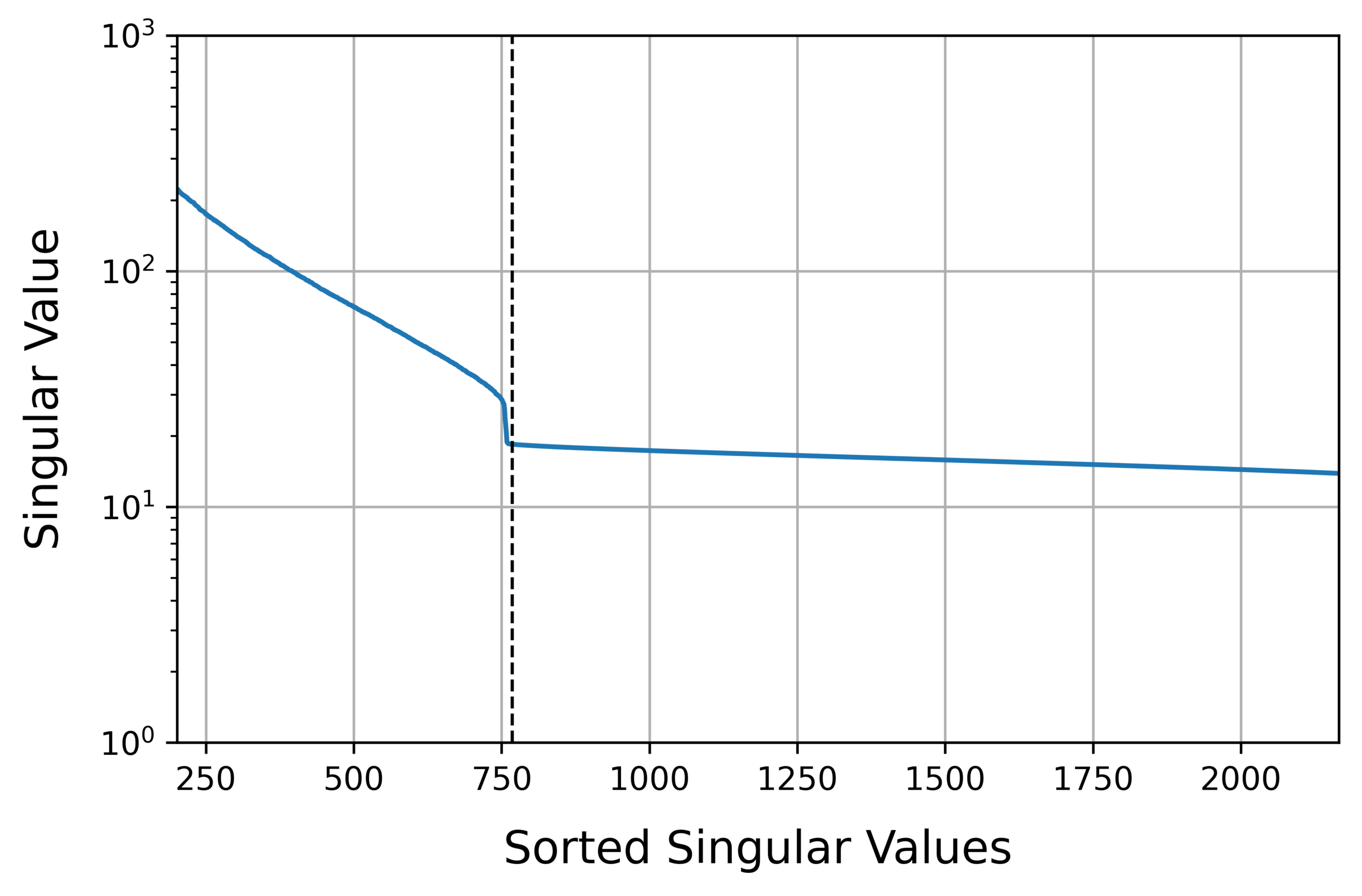}
    \end{subfigure}
    \begin{subfigure}{0.24\textwidth}
        \centering
        \includegraphics[width=\textwidth]{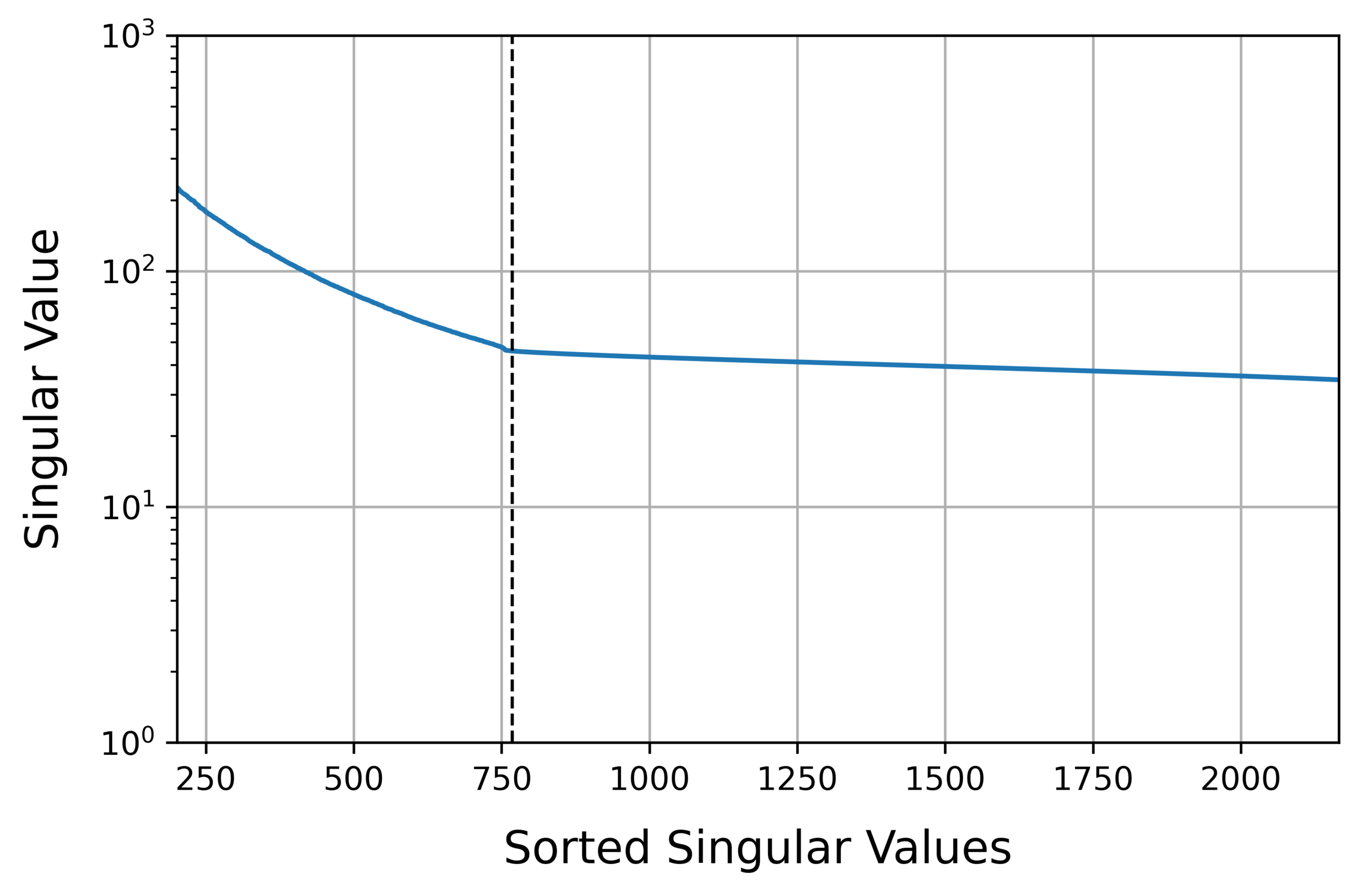}
    \end{subfigure}
    \begin{subfigure}{0.24\textwidth}
        \centering
        \includegraphics[width=\textwidth]{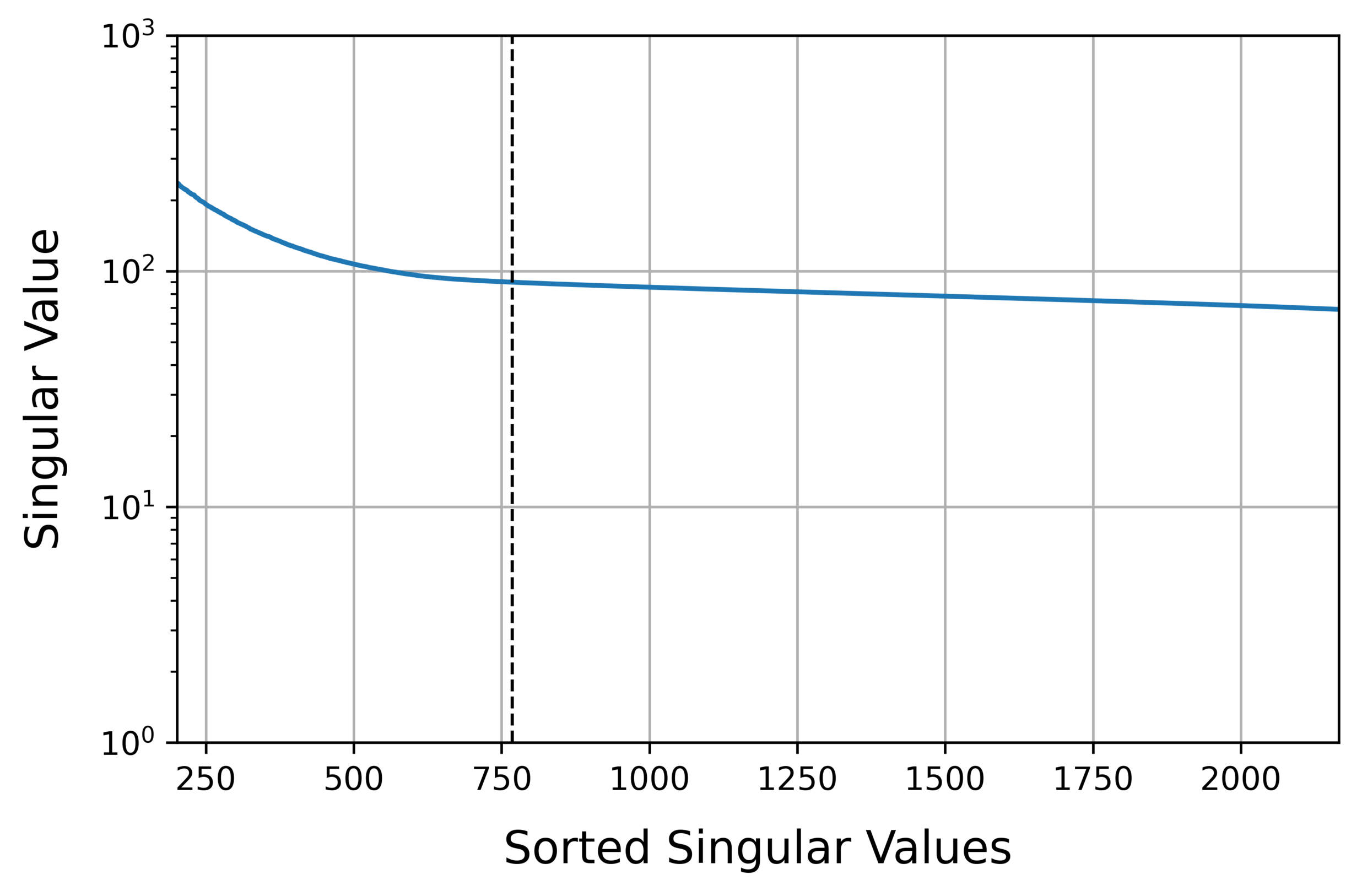}
    \end{subfigure}
    \begin{subfigure}{0.24\textwidth}
        \centering
        \includegraphics[width=\textwidth]{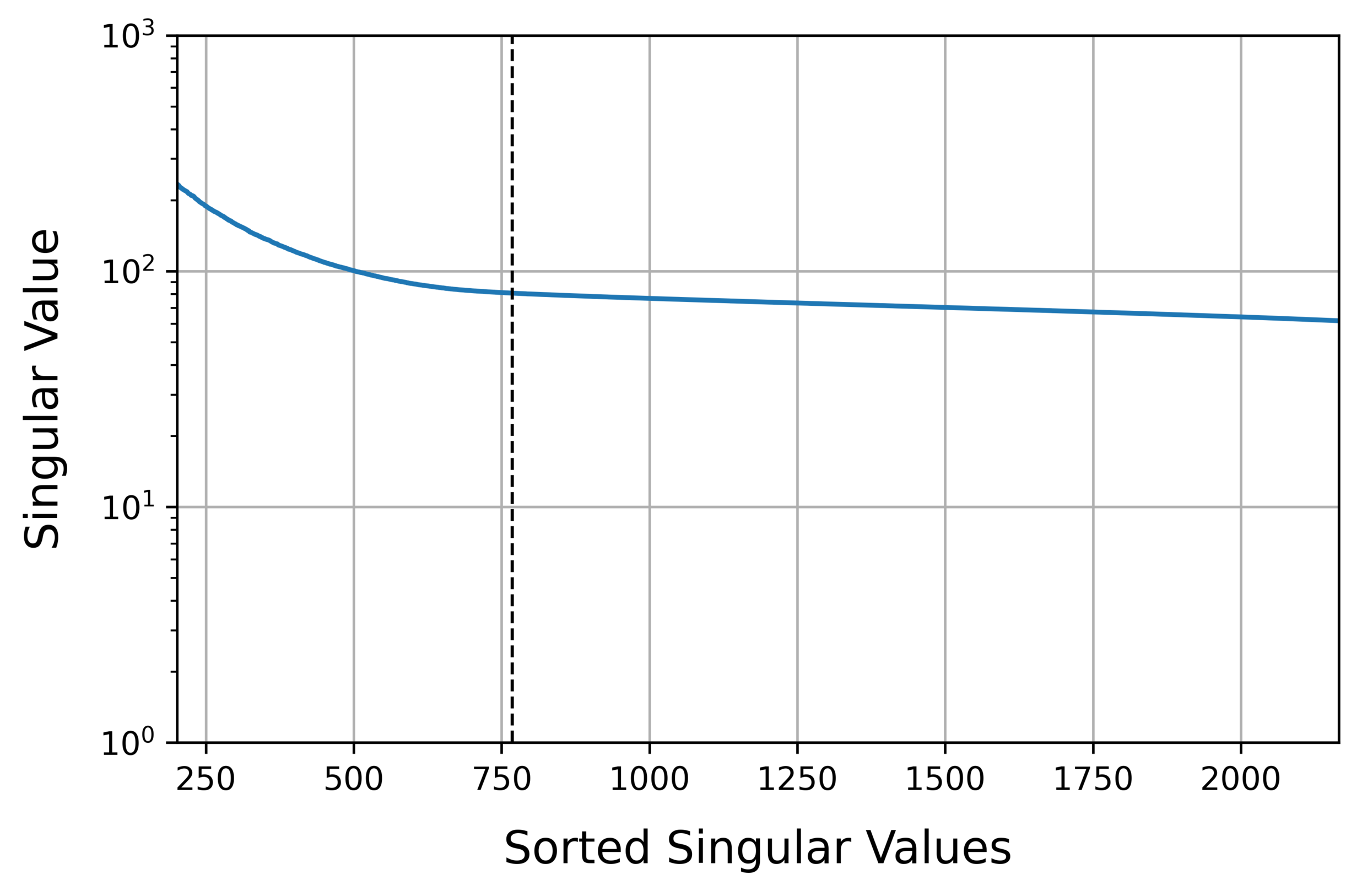}
    \end{subfigure}


    \begin{subfigure}{0.24\textwidth} 
        \centering
        \includegraphics[width=\textwidth]{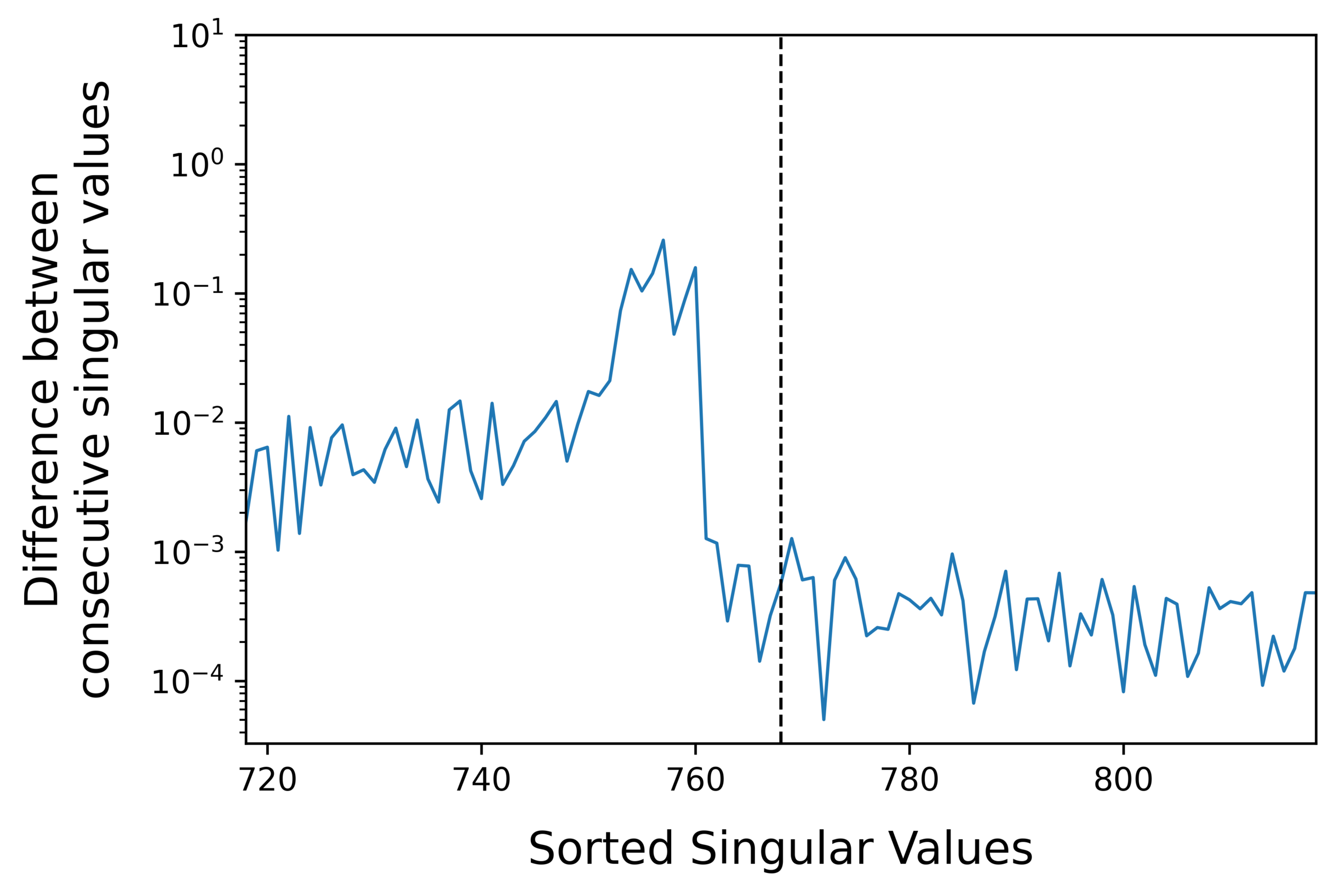}
        \caption{$\text{std}=0.1$, $\gamma=0.1$}
    \end{subfigure}
    \begin{subfigure}{0.24\textwidth}
        \centering
        \includegraphics[width=\textwidth]{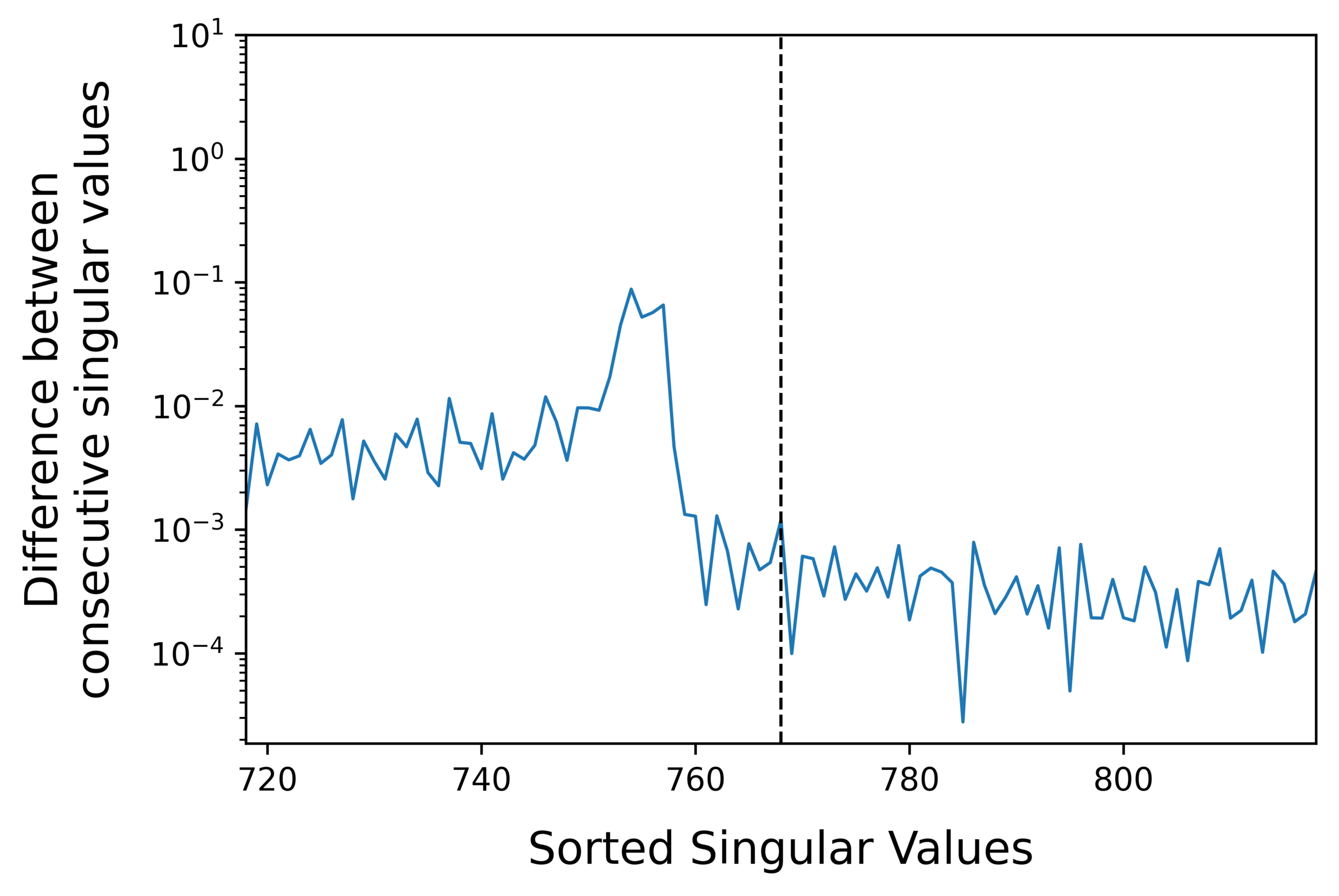}
        \caption{$\text{std}=0.25$, $\gamma=0.1$}
    \end{subfigure}
    \begin{subfigure}{0.24\textwidth}
        \centering
        \includegraphics[width=\textwidth]{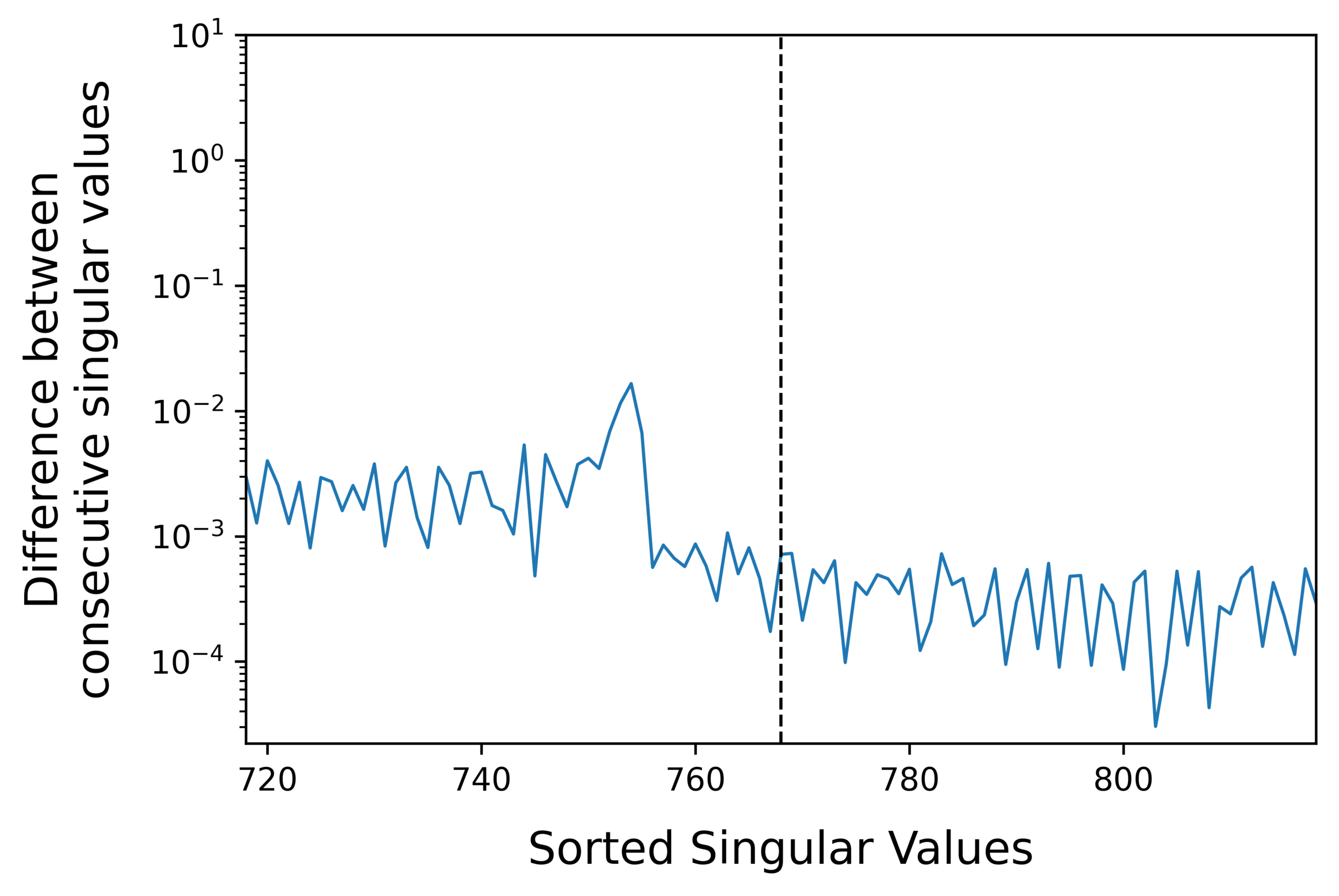}
        \caption{$\text{std}=0.5$, $\gamma=0.1$}
    \end{subfigure}
    \begin{subfigure}{0.24\textwidth}
        \centering
        \includegraphics[width=\textwidth]{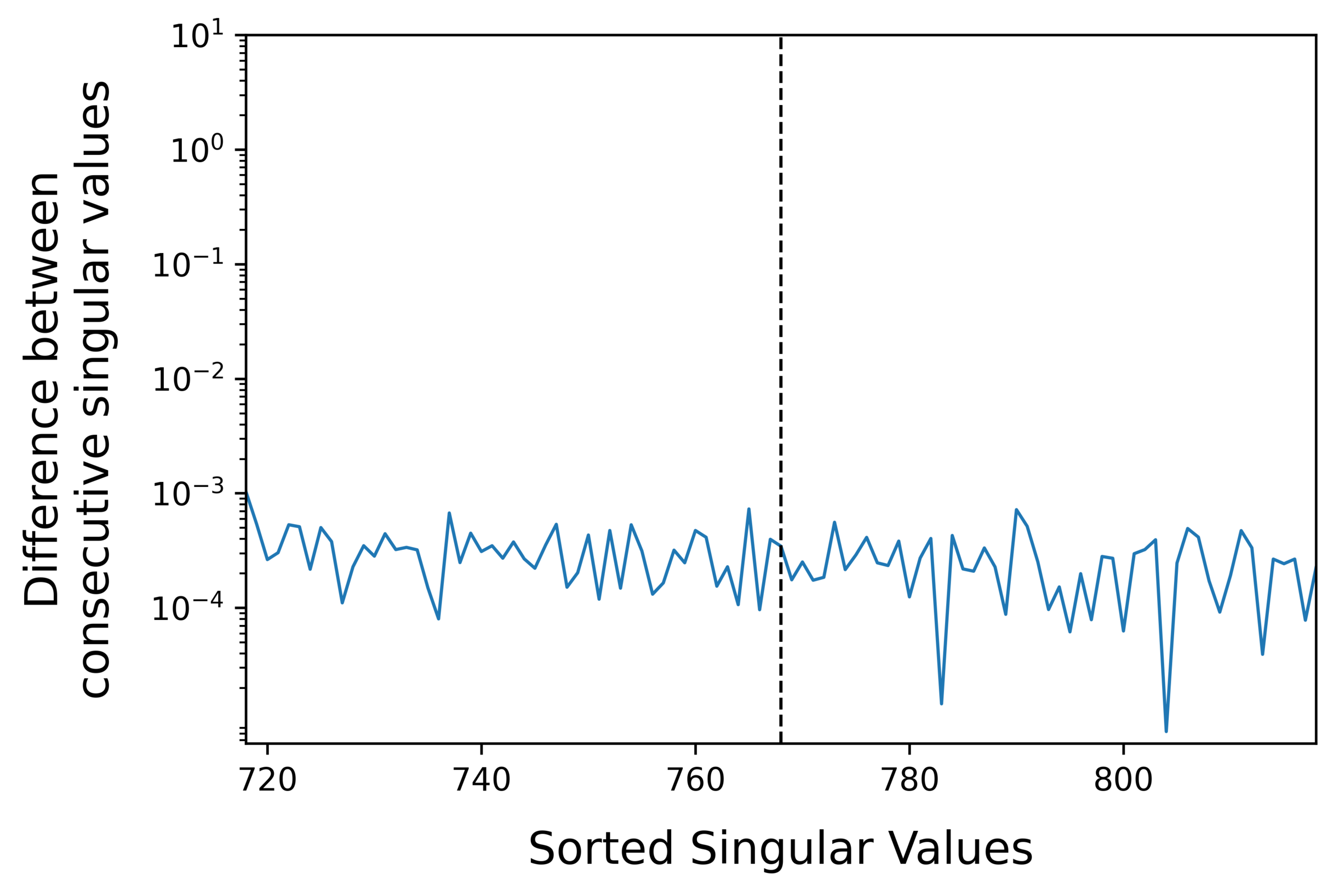}
        \caption{$\text{std}=1.0$, $\gamma=0.1$}
    \end{subfigure}


\vspace{.2in}

    \begin{subfigure}{0.24\textwidth} 
        \centering
        \includegraphics[width=\textwidth]{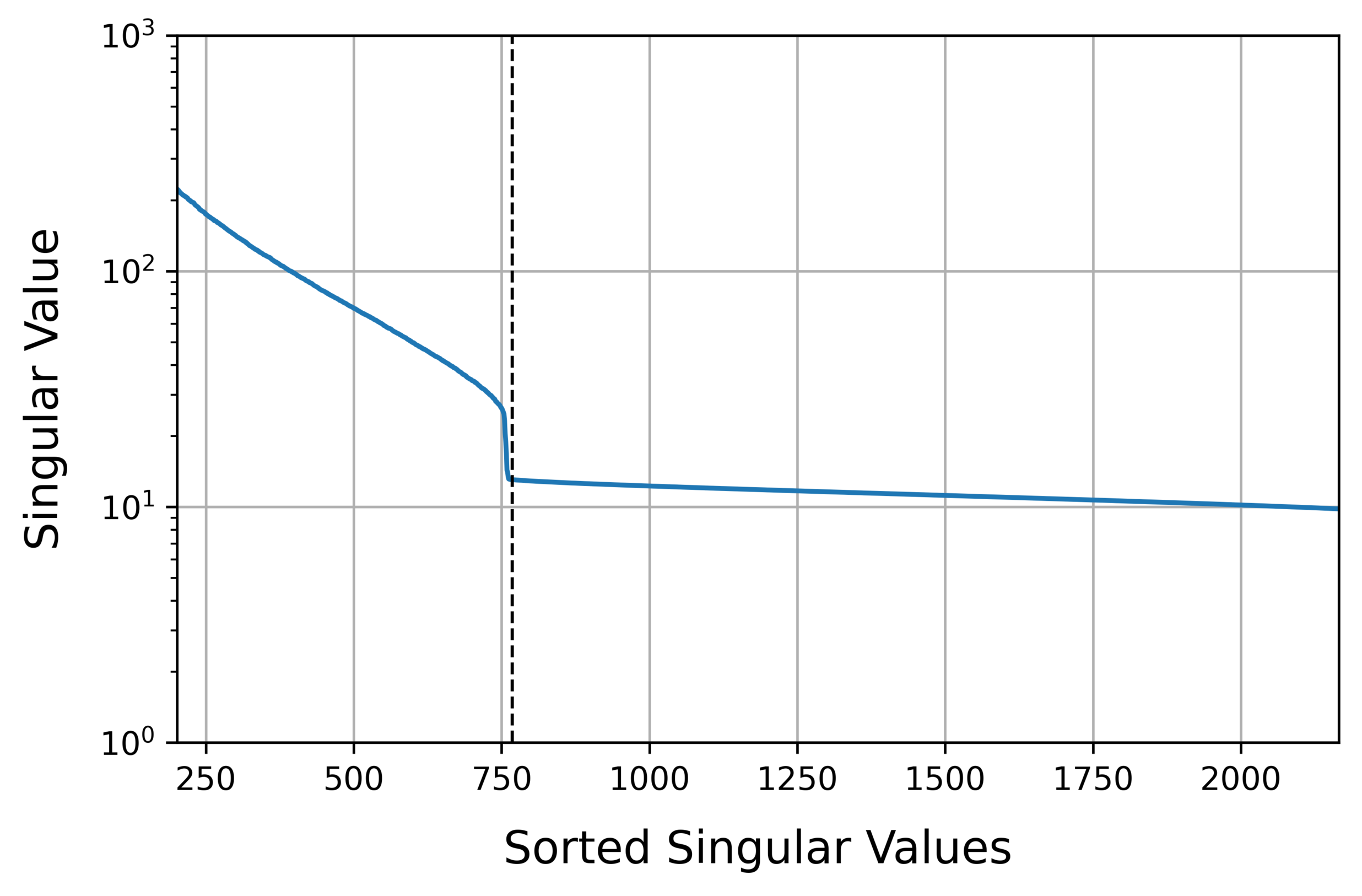}
    \end{subfigure}
    \begin{subfigure}{0.24\textwidth}
        \centering
        \includegraphics[width=\textwidth]{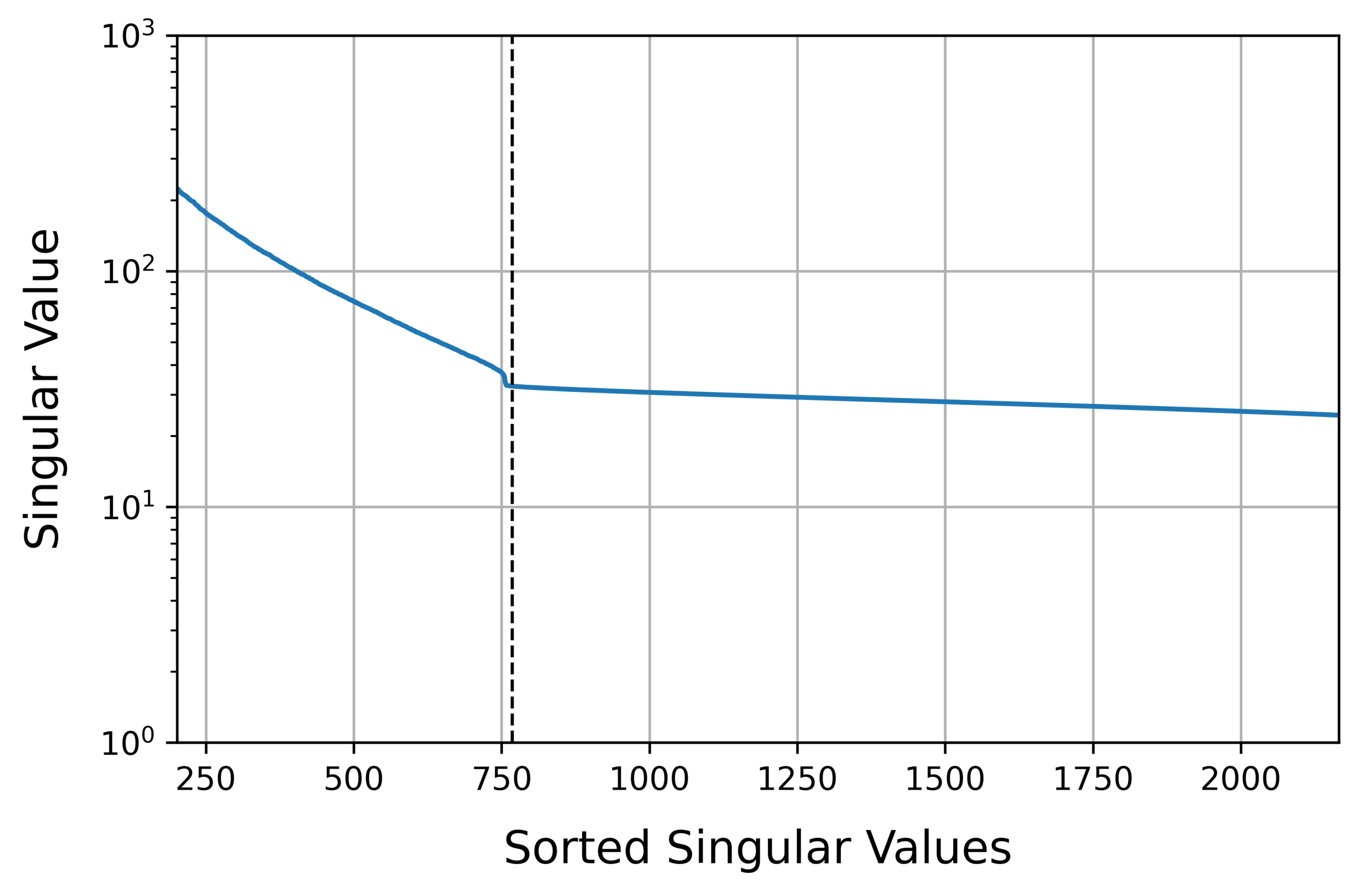}
    \end{subfigure}
    \begin{subfigure}{0.24\textwidth}
        \centering
        \includegraphics[width=\textwidth]{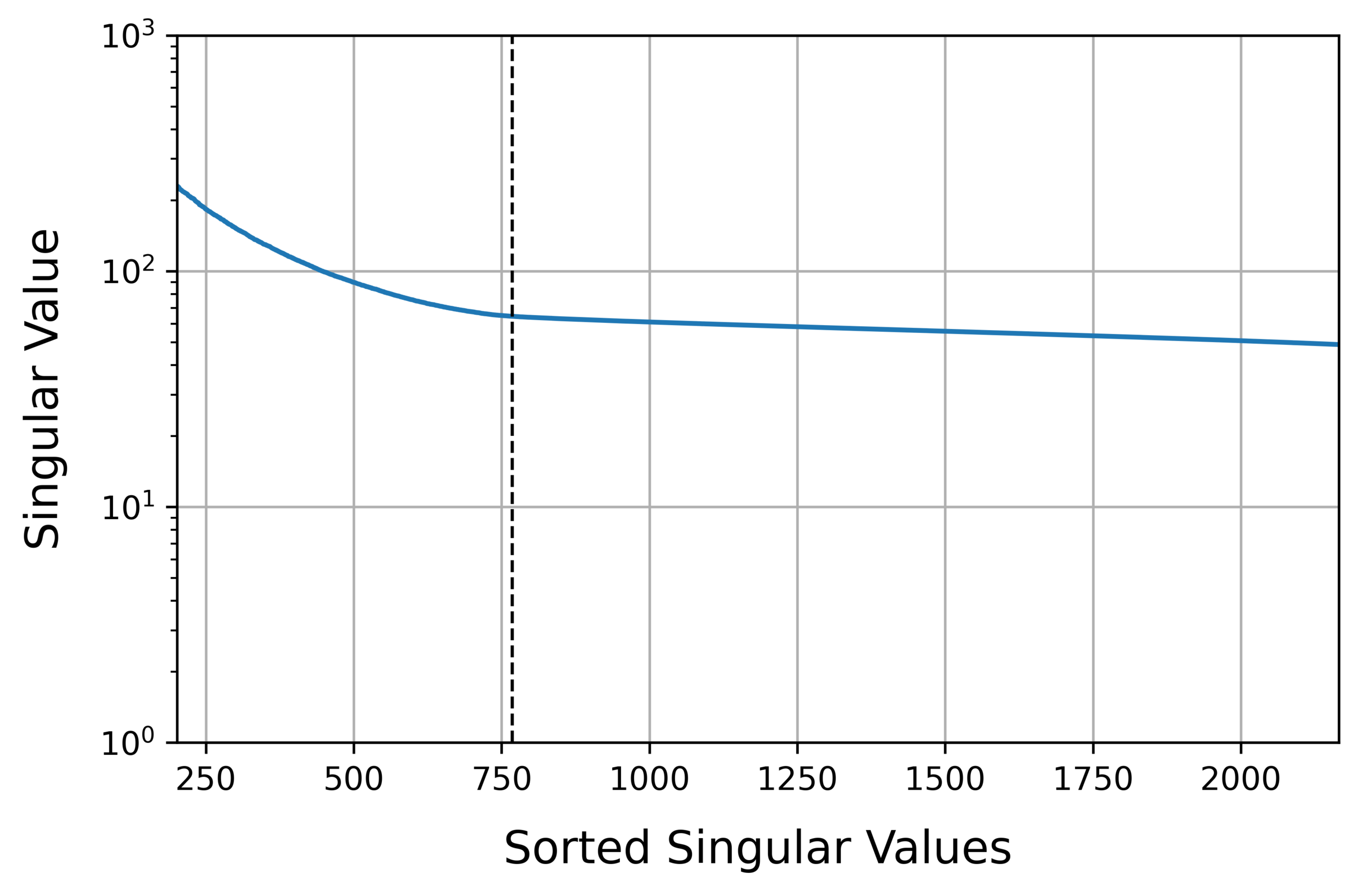}
    \end{subfigure}
    \begin{subfigure}{0.24\textwidth}
        \centering
        \includegraphics[width=\textwidth]{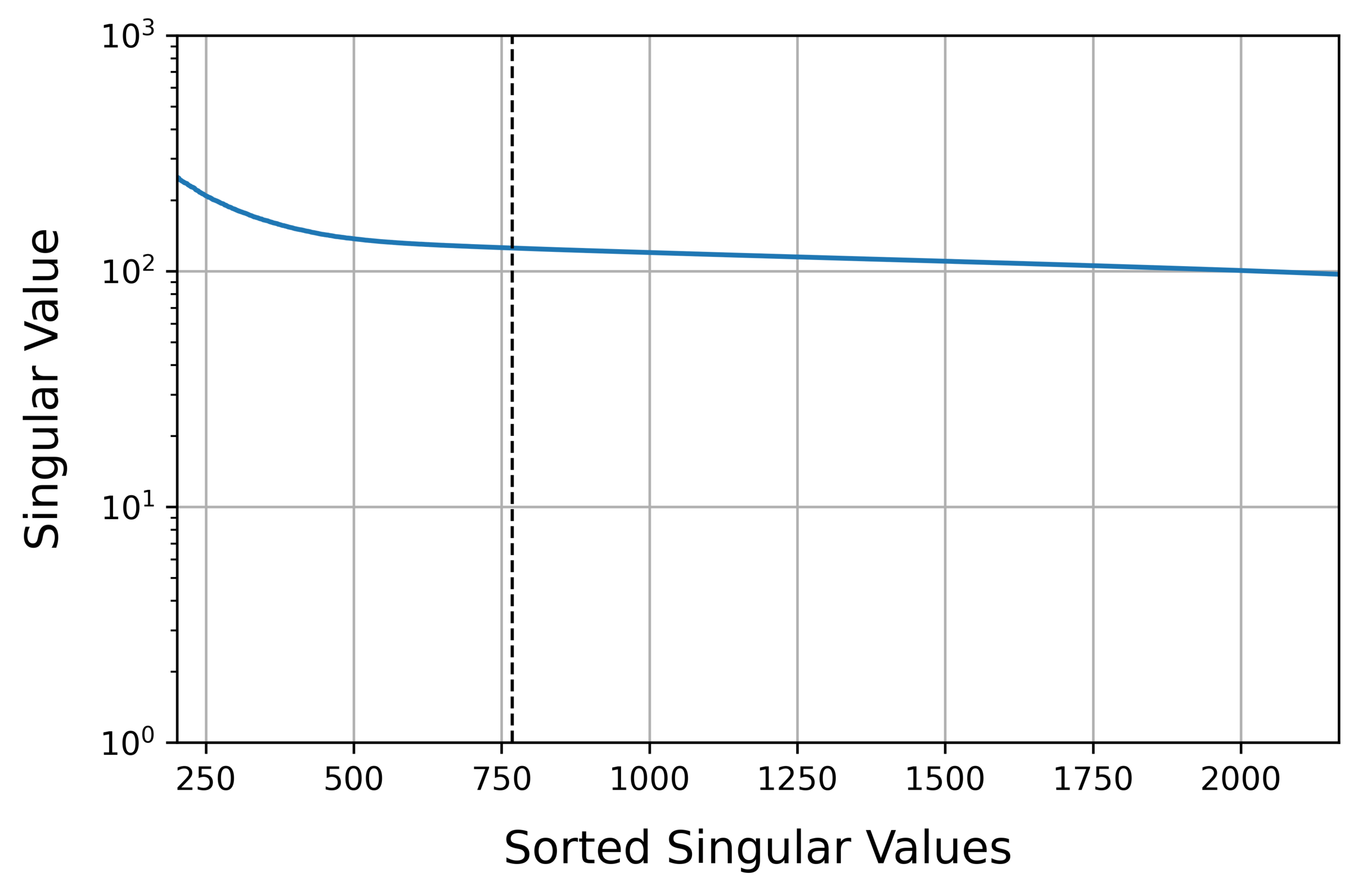}
    \end{subfigure}


    \begin{subfigure}{0.24\textwidth} 
        \centering
        \includegraphics[width=\textwidth]{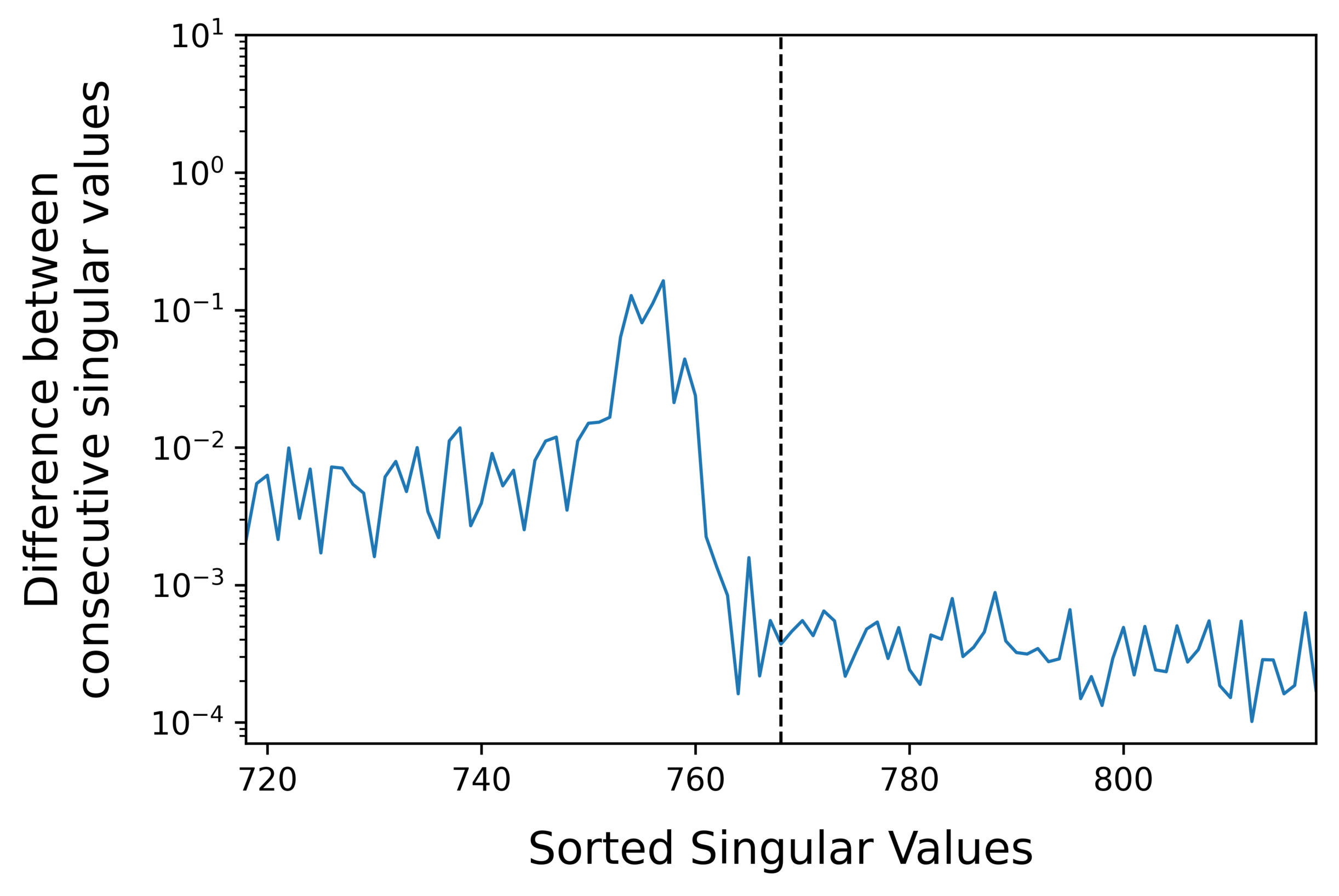}
        \caption{$\text{std}=0.1$, $\gamma=0.25$}
    \end{subfigure}
    \begin{subfigure}{0.24\textwidth}
        \centering
        \includegraphics[width=\textwidth]{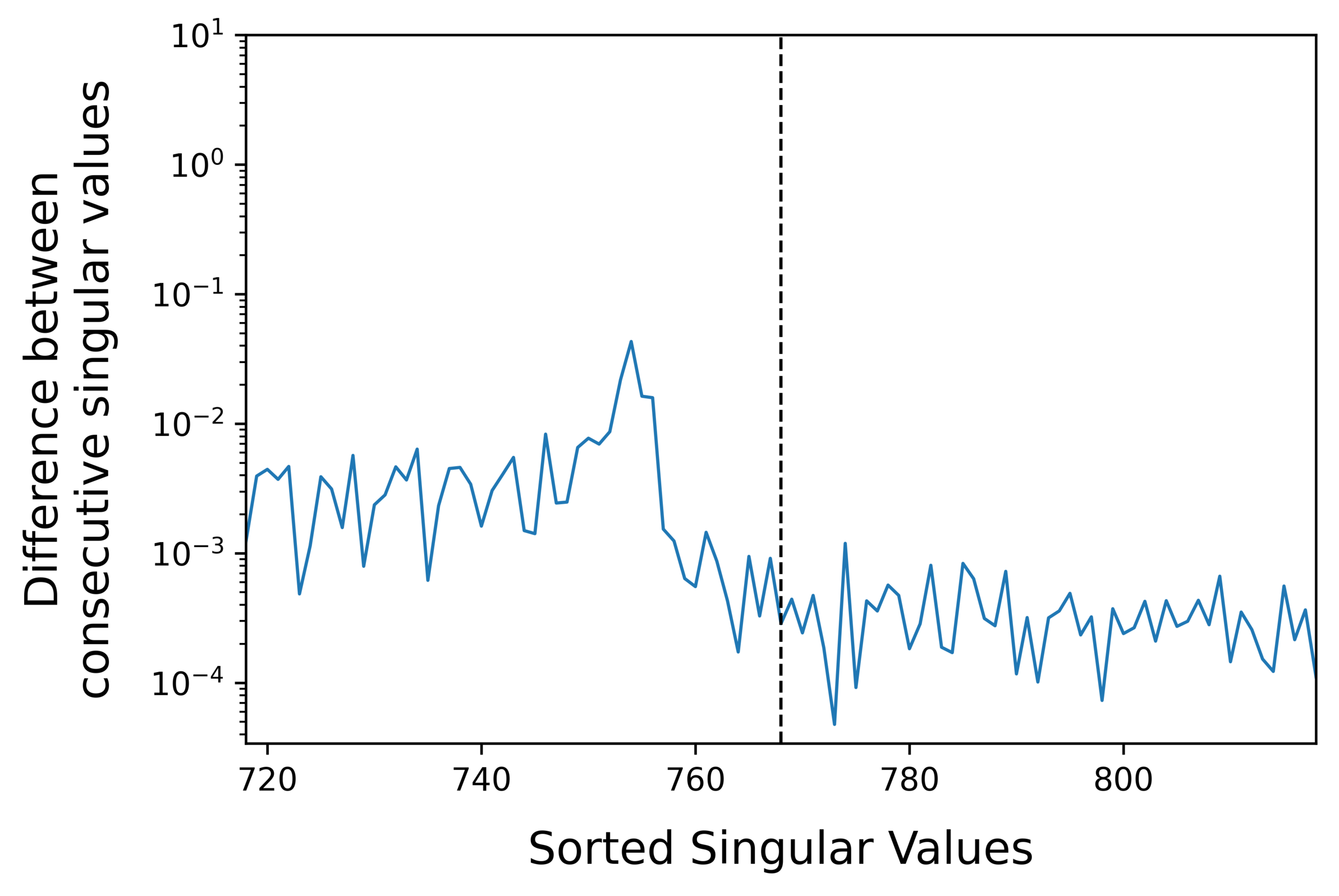}
        \caption{$\text{std}=0.25$, $\gamma=0.25$}
    \end{subfigure}
    \begin{subfigure}{0.24\textwidth}
        \centering
        \includegraphics[width=\textwidth]{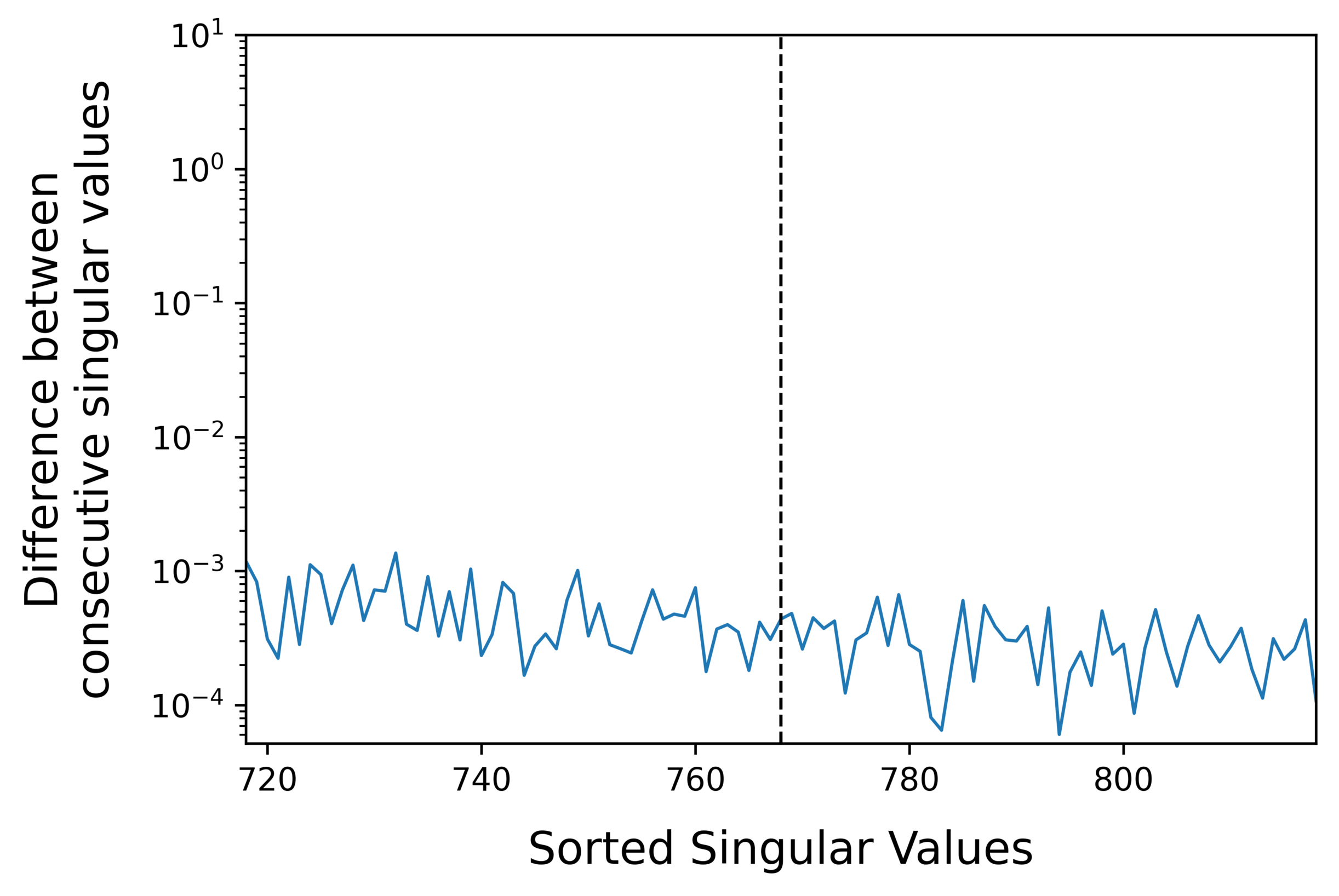}
        \caption{$\text{std}=0.5$, $\gamma=0.25$}
    \end{subfigure}
    \begin{subfigure}{0.24\textwidth}
        \centering
        \includegraphics[width=\textwidth]{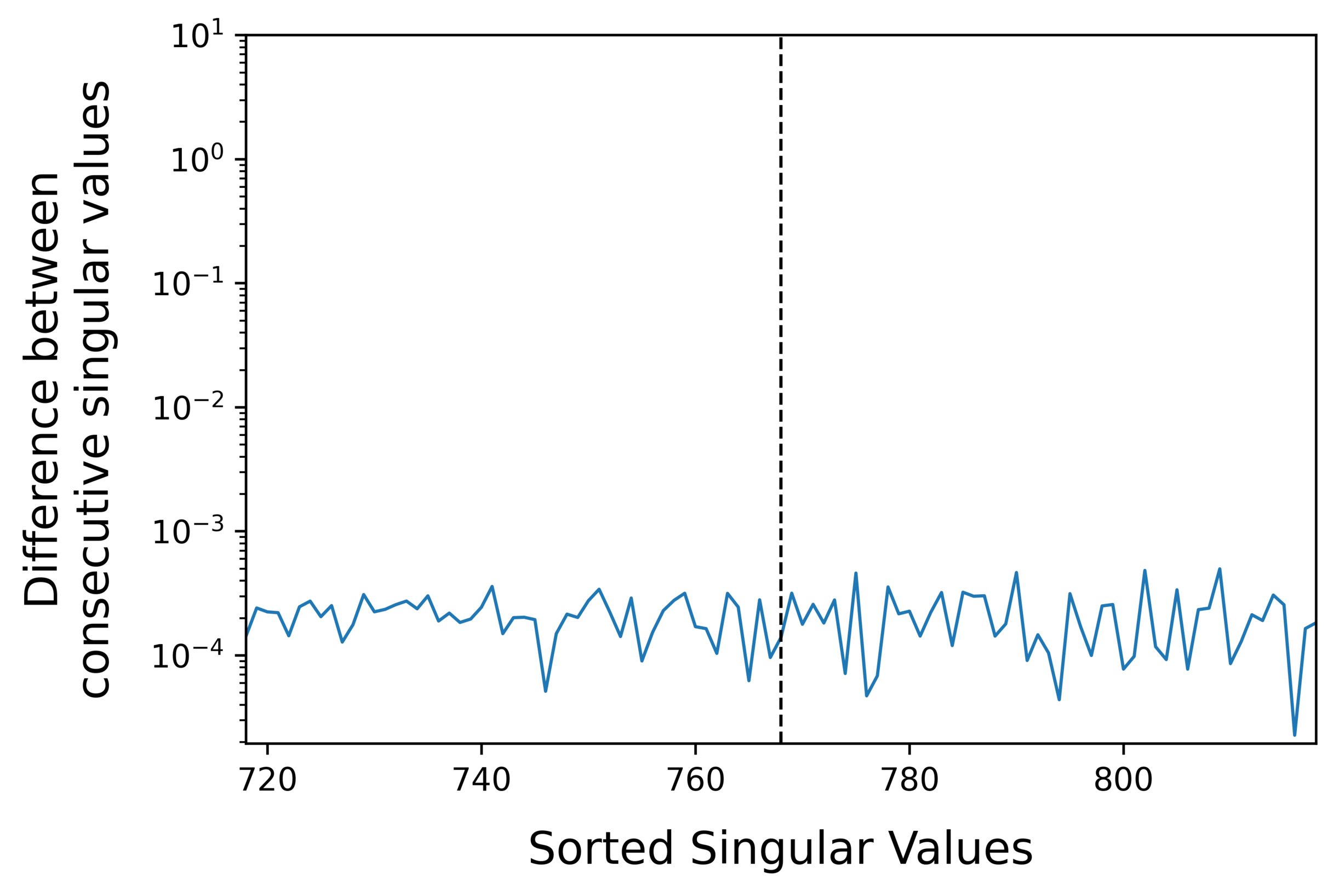}
        \caption{$\text{std}=1.0$, $\gamma=0.25$}
    \end{subfigure}


\vspace{.2in}

    \begin{subfigure}{0.24\textwidth} 
        \centering
        \includegraphics[width=\textwidth]{images/gpt2_sparsity_noise/gpt2_hidden_dim_gamma_0.5_std_0.1_trimmed_values.png}
    \end{subfigure}
    \begin{subfigure}{0.24\textwidth}
        \centering
        \includegraphics[width=\textwidth]{images/gpt2_sparsity_noise/gpt2_hidden_dim_gamma_0.5_std_0.25_trimmed_values.png}
    \end{subfigure}
    \begin{subfigure}{0.24\textwidth}
        \centering
        \includegraphics[width=\textwidth]{images/gpt2_sparsity_noise/gpt2_hidden_dim_gamma_0.5_std_0.5_trimmed_values.png}
    \end{subfigure}
    \begin{subfigure}{0.24\textwidth}
        \centering
        \includegraphics[width=\textwidth]{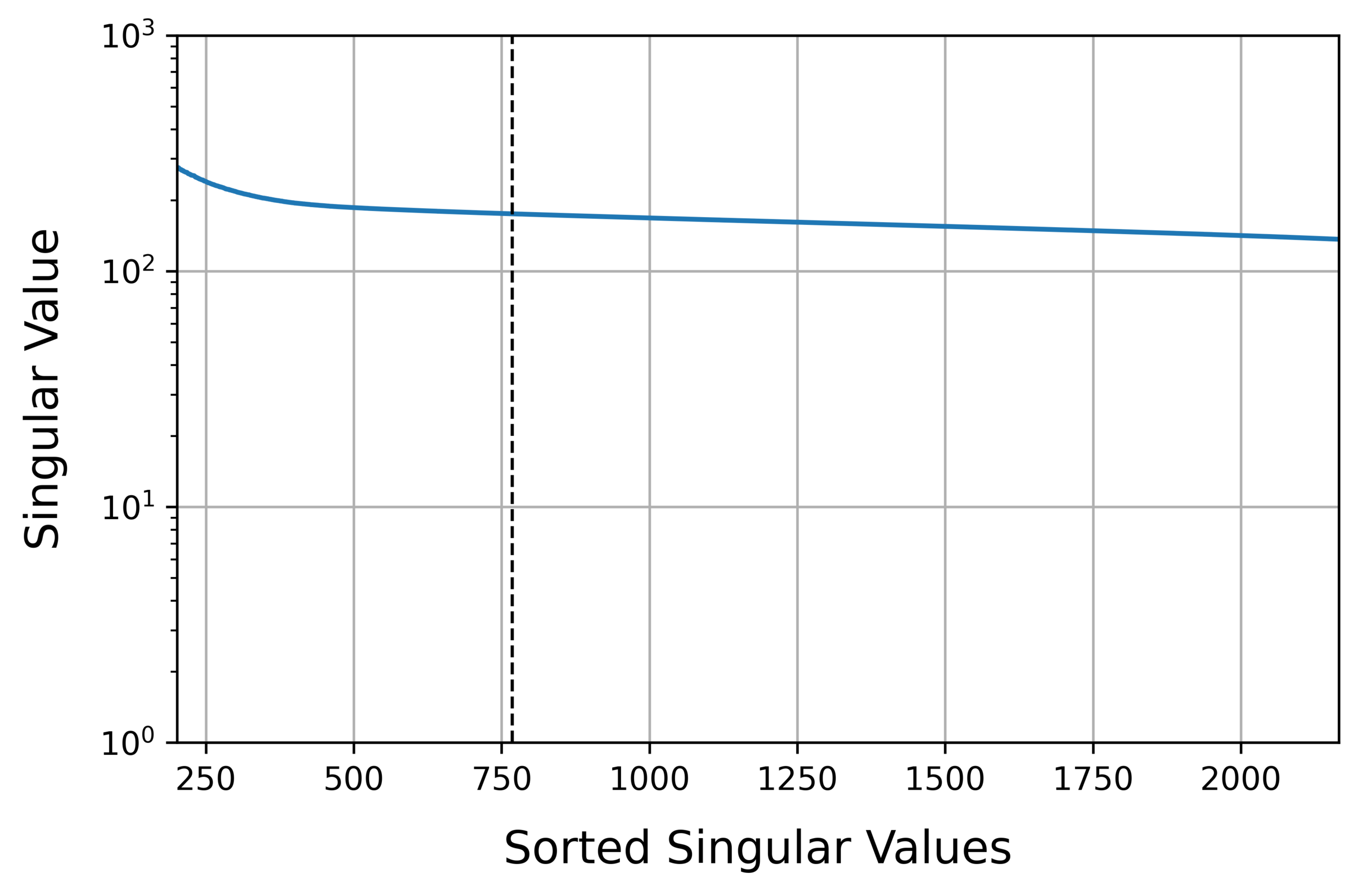}
    \end{subfigure}


    \begin{subfigure}{0.24\textwidth} 
        \centering
        \includegraphics[width=\textwidth]{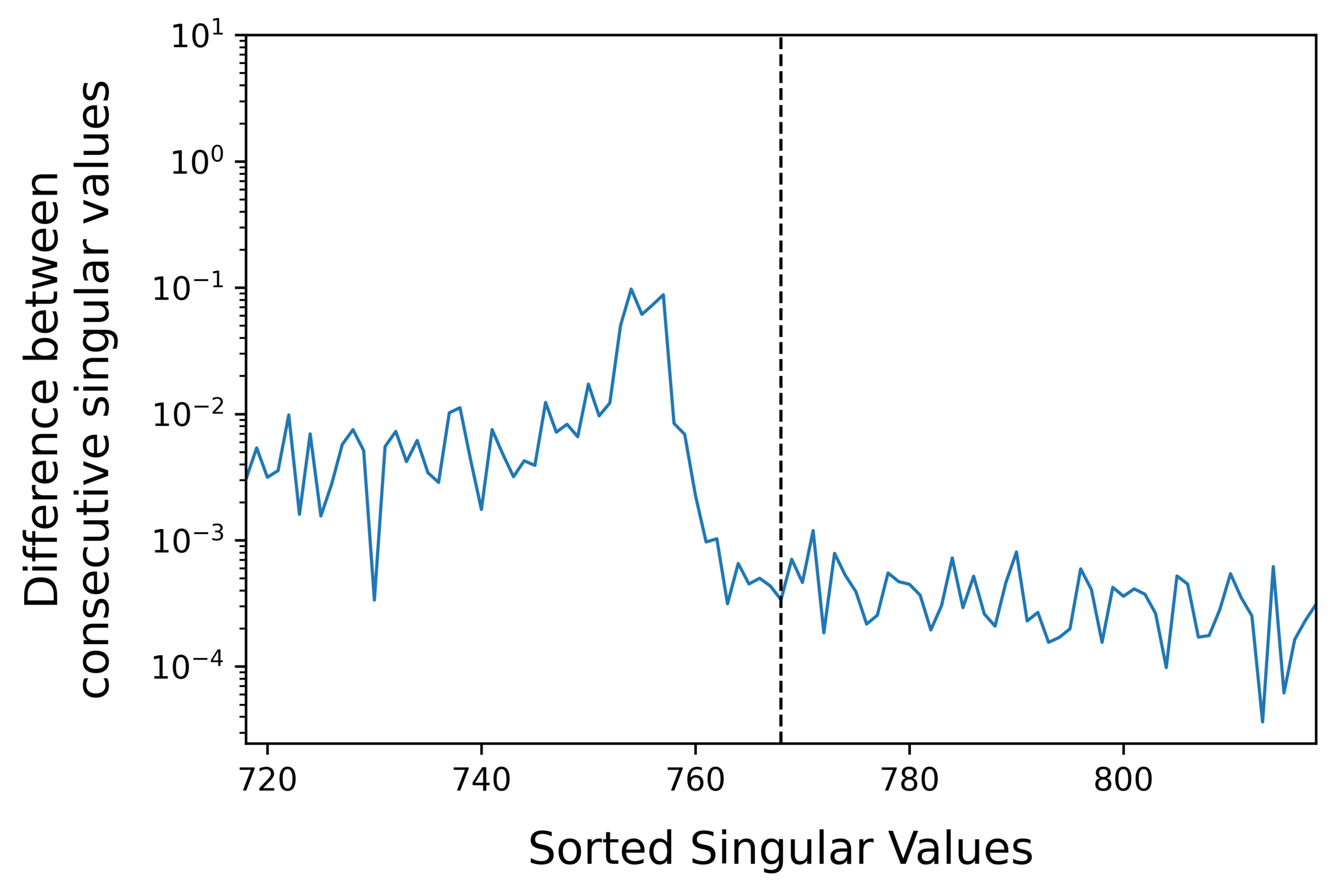}
        \caption{$\text{std}=0.1$, $\gamma=0.5$}
    \end{subfigure}
    \begin{subfigure}{0.24\textwidth}
        \centering
        \includegraphics[width=\textwidth]{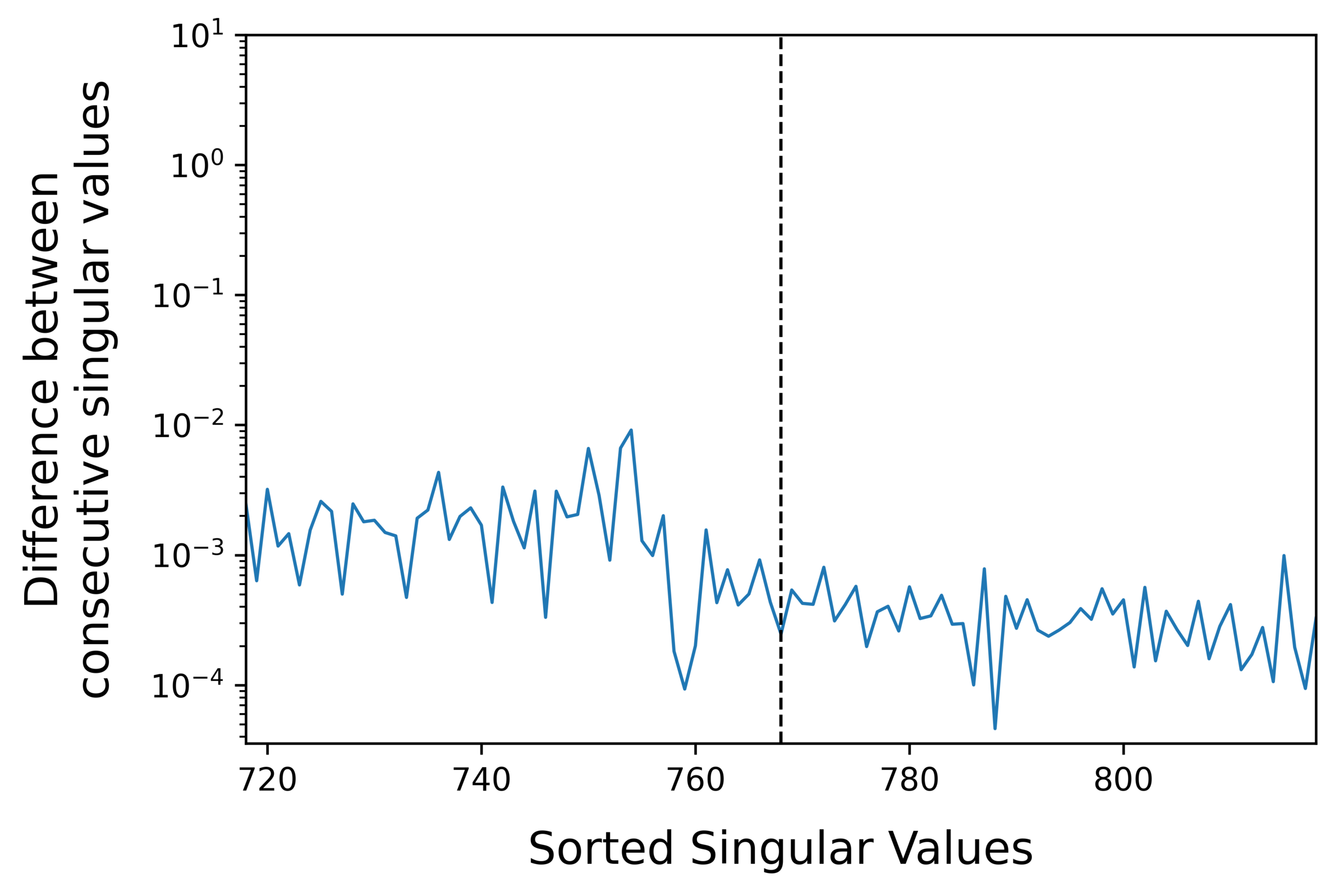}
        \caption{$\text{std}=0.25$, $\gamma=0.5$}
    \end{subfigure}
    \begin{subfigure}{0.24\textwidth}
        \centering
        \includegraphics[width=\textwidth]{images/gpt2_sparsity_noise/gpt2_hidden_dim_gamma_0.5_noise_1.0_windowed.png}
        \caption{$\text{std}=0.5$, $\gamma=0.5$}
    \end{subfigure}
    \begin{subfigure}{0.24\textwidth}
        \centering
        \includegraphics[width=\textwidth]{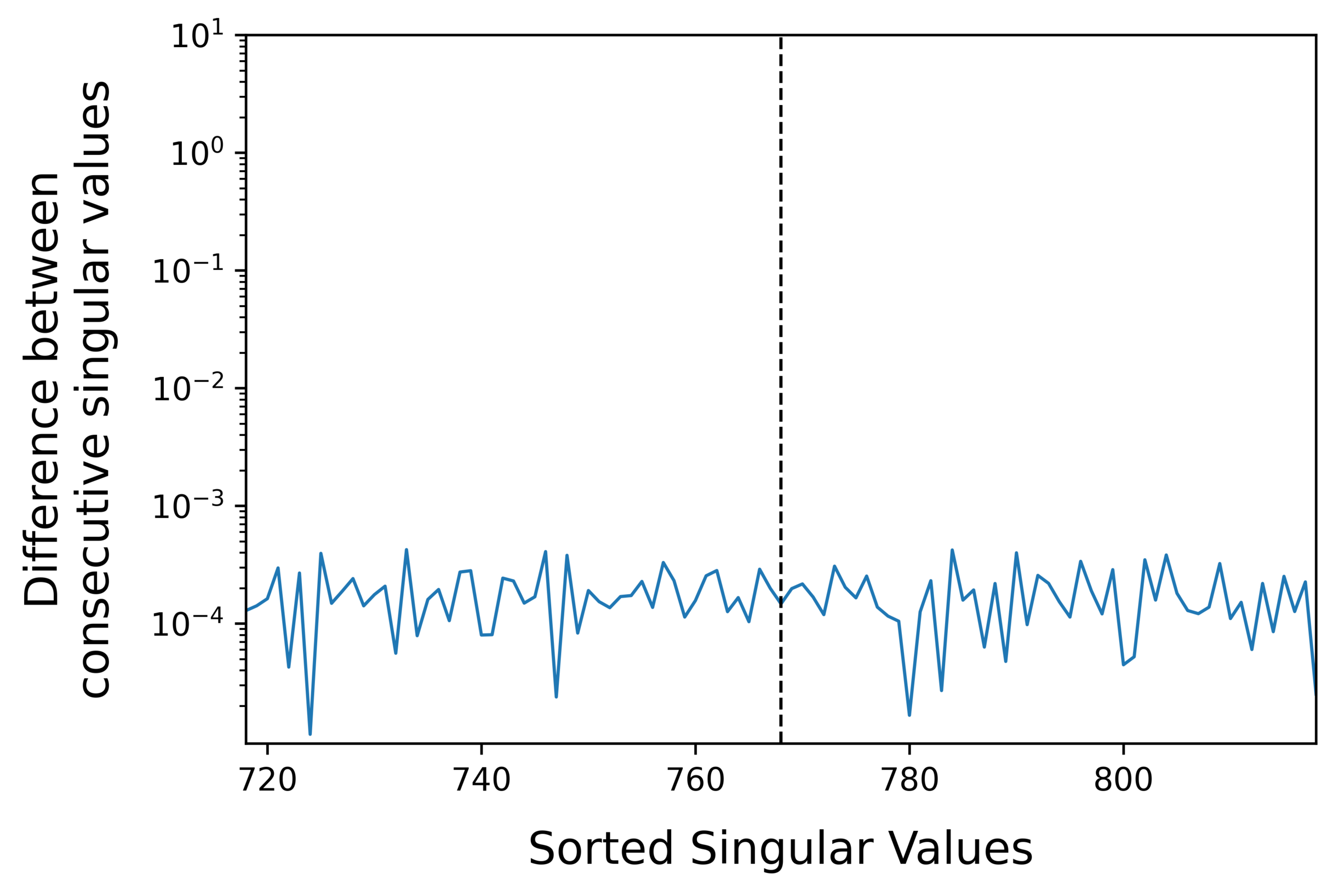}
        \caption{$\text{std}=1.0$, $\gamma=0.5$}
    \end{subfigure}

    \caption{(GPT-2 Gaussian Noise). Dimension extraction attack on GPT-2 by SVD singular values for {\em Gaussian} noise addition with varying sparsity $\gamma \in \{0.1, 0.25, 0.5\}$ and standard deviation $\text{std} \in \{0.1, 0.25, 0.5, 1.0\}$. Each top plot depicts the sorted singular values of the corresponding noisy logits matrix, where a jump at correct position 768 yields a successful dimension extraction. Each bottom plot provides a zoomed in picture of the difference between consecutive singular values; the attack outputs the index for which this difference is maximized.}
    \label{fig:GPT2-many-noise-Gaussian}
\end{figure*}

Figure~\ref{fig:GPT2-many-noise-discrete} provides the results for GPT-2 with {\em discrete} noise for varying sparsity $\gamma \in \{0.1, 0.25, 0.5\}$ and magnitude $\in \{0.1, 0.25, 0.5, 1.0\}$.  Figure~\ref{fig:GPT2-many-noise-Gaussian} provides the comparable results for GPT-2 with {\em Gaussian} noise, for the same sparsity values, and standard deviations $\text{std} \in \{0.1, 0.25, 0.5, 1.0\}$.

Finally, we perform an evaluation of the perplexity of GPT-2 with and without noise perturbation, again using Meta-Llama-3-8B as the larger oracle model. The corresponding perplexity values are given in Table~\ref{tab:gpt2_perplexity_tab}.

\end{document}